\documentclass[fleqn,usenatbib]{mnras}

\usepackage{newtxtext,newtxmath}
\usepackage{float}
\usepackage[T1]{fontenc}
\DeclareRobustCommand{\VAN}[3]{#2}
\let\VANthebibliography\thebibliography
\def\thebibliography{\DeclareRobustCommand{\VAN}[3]{##3}\VANthebibliography}
\usepackage{graphicx}
\usepackage{caption}
\usepackage{tabularx}
\usepackage{amsmath}
\usepackage{scalerel}
\usepackage{natbib}
\usepackage{url}
\usepackage[flushleft]{threeparttable}
\usepackage{balance}

\usepackage{microtype}
\usepackage{media9}
\usepackage{rotating}
\usepackage{microtype}
\usepackage[dvipsnames]{xcolor}
\defcitealias{2019Natur.565..460D}{DHO19}
\defcitealias{2005ApJ...623..398Y}{Y05}

\title[Observability of Recurrent Nova Super-Remnants]{On the Observability of Recurrent Nova Super-Remnants}

\author[M. W. Healy-Kalesh et al.]{M. W. Healy-Kalesh,$^{1}$\thanks{E-mail: M.W.HealyKalesh@ljmu.ac.uk (MWH-K)} M. J.  Darnley,$^{1}$\thanks{E-mail: M.J.Darnley@ljmu.ac.uk (MJD)} {\'E}. J. Harvey,$^{1}$ C. M. Copperwheat,$^{1}$ \newauthor  P. A. James,$^{1}$ T. Andersson,$^{1}$ M. Henze,$^{2}$ and T. J. O'Brien$^{3}$  \\
$^{1}$Astrophysics Research Institute, Liverpool John Moores University, Liverpool, L3 5RF, UK \\
$^{2}$Department of Astronomy, San Diego State University, San Diego, CA 92182, USA \\
$^{3}$Jodrell Bank Centre for Astrophysics, University of Manchester, Manchester, UK
}

\date{Accepted 2023 February 22. Received 2023 February 21; in original form 2022 October 12}

\pubyear{2023}

\begin{document}
\label{firstpage}
\pagerange{\pageref{firstpage}--\pageref{lastpage}}
\maketitle

\begin{abstract}
The nova super-remnant (NSR) surrounding M\,31N 2008-12a (12a), the annually erupting recurrent nova (RN), is the only known example of this phenomenon. As this structure has grown as a result of frequent eruptions from 12a, we might expect to see NSRs around other RNe; this would confirm the RN--NSR association and strengthen the connection between novae and type Ia supernovae (SN Ia) as NSRs centered on SN Ia provide a lasting, unequivocal signpost to the single degenerate progenitor type of that explosion. The only previous NSR simulation used identical eruptions from a static white dwarf (WD). In this Paper, we simulate the growth of NSRs alongside the natural growth/erosion of the central WD, within a range of environments, accretion rates, WD temperatures, and initial WD masses. The subsequent evolving eruptions create dynamic NSRs tens of parsecs in radius comprising a low-density cavity, bordered by a hot ejecta pile-up region, and surrounded by a cool high-density, thin, shell. Higher density environments restrict NSR size, as do higher accretion rates, whereas the WD temperature and initial mass have less impact. NSRs form around growing or eroding WDs, indicating that NSRs also exist around old novae with low-mass WDs. Observables such as X-ray and H$\alpha$ emission from the modelled NSRs are derived to aid searches for more examples; only NSRs around high accretion rate novae will currently be observable. The observed properties of the 12a NSR can be reproduced when considering both the dynamically grown NSR and photoionisation by the nova system.
\end{abstract}


\section{Introduction}\label{Introduction}
Recurrent novae (RNe) are a subclass of the cataclysmic variables that experience repeated thermonuclear eruptions on timescales of a human lifetime. Like classical novae (CNe) -- systems observed in eruption just once -- RNe are interacting binary systems \citep{1954PASP...66..230W,1995CAS....28.....W} containing a white dwarf (WD) and a main-sequence, subgiant, or red giant donor \citep{2012ApJ...746...61D}. Hydrogen-rich material is expelled from the outer layers of the donor through stellar winds or Roche lobe overflow, following which it accumulates on the surface of the WD usually via an accretion disc. At the base of the accreted layer, compression and heating continually increase until the critical pressure for a thermonuclear runaway \citep[TNR;][]{1972ApJ...176..169S,1976IAUS...73..155S,2020ApJ...895...70S} is reached. Once degeneracy is lifted, the accreted envelope is driven upwards by radiation pressure and expands violently, with material travelling faster than the escape velocity of the WD ejected into the surrounding environment as the nova eruption \citep[see, for example,][]{1976IAUS...73..155S,2020ApJ...895...70S}. Mass accretion then continues after \citep[and possibly during;][]{2017ApJ...838..153K,2018ApJ...857...68H} the eruption, leading to successive RN eruptions, separated by a recurrence period ($P_\mathrm{rec}$) which can vary.

Novae with carbon-oxygen WDs present a compelling single degenerate (SD) pathway to type Ia supernovae \citep[SN Ia;][]{1973ApJ...186.1007W,1999ApJ...519..314H,1999ApJ...522..487H,2000ARA&A..38..191H}. Multi-cycle nova simulations (\citealt{2005ApJ...623..398Y}, hereafter \citetalias{2005ApJ...623..398Y}; \citealt{2007ApJ...659L.153H,2015ApJ...808...52K,2015MNRAS.446.1924H,2016ApJ...819..168H,2021gacv.workE..30S}) show that a substantial amount of accreted material is retained on the WD's surface post-eruption, ultimately growing the WD to the \citet{1931ApJ....74...81C} limit (M$_{\text{Ch}}$) in ${\sim}10^{7 - 8}$ years \citep{2016ApJ...819..168H}. The other leading SN Ia pathway is the double degenerate (DD) scenario with two merging WDs \citep{1984ApJ...277..355W,1984ApJS...54..335I} yet within both the SD and DD pathways, novae are the brightest proposed progenitor, even at quiescence \citep{2021gacv.workE..44D}. Therefore, extragalactic nova population studies can link environmental effects such as star formation and metallicity with SN Ia sub-classes. Alternatively, if the donor evolves such that no donatable material remains in the envelope, then the WD will cease growing and thereby never reach the M$_{\text{Ch}}$, resulting in an extinct RN \citep{2021gacv.workE..44D}.

A CN eruption will eject approximately ${\sim}10^{-4}$ $\text{M}_{\odot}$ of material into its surroundings with typical velocities ranging from a few hundred to several thousand km\,s$^{-1}$ \citep{2001IAUS..205..260O}. The interaction of ejecta with different velocities \citep{2020ApJ...905...62A} will shock heat the gas leading to X-ray and radio emission such as that seen in RS\,Ophiuchi \citep{1985MNRAS.217..205B,1992MNRAS.255..683O} and V838 Her \citep{1994MNRAS.271..155O}. This ejected material then goes on to form a nova shell \citep[see, e.g.,][]{2014ASPC..490.....W,2020MNRAS.499.2959H,2020ApJ...892...60S,2022arXiv220213946S}. For the ${\sim}10\%$ of Galactic novae with observed shells \citep[see, e.g.,][]{1990LNP...369..179W,1995MNRAS.276..353S,1998MNRAS.300..221G,2019MNRAS.483.3773S,2022arXiv220213946S}, their morphologies can inform us of the underlying configuration of the binary. In particular, nova shells are structured with an equatorial waist and polar cones of emission \citep{1972MNRAS.158..177H}. This structure forms from the originally near-spherically symmetrical nova ejecta interacting with the material in the orbital plane lost by the donor \citep[see, e.g.,][]{2013ASPC..469..323M}. Polar blobs, equatorial (and/or tropical) rings as well as knots are common to almost all nova shells; see, for example, DQ Her \citep{1978ApJ...224..171W}, HR Del \citep{2003MNRAS.344.1219H}, DO Aql and V4362 Sgr \citep{2020MNRAS.499.2959H} as well as V5668 Sgr \citep{2022MNRAS.511.1591T}. In addition, due to the repeating nature of RNe, we have an example of interacting ejecta from successive eruptions producing clumping and shock heating around the RN T Pyxidis \citep{1997AJ....114..258S,2013ApJ...768...48T}.

Even though the accretion disk surrounding the WD can be altered \citep{2018ApJ...857...68H} to the point of removal in many cases \citep{2010ApJ...720L.195D,2018A&A...613A...8F}, it will re-establish after the nova outburst \citep{2007MNRAS.379.1557W} in preparation for future eruptions. Consequently, all nova systems are predicted to experience repeated outbursts with substantial variation in recurrence period between systems \citepalias{2005ApJ...623..398Y}. Yet, only the recurrence periods for the {\it known} RNe, all contained within the Galaxy \citep[10;][]{2010ApJS..187..275S,2021gacv.workE..44D}, the Large Magellanic Cloud (4) and M31 \citep[19;][]{2020AdSpR..66.1147D}, have been determined, ranging from 98 years \citep{2009AJ....138.1230P} down to 1 year \citep{2015A&A...582L...8H,2018ApJ...857...68H,2020AdSpR..66.1147D}. Such short inter-eruption intervals are powered by a combination of a massive WD and a high mass accretion rate \citep{1988ApJ...325L..35S}.

The most rapidly recurring nova known is M31\,N 2008-12a, or simply `12a', \citep[see, e.g.,][and references therein]{2016ApJ...833..149D,2018ApJ...857...68H,2020AdSpR..66.1147D,2021gacv.workE..44D}. This extreme example erupts annually ($\overline{P}_{\text{rec}} = 0.99 \pm 0.02$ years; \citealt{2020AdSpR..66.1147D}) and has the most massive WD known \citep[$\simeq 1.38\,\mathrm{M}_{\odot};$][]{2015ApJ...808...52K}, likely CO in composition \citep{2017ApJ...847...35D}, accreting with a substantial mass accretion rate of ($0.6 \leq \dot M \leq 1.4) \times 10^{-6}\,\mathrm{M}_{\odot}\,\mathrm{yr}^{-1}$ from a red giant (or clump) companion \citep{2014A&A...563L...9D,2017ApJ...849...96D}.

First associated with 12a by \citet{2015A&A...580A..45D}, this RN is surrounded by a vastly extended nebulosity. Compared to some of the largest Galactic CN shells known such as GK Persei \citep[${\sim}0.5$ pc;][]{2004ApJ...600L..63B,2016A&A...595A..64H}, Z Camelopardalis \citep[${\sim}0.7$ pc;][]{2007Natur.446..159S} and AT Cancri \citep[0.2 pc;][]{2012ApJ...758..121S}, 12a's shell has semi-major and -minor axes of 67 and 45 pc, respectively, justifying a nova super-remnant (NSR; \citealt{2019Natur.565..460D}, hereafter \citetalias{2019Natur.565..460D}) status. \citetalias{2019Natur.565..460D} ruled out the possibility of the shell being a SN remnant, a superbubble or a fossil H {\sc ii} region with H$\alpha +$[N{\sc ii}] imaging and deep low-resolution spectroscopy. Instead, the NSR's existence was attributed to the cumulative sweeping up of ${\sim}10^{5-6} \text{ M}_{\odot}$ \citepalias{2019Natur.565..460D} of local interstellar medium (ISM) from many previous nova eruptions.

To test the viability of a RN origin for 12a's NSR, \citetalias{2019Natur.565..460D} utilised \texttt{Morpheus} \citep{2007ApJ...665..654V} to perform 1D hydrodynamical simulation of $10^5$ 12a eruptions. Each of these eruptions ejected $5 \times 10^{-8} \ \text{M}_{\odot}$ at a terminal velocity of 3000 km s$^{-1}$ over seven days, repeating every 350 days \citepalias{2019Natur.565..460D}. We assign the \citetalias{2019Natur.565..460D} simulation as Run 0 and it will be used as a comparison throughout this work.

Self- and ISM-interaction of the ejecta from each Run 0 eruption formed a huge cavity surrounded by an expanding shell with relative thickness of 22\%. An unavoidable consequence of continual eruptions from a central system is the formation of a dynamical structure, be that a nova shell or larger remnant. However, the existence of a dynamical NSR does not necessarily signify a NSR that is observable. Nevertheless, the simulated dynamic remnant of Run 0 was found to be consistent with observations of the 12a NSR \citepalias{2019Natur.565..460D}.

A unique feature of a structure formed from repeatedly interacting eruptions is a continuously shock-heated region located inside the outer shell \citepalias{2019Natur.565..460D}. Extrapolating the growth rate from these simulations to the observed size of the super-remnant, \citetalias{2019Natur.565..460D} suggested an age of $6 \times 10^6$ yrs. Importantly, the mechanism driving the NSR formation is also growing the 12a CO WD, which \citet{2017ApJ...849...96D} predict will surpass the Chandrasekhar limit and explode as a SN Ia in $<$ 20,000 years.

In this paper, we build upon the NSR hydrodynamic modelling presented by \citetalias{2019Natur.565..460D} through consideration of the complete eruption history of a nova system as the WD mass grows from its formation toward the Chandrasekhar mass. We also explore a number of factors, both intrinsic and extrinsic to the nova system that might impact NSR formation, to aid the search for more NSRs. This will be the first attempt to determine if the NSR associated with 12a is unique or whether it is simply the first of the phenomena to be found.

In Section~\ref{Generating Nova Ejecta Properties} we describe the eruption model used to generate input parameters. We describe the \texttt{Morpheus} hydrodynamic code employed in this paper in Section~\ref{Hydrodynamical Simulations} before outlining each of the separate runs of our main simulations. Various tests conducted after the main simulations are presented in Section~\ref{Additional tests}. We explore the observability of NSRs in Section~\ref{Observational predictions} by modelling emission from the simulations and then compare our simulations to observations of the 12a NSR in Section~\ref{Comparing simulations and observations}, before concluding our paper in Section~\ref{Conclusions}.

\section{Generating Nova Ejecta Properties}\label{Generating Nova Ejecta Properties}

The \citetalias{2019Natur.565..460D} simulations of the 12a NSR utilised $10^5$ identical eruptions with a fixed recurrence period. While a good approximation for this system during its recent evolution, identical eruptions do not match the expected long term evolution of such a system, whereby the characteristics of the ejecta evolve with the changing WD mass. Therefore, to obtain the properties of a nova system with incrementally changing nova eruptions, we were required to grow a WD (see Section~\ref{Growing a white dwarf}). We will only describe the model we used to grow the WD for a `reference simulation' as an illustration, however this model was utilised for each of the different WD temperatures and accretion rates. As a reference simulation corresponding to the 12a system, we chose to grow a $10^7$ K WD with $\dot{M}=10^{-7} \text{ M}_{\odot} \, \text{yr}^{-1}$ (see Section~\ref{Parameter Space} for details), which we then placed within an environment with a hydrogen-only ISM density of $1.67 \times 10^{-24} \ \text{g cm}^{-3}$ (1 H atom per cubic centimetre). We refer to this ISM density throughout the paper by the number density $n = 1\,\text{cm}^{-3}$ (but drop the units for clarity).

\subsection{Parameter space}\label{Parameter Space}

\citetalias{2005ApJ...623..398Y} provides a parameter space for the characteristics of a nova envelope and the outburst characteristics for an extended grid of nova models with varying WD mass, temperature and accretion rate. This grid runs through all permutations of these parameters and outputs various eruption characteristics such as the mass accreted onto the WD which ignites during the TNR ($m_{\text{acc}}$), the mass ejected from the WD during the nova eruption ($m_{\text{ej}}$) and the duration of the mass-loss phase ($t_{\text{ml}}$) i.e., the timescale of each eruption.

\begin{figure*}
\centering
\includegraphics[width=\textwidth]{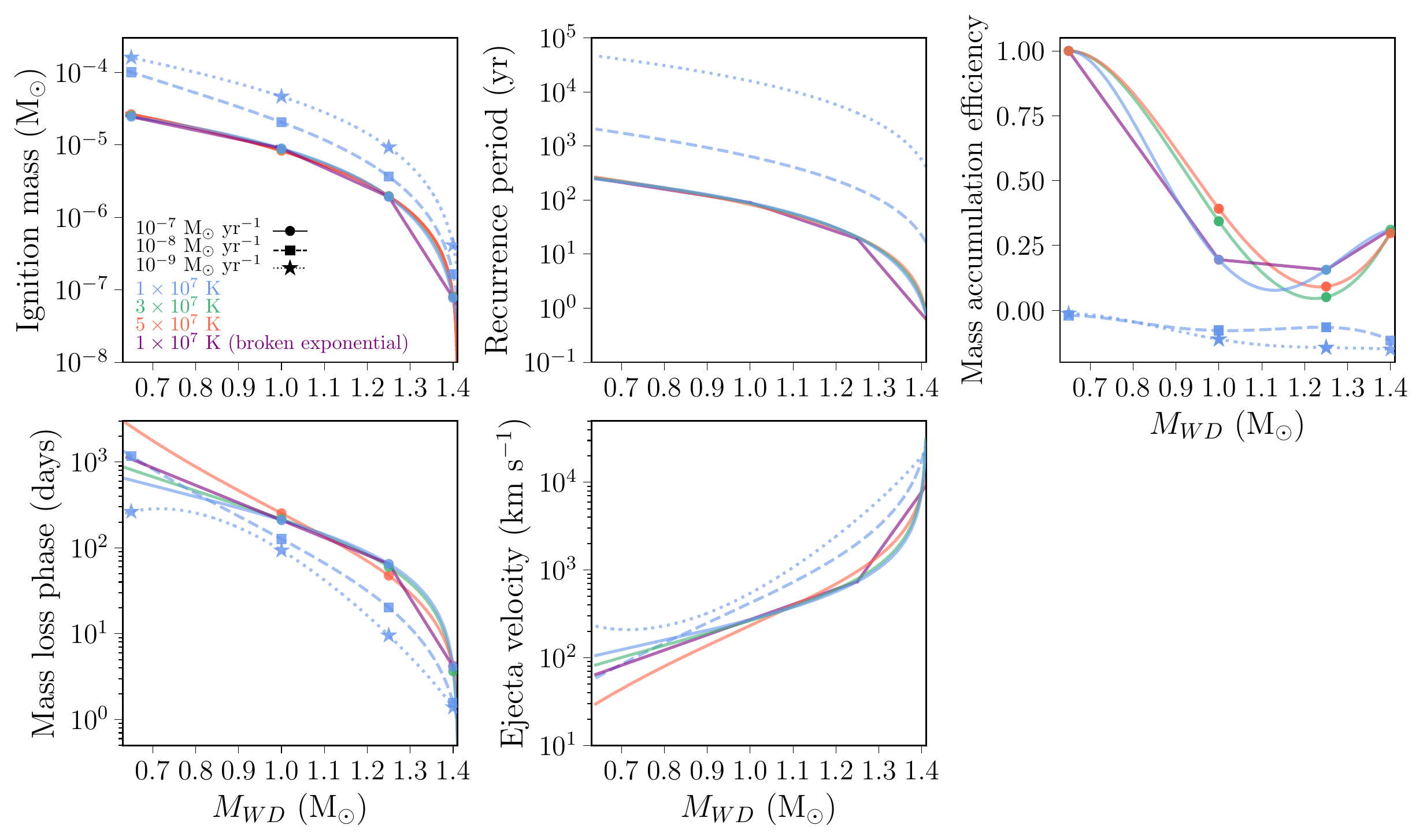}
\caption{Top left: Ignition mass ($m_{\text{ig}}$) as a function of WD mass ($M_{\text{WD}}$) derived from fitting to the output characteristics for $m_{\text{acc}}$ (circles, squares and stars) from \citetalias{2005ApJ...623..398Y}. Top middle: Recurrence period ($P_{\text{rec}}$) as a function of WD mass found by dividing the ignition mass in the $m_{\text{ig}}-M_{\text{WD}}$ relation (top left panel) by $\dot{M} = 10^{-7} \ \text{M}_{\odot} \ \text{yr}^{-1}$. Top right: Mass accumulation efficiency ($\eta$) as a function of WD mass derived from fitting to the output characteristics for $(m_{\text{ig}} - m_{\text{ej}}) / m_{\text{ig}}$ (circles, squares and stars) from \citetalias{2005ApJ...623..398Y}. We set $\eta = 1$ for all $1 \times 10^{-7}$ points at $M_{\text{WD}} = 0.65 \, \text{M}_{\odot}$ as \citetalias{2005ApJ...623..398Y} indicated that there were no eruptions (no mass ejected) for these models. Bottom left: Mass loss phase ($t_{\text{ml}}$) as a function of WD mass derived from fitting to the output characteristics for $t_{\text{ml}}$ (circles, squares and stars) from \citetalias{2005ApJ...623..398Y}. Bottom middle: Terminal ejecta velocity ($v_{\text{ej}}$) as a function of WD mass derived from relations presented in \citet{1995CAS....28.....W} and \citet{2014A&A...563A...2H} to the $t_{\text{ml}}-M_{\text{WD}}$ relation (bottom left panel). Purple lines indicate broken exponential (or linear for $\eta$) fits to the data as described in Section~\ref{Fitting a broken exponential to find system parameters}.}
\label{Five Relationships}
\end{figure*}

For this study, we use values of $m_{\text{acc}}$, which we equate to the ignition mass ($m_{\text{ig}}$), $m_{\text{ej}}$, and $t_{\text{ml}}$ for WDs with masses 0.65, 1.0, 1.25 and 1.4\,$\text{M}_{\odot}$\footnote{These WD masses were chosen from the set of WD masses given in \citetalias{2005ApJ...623..398Y}, as were the WD temperatures and accretion rates used in our study. We were limited to these accretion rates by the eruption models of \citetalias{2005ApJ...623..398Y} whereby no $t_{\text{ml}}$ is provided for $\dot{M}=10^{-6} \, \mathrm{M}_{\odot} \, \mathrm{yr}^{-1}$.}, three temperatures of 10\,MK, 30\,MK and 50,MK, and three accretion rates of $10^{-7} \text{ M}_{\odot} \, \text{yr}^{-1}$, $10^{-8} \text{ M}_{\odot} \, \text{yr}^{-1}$ and $10^{-9} \text{ M}_{\odot} \, \text{yr}^{-1}$ (we consider three temperatures for $\dot{M}=10^{-7} \text{ M}_{\odot} \, \text{yr}^{-1}$ but only 10\,MK for the other accretion rate values). To interpolate and extrapolate these points for a continuous set of values for our WD growth model, we required a function that evolved smoothly, behaved as a power law for lower masses, yet which became asymptotic as the Chandrasekhar mass was approached (see Section~\ref{Fitting a broken exponential to find system parameters} for an alternative approach). The functions we fit to $m_{\text{ig}}$ and $t_{\text{ml}}$ are shown in Figure~\ref{Five Relationships}, as well as the continuous function for $P_{\text{rec}}$ (the ratio of $m_\mathrm{ig}$ and accretion rate). As we also wish to be consistent with observed characteristics of the nova eruption, we utilised observationally determined relations from \citet{1995CAS....28.....W} and \citet{2014A&A...563A...2H} to determine a function for the terminal ejecta velocity of the outburst.

\subsection{Growing a white dwarf}\label{Growing a white dwarf}
We grew a $1\,\text{M}_{\odot}$ WD to a $\text{M}_{\text{Ch}}$ WD by accumulating the retained mass from iterated nova eruptions and using the interpolated relationships given in Section~\ref{Parameter Space} to obtain properties of each eruption. For this example, a $1 \text{ M}_{\odot}$ WD with a temperature of 10\,MK experiences approximately 1,900,000 eruptions while growing from $1 \text{ M}_{\odot}$ to $1.4 \text{ M}_{\odot}$, reaching a recurrence period lower limit of ${\sim}282$ days. This WD mass upper limit of $1.4 \text{ M}_{\odot}$ is assumed for all WD scenarios, which we equate to the Chandrasekhar mass ($\mathrm{M}_\mathrm{Ch}$) for this study.

A WD is grown (or eroded) according to the amount of accreted material retained (or removed) between eruptions. To model the evolution of the mass accumulation efficiency ($\eta$) over the evolution of a WD, we utilised the values of $m_{\text{ig}}$ and $m_{\text{ej}}$ from \citetalias{2005ApJ...623..398Y} such that $\eta = (m_{\text{ig}} - m_{\text{ej}}) / m_{\text{ig}}$ and interpolated between these points for a continuous set of values (see top right panel of Figure~\ref{Five Relationships}). The changing mass of the WD can thus be described as:

\begin{equation}
M_{\text{WD},  i + 1} = M_{\text{WD}, i} + \left(m_{\text{ig}, i} \times \eta_{i}\right),
\label{grow WD}
\end{equation}

\noindent where $M_{\text{WD}, i}$ is the pre-eruption mass of the WD, $m_{\text{ig}, i}$ is the mass accreted by the WD before the eruption, $\eta_{i}$ is the evolving mass accumulation efficiency, and $M_{\text{WD}, i + 1}$ is the post-eruption mass of the WD. With the initial WD mass being $1\,\text{ M}_{\odot}$, we utilised the relationships found in Section~\ref{Parameter Space} to give the associated $m_{\text{ig}}$ value for equation~\ref{grow WD}. The post-eruption mass was then used as the $M_{\text{WD}}$ value in the next iteration and we continued this until we reached the limiting mass stated previously. We used the output parameters from this iterative model in our simulations. With each iteration, we were also able to use the relationships found in Section~\ref{Parameter Space} to illustrate the evolution of a number of parameters including ejecta kinetic energy and momentum in terms of WD mass, recurrence period, elapsed time (from the first eruption), and the number of eruptions.

Utilising the WD growth model, we generated nova ejecta with incrementally changing properties. As the mass of the WD increases, eruptions become more frequent, and ejecta become less massive but with higher velocity in response to the increasing WD surface gravity.

\section{Hydrodynamical Simulations}\label{Hydrodynamical Simulations}

As the net mass loss rate from the WD varies as the WD mass grows, an analytic relation for the growth of the NSR shell cannot be derived. As such, full hydrodynamic simulations are a necessity if we are to understand the evolution of NSRs and their emission characteristics.

As in \citetalias{2019Natur.565..460D}, the hydrodynamical simulations in this work were performed with \texttt{Morpheus} \citep{2007ApJ...665..654V} -- developed by the Nova Groups from the University of Manchester and Liverpool John Moores University. \texttt{Morpheus} brings together one-dimensional \citep[Asphere; see][]{2007ApJ...665..654V}, two-dimensional \citep[Novarot; see][]{1997MNRAS.284..137L} and three-dimensional \citep[CubeMPI; see][]{2006MNRAS.366..387W} codes to form an MPI-OpenMP Eulerian second-order Godunov simulation code that functions with Cartesian, spherical or cylindrical coordinates, and includes radiative cooling and gravity.

The configuration of the nova systems in this work are modelled in the same manner as given in \citetalias{2019Natur.565..460D} such that the mass donor is a red giant exhibiting a continuous wind mass loss rate (after accretion) of $2.6 \times 10^{-8} \text{ M}_{\odot} $yr$^{-1}$ with a terminal velocity of 20 km\,s$^{-1}$. These values are assumed to be consistent with the donor in the RS Ophuichi system \citep{1985MNRAS.217..205B}, thus are used as representative values with the red giant wind having negligible influence on the NSR evolution. The nova eruption is represented by an instantaneous increase in mass loss and ejecta velocity (the red giant wind's contribution becomes negligible here) followed by a quiescent period in which only the red giant wind (with decreased mass loss and lower ejecta velocity) is present. Furthermore, unless otherwise stated, each ejection is modelled as a wind with a mass-loss rate and velocity that incrementally increase throughout the simulation as governed by the relationships determined from \citetalias{2005ApJ...623..398Y} models (see Section~\ref{Generating Nova Ejecta Properties} for details and Figure~\ref{Five Relationships}). The eruptions are separated by incrementally decreasing recurrence periods also governed by the aforementioned relationships. True nova ejecta are not spherically symmetric, however largely for computational reasons, we have assumed one-dimensional spherical symmetry for these simulations, effectively modelling the bulk equatorial ejecta \citep[see, e.g.,][]{2013ASPC..469..323M}. The spatial resolution of the full simulations ($\geq$200 AU/cell) is larger than the expected orbital separation of the WD and the donor \citep[for example, the orbital separation for 12a is ${\sim}1.6$ AU;][]{2018ApJ...857...68H} so we assume that both are located at the origin. Therefore, interaction between the ejecta and the donor or accretion disk is ignored.

Ideally, we would want to run each complete simulation at a  high spatial resolution, however, this is not feasible with temporal and computing constraints. Running the reference simulation (see Section~\ref{Reference simulation}) several times with varying spatial resolution (and varying number of eruptions), we found that running its full 1,900,750 eruptions at 200 AU/cell would have the same long term structure as a simulation with resolution of 1 AU/cell (the resolution of a test run with 100 eruptions). Consequently, we set a spatial resolution of 200 AU/cell for most of our simulations, while those with lower spatial resolution (as indicted in Table~\ref{Runs}) are set in response to the infrequency of eruptions, and therefore lessened impact on resolving the gross NSR structure, within those particular runs.

\subsection{Incorporating radiative cooling}\label{Incorporating radiative cooling}

Nova ejecta lose energy through radiative cooling, which affects the evolution of any NSR. Therefore, the effects of cooling were tested in \citetalias{2019Natur.565..460D}, with a NSR grown from $10^3$ eruptions with the inclusion of the radiative cooling module in \texttt{Morpheus}. 
The cooling model utilised in \texttt{Morpheus} was taken from \citet[][their Figure 1]{1976ApJ...204..290R}. The cooling rate is given as a function of gas temperature of an optically thin plasma, with no dust or molecules, made up of H, He, C, N, O, Ne, Mg, Si, S, Ca, Fe and Ni. Radiative cooling becomes ineffective below a temperature of $10^4$\,K. Above $10^8$ K, the gas is ionised and only radiates through free-free Bremsstrahlung \citep{2007ApJ...665..654V}. Between these limits, cooling is dominated by line-cooling from the metals within the gas \citep{Vaytet_Thesis}.

\citetalias{2019Natur.565..460D} demonstrated that there was no significant difference between the Run 0 NSR structure with or without cooling (see their Extended Data Figure 4). Cooling was suppressed in the Run 0 NSR as the recurrence period was much shorter than the cooling timescale. Hence, radiative cooling in the full simulation of Run 0 was not included.

In all cases, the NSR evolution presented in this work begins with high mass and low velocity ejecta (see Section~\ref{Generating Nova Ejecta Properties}) leading to less energetic eruptions and, crucially, with long gaps between consecutive eruptions. Therefore, at early times, the recurrence period will be longer than the cooling timescale and, as such, we incorporate radiative cooling in all simulations.

\subsection{Reference simulation --- Run 1}\label{Reference simulation}

Our reference simulation, Run 1, models nova eruptions from a growing WD with a temperature $T_\mathrm{WD}=10^7$ K, with $\dot{M}=10^{-7} \, \text{M}_{\odot} \, \text{yr}^{-1}$, and within a low density ISM ($n = 1$). With the varying mass accumulation efficiency, it would take ${\sim}$31\,Myr (1,900,750 eruptions) for this WD  to grow from $1 \text{ M}_{\odot}$ to $\text{M}_\mathrm{Ch}$. Run 1 has a spatial resolution of 200\,AU. This information, including the total kinetic energy released, is summarised in Table~\ref{Runs} for all simulations in this paper.

\begin{table*}
\caption[Each run with associated parameters]{Parameters for each run. Columns record the simulation number, initial WD mass, WD temperature, accretion rate, ISM density, spatial resolution, number of eruptions to grow the WD to $\text{M}_\mathrm{Ch}$ or for the simulation to reach the temporal upper limit of $10^8$ years, the cumulative time of the simulation, and the total kinetic energy released. Run 0 relates to the $10^5$ identical eruptions as modelled by \citetalias{2019Natur.565..460D}. Ejecta characteristics for Run 1$^\dag$ used a broken exponential/linear interpolation (see Section~\ref{Fitting a broken exponential to find system parameters}). Runs $1^\star$, $2^\star$, $5^\star$ and $7^\star$ have the same ejecta characteristics as Runs 1, 2, 5 and 7, respectively, but do not include radiative cooling. Run 22 contains the same nova system as Run 1 but tuned with an ISM density of $n=1.278$ to match the ISM predicted in Section~\ref{Photoionisation region?} for the reference simulation WD to grow a NSR to the size (67 pc) of the observed NSR around M\,31N 2008-12a.}
\label{Runs}
\begin{center}
\begin{tabular}{ccccccccc}
\hline
Run \# & $\text{M}_{\text{WD}}$ & $T_{\text{WD}}$ & $\dot M$ & ISM density & Spatial resolution & Number & Cumulative time & Total Kinetic Energy\\
 & ($\text{M}_{\odot}$) & (K) & ($\text{M}_{\odot} \text{yr}^{-1}$) & ($1.67 \times 10^{-24} \text{ g cm}^{-3}$) & (AU/cell) & of eruptions & (years) & (erg) \\
\hline
0 & \textit{n/a} & \textit{n/a} & $1.6 \times 10^{-7}$ & 1 & \phantom{000}4 & \phantom{0,}100,000 & $1.0 \times 10^5$ & $4.5 \times 10^{47}$ \\
1 & 1 & $1 \times 10^7$ & $1 \times 10^{-7}$ & 1 & \phantom{0}200 & 1,900,750 & $3.1 \times 10^7$ & $2.4 \times 10^{49}$ \\
2 & 1 & $1 \times 10^7$ & $1 \times 10^{-7}$ & 0.1 & \phantom{0}200 & 1,900,750 & $3.1 \times 10^7$ & $2.4 \times 10^{49}$ \\
3 & 1 & $1 \times 10^7$ & $1 \times 10^{-7}$ & 0.316 & \phantom{0}200 & 1,900,750 & $3.1 \times 10^7$ & $2.4 \times 10^{49}$ \\
4 & 1 & $1 \times 10^7$ & $1 \times 10^{-7}$ & 3.16 & \phantom{0}200 & 1,900,750 & $3.1 \times 10^7$ & $2.4 \times 10^{49}$ \\
5 & 1 & $1 \times 10^7$ & $1 \times 10^{-7}$ & 10 & \phantom{0}200 & 1,900,750 & $3.1 \times 10^7$ & $2.4 \times 10^{49}$ \\
6 & 1 & $1 \times 10^7$ & $1 \times 10^{-7}$ & 31.6 & \phantom{0}200 & 1,900,750 & $3.1 \times 10^7$ & $2.4 \times 10^{49}$ \\
7 & 1 & $1 \times 10^7$ & $1 \times 10^{-7}$ & 100 & \phantom{0}200 & 1,900,750 & $3.1 \times 10^7$ & $2.4 \times 10^{49}$ \\
8 & 1 & $1 \times 10^7$ & $1 \times 10^{-8}$ & 1 & \phantom{0}200 & \phantom{0,0}40,343 & $1.0 \times 10^8$ & $1.3 \times 10^{48}$ \\
9 & 1 & $1 \times 10^7$ & $1 \times 10^{-8}$ & 10 & \phantom{0}200 & \phantom{0,0}40,343 & $1.0 \times 10^8$ & $1.3 \times 10^{48}$ \\
10 & 1 & $1 \times 10^7$ & $1 \times 10^{-8}$ & 100 & \phantom{0}200 & \phantom{0,0}40,343 & $1.0 \times 10^8$ & $1.3 \times 10^{48}$ \\
11 & 1 & $1 \times 10^7$ & $1 \times 10^{-9}$ & 1 & \phantom{0}400 & \phantom{0,00}2,094 & $1.0 \times 10^8$ & $3.1 \times 10^{47}$ \\
12 & 1 & $1 \times 10^7$ & $1 \times 10^{-9}$ & 10 & \phantom{0}400 & \phantom{0,00}2,094 & $1.0 \times 10^8$ & $3.1 \times 10^{47}$ \\
13 & 1 & $1 \times 10^7$ & $1 \times 10^{-9}$ & 100 & 4000 & \phantom{0,00}2,094 & $1.0 \times 10^8$ & $3.1 \times 10^{47}$ \\
14 & 1 & $3 \times 10^7$ & $1 \times 10^{-7}$ & 1 & \phantom{0}200 & 2,770,545 & $4.1 \times 10^7$ & $4.6 \times 10^{49}$ \\
15 & 1 & $5 \times 10^7$ & $1 \times 10^{-7}$ & 1 & \phantom{0}200 & 2,029,154 & $2.7 \times 10^7$ & $5.0 \times 10^{49}$ \\
16 & 0.65 & $1 \times 10^7$ & $1 \times 10^{-7}$ & 1 & \phantom{0}200 & 1,953,955 & $3.7 \times 10^7$ & $2.5 \times 10^{49}$ \\
17 & 0.8 & $1 \times 10^7$ & $1 \times 10^{-7}$ & 1 & \phantom{0}200 & 1,945,717 & $3.6 \times 10^7$ & $2.5 \times 10^{49}$ \\
18 & 0.9 & $1 \times 10^7$ & $1 \times 10^{-7}$ & 1 & \phantom{0}200 & 1,933,696 & $3.4 \times 10^7$ & $2.5 \times 10^{49}$ \\
19 & 1.1 & $1 \times 10^7$ & $1 \times 10^{-7}$ & 1 & \phantom{0}200 & 1,779,622 & $2.2 \times 10^7$ & $2.4 \times 10^{49}$ \\
20 & 1.2 & $1 \times 10^7$ & $1 \times 10^{-7}$ & 1 & \phantom{0}200 & 1,494,979 & $1.0 \times 10^7$ & $2.1 \times 10^{49}$ \\
21 & 1.3 & $1 \times 10^7$ & $1 \times 10^{-7}$ & 1 & \phantom{0}200 & 1,149,284 & $3.7 \times 10^6$ & $1.8 \times 10^{49}$ \\
22 & 1 & $1 \times 10^7$ & $1 \times 10^{-7}$ & 1.278 & \phantom{0}200 & 1,900,750 & $3.1 \times 10^7$ & $2.4 \times 10^{49}$ \\
\hline
1$^\dag$ & 1 & $1 \times 10^7$ & $1 \times 10^{-7}$ & 1 & \phantom{0}200 & 2,591,344 & $2.1 \times 10^7$ & $5.3 \times 10^{49}$ \\
1$^\star$ & 1 & $1 \times 10^7$ & $1 \times 10^{-7}$ & 1 & \phantom{0}200 & 1,900,750 & $3.1 \times 10^7$ & $2.4 \times 10^{49}$ \\
2$^\star$ & 1 & $1 \times 10^7$ & $1 \times 10^{-7}$ & 0.1 & \phantom{0}200 & 1,900,750 & $3.1 \times 10^7$ & $2.4 \times 10^{49}$ \\
5$^\star$ & 1 & $1 \times 10^7$ & $1 \times 10^{-7}$ & 10 & \phantom{0}200 & 1,900,750 & $3.1 \times 10^7$ & $2.4 \times 10^{49}$ \\
7$^\star$ & 1 & $1 \times 10^7$ & $1 \times 10^{-7}$ & 100 & \phantom{0}200 & 1,900,750 & $3.1 \times 10^7$ & $2.4 \times 10^{49}$ \\
\hline
\end{tabular}
\end{center}
\end{table*}

\begin{figure*}
\includegraphics[width=\textwidth]{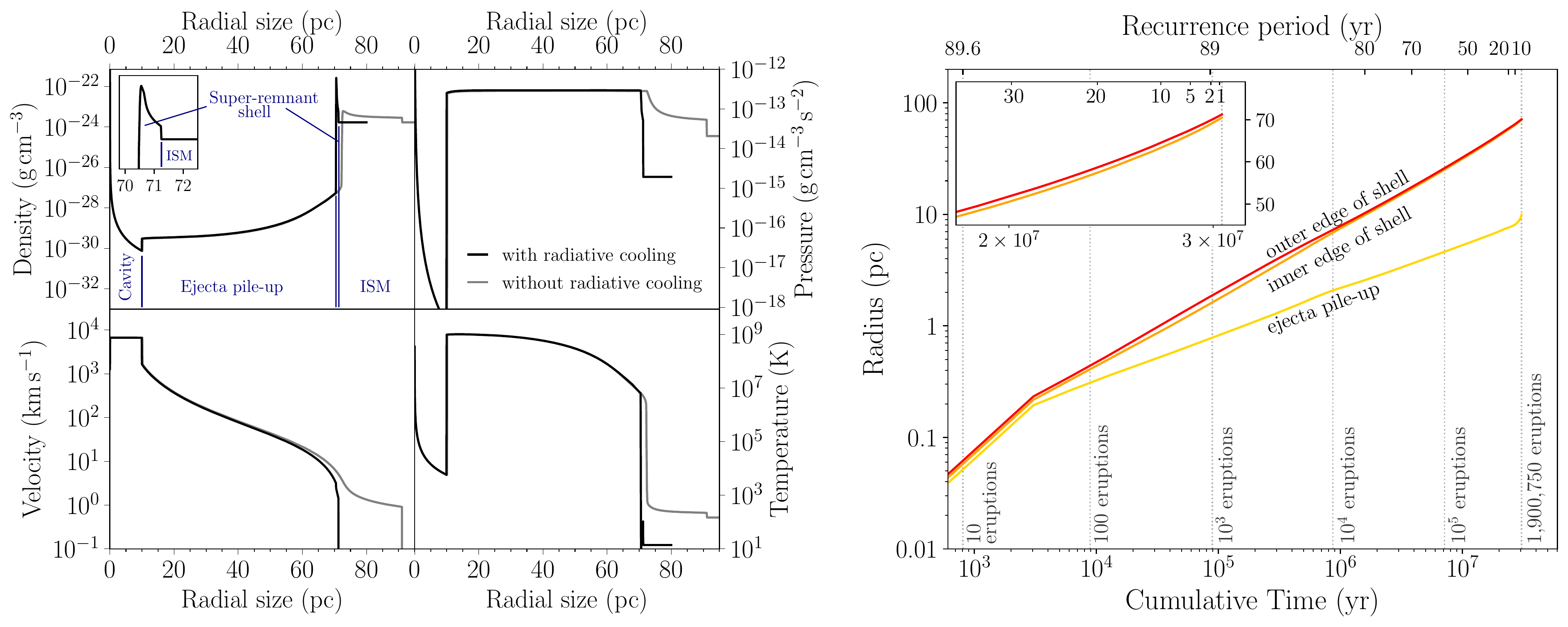}
\caption{Left: The dynamics of the Run 1 (with radiative cooling; black) and the Run 1$^\star$ (without radiative cooling; grey) NSR with $\dot M = 10^{-7} \ \text{M}_{\odot} \ \text{yr}^{-1}$ and $n = 1$ after 1,900,750 eruptions with 200\,AU resolution. Note that the finite simulated ISM can cool over these long timescales. Regions of interest are labelled and the inset highlights the thin NSR shell. Right: Evolution of the inner and outer edges of the NSR shell and the inner edge of the ejecta pile-up region with respect to cumulative time and recurrence period.}
\label{1e-7 1ISM cool full}
\end{figure*}

Run 1 is presented in Figure~\ref{1e-7 1ISM cool full}:\ the left-hand plot shows the density, pressure, velocity and temperature characteristics of the NSR after all ${\sim}$1,900,000 eruptions; the right-hand plot shows the evolution of the NSR shell outer edge and the inner edge, and the inner edge of the ejecta pile-up boundary (regions of the NSR are outlined in the top left panel of Figure~\ref{1e-7 1ISM cool full}).

\begin{figure}
\includegraphics[width=\columnwidth]{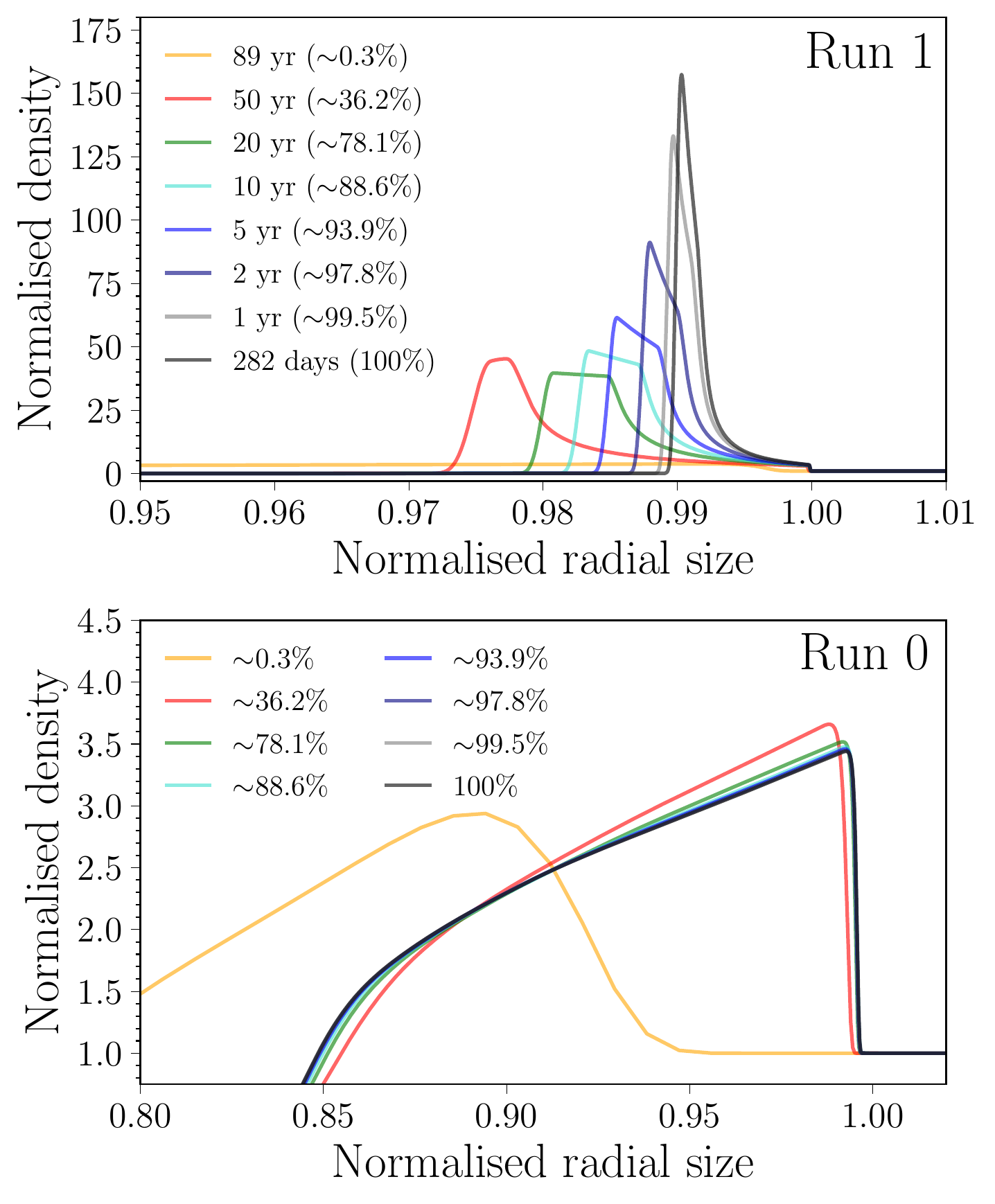}
\caption{NSR shell thickness evolution comparison between Run 1 (top) and Run 0 (bottom). Percentages indicate progress through each simulation, with the recurrence period given for Run 1; for Run 0 $P_\mathrm{rec}=1$ throughout. Radii are normalised to the outer edge of the NSR at each epoch, density is normalised to the ISM. Note the range of radial size is different in each panel.}
\label{shell evolution comparison Run 0 and Run 1}
\end{figure}

In the top-left panel of the left-hand plot of Figure~\ref{1e-7 1ISM cool full}, we see that the inner and outer edges of the dynamical NSR shell extend to ${\sim}70.5$ and ${\sim}71.3$\,pc, respectively -- a shell thickness of 1.1\%. As can be seen in the right-hand plot of Figure~\ref{1e-7 1ISM cool full}, shell thickness varies over the NSR evolution. For example, the shell compresses from $2.72\%$ ($P_{\text{rec}} = 50$ years) to $1.14\%$ ($P_{\text{rec}} = 1$ year) to $1.10\%$ ($P_{\text{rec}} = 282$ days). At all times, this is much thinner than the 12a NSR shell \citepalias{2019Natur.565..460D}, which is $22\%$ from observations and remained at this thickness throughout Run 0 (see Figure~\ref{shell evolution comparison Run 0 and Run 1}). The shell thickness evolution during Run 1 is directly related to energy losses via cooling and to the evolution of eruption properties whereby the increasing frequency and kinetic energy of the ejecta drive a compression through the NSR shell.

In Run 1, the higher density found at the NSR shell inner edge ($n\,{\simeq}\,160$) compared to the outer edge ($n\,{\simeq}\,3$), seen in the top panel of Figure~\ref{shell evolution comparison Run 0 and Run 1}, is attributed to the contribution from the more recent, more frequent and more energetic eruptions -- the rate of change of eruption properties surpasses the dynamic time-scale of the NSR shell at later times. The rate of propagation of the NSR shell into the surrounding ISM, and therefore the outer edge of the shell, remains largely based upon the combined properties of the entire eruption history, whereas the inner edge is shaped by newly arriving material. 

As evident in the bottom left panel of the left-hand plot of Figure~\ref{1e-7 1ISM cool full}, the velocity of material in the inner cavity is high (${\sim}6.7 \times 10^{3}$ km s$^{-1}$) as it is essentially in free expansion. The velocity then drops substantially as the ejecta pile-up region is encountered, with the resultant shock-heating increasing temperatures by five orders of magnitude (see bottom right panel of the left-hand plot). The velocity and temperature in the ejecta pile-up region declines continuously out to the NSR shell as the ejecta encounter previously ejected material and reverse shocks (from the pile-up/inner shell boundary), with the cool outer edge expanding at a relatively low ${\sim}1$ km s$^{-1}$.

Figure~\ref{1e-7 1ISM cool full} provides a comparison between Run 1 and Run 1$^\star$ (with and without radiative cooling, respectively), to illustrate the significant difference in the NSR size and shell structure. The outer edge of the NSR in Run 1 extends to 71.3 pc yet, without radiative cooling in Run 1$^\star$, the NSR extends to ${\sim}$90 pc (having swept up around twice as much ISM). This substantial reduction in size can only be attributed to radiative losses within the NSR. Additionally, the radiatively cooled NSR shell from Run 1 is much thinner (${\sim}$1\%) than the uncooled equivalent in Run 1$^\star$ (${\sim}$21\%; see Figure~\ref{1e-7 1ISM cool full}). This results from the material in the early NSR shell losing energy via radiative cooling and therefore lacking the necessary pressure to maintain its size. This suppresses the early NSR shell formation such that when shell compression takes effect at later times (as increasingly energetic ejecta collide with the inner edge of the shell), the starting point is a thinner shell. 

The NSR cavity and ejecta pile up boundary at ${\sim}10$\,pc have similar density, pressure, velocity and temperature in Run 1 and Run 1$^\star$. At later stages, the increased frequency and energy of the eruptions results in the scenario that tends toward the Run 0 regime, whereby there is not enough time for the ejecta or remnant to cool radiatively between consecutive eruptions. Consequently, we see the effects of radiative cooling at the outer edge of the remnant, a relic of the earlier spaced out less energetic eruptions, and the centre of the NSR reflecting the later frequent eruptions. Furthermore, this point can be extended to all of the simulations conducted throughout this paper, whereby the growth and subsequent size of the nova super-remnant is shaped heavily by its early evolution.

\begin{figure}
\centering
\includegraphics[width=\columnwidth]{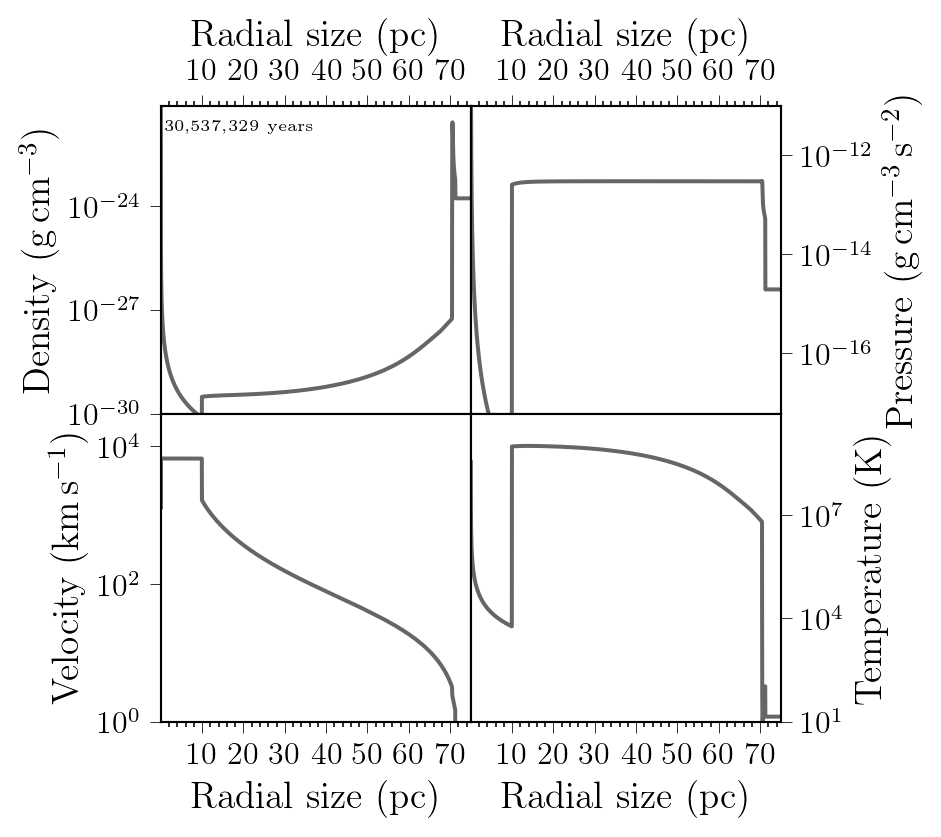}
\caption{Animated evolution of density, pressure, velocity, and temperature for Run 1.}
\label{reference animation}
\end{figure}

So far, we have only considered the final epoch of Run 1, after the full 1,900,750 eruptions (Figure~\ref{1e-7 1ISM cool full}). However, to appreciate the changing structure and characteristics of the NSR, we have provided an animation of the Run 1 in Figure~\ref{reference animation}.

\begin{figure*}
\includegraphics[width=\textwidth]{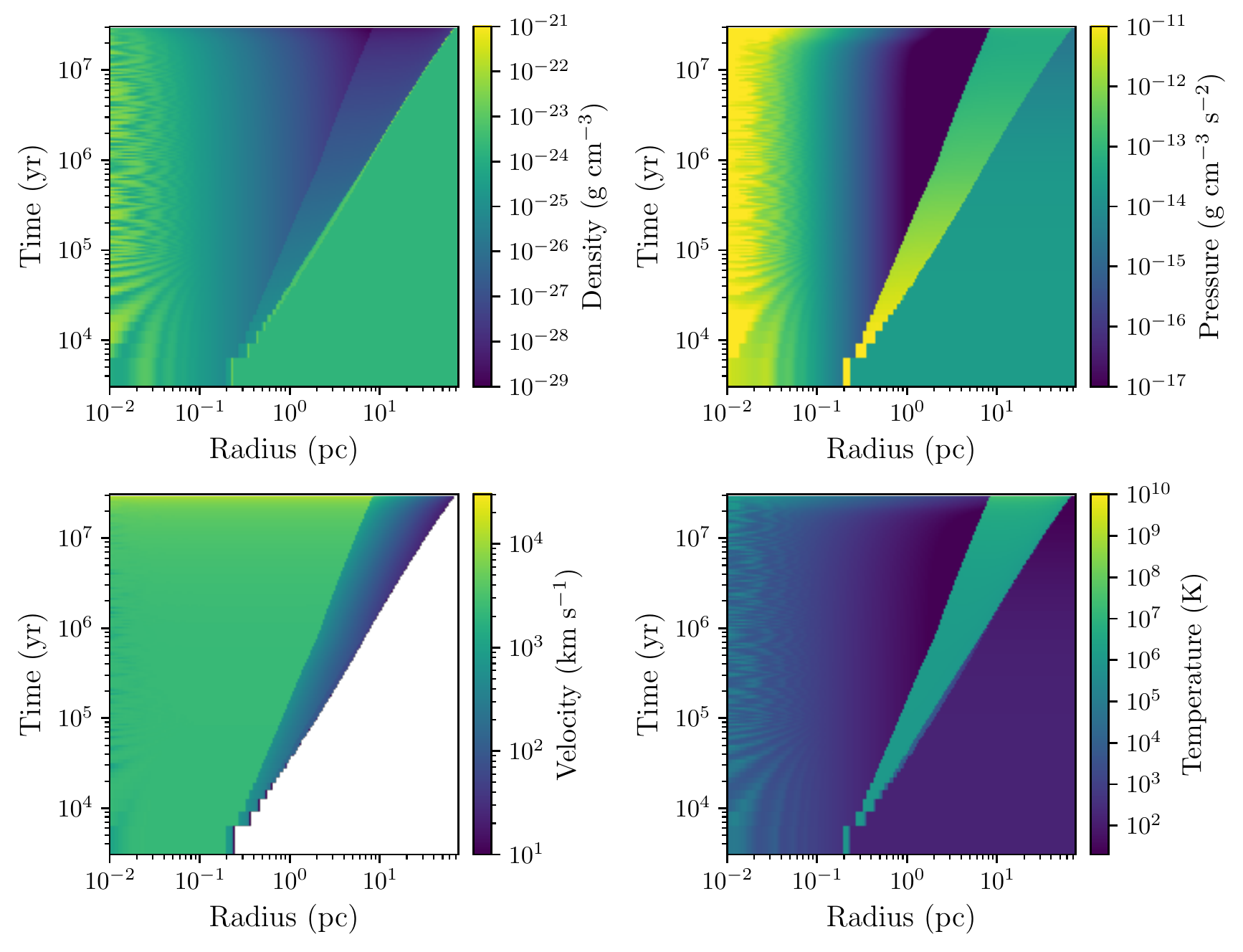}
\caption{Run 1 spatiotemporal evolution of density, pressure, velocity, and temperature. The structure apparent $\lesssim 0.1$ pc is associated with individual eruptions. At early times ($t \lesssim 3 \times 10^4$ years), the temporal resolution becomes evident. As shown in the bottom left panel, the velocity of the ISM is negligible ($\ll 10$km s$^{-1}$).}
\label{Run 1 Spatiotemporal evolution}
\end{figure*}

We illustrate in Figure~\ref{Run 1 Spatiotemporal evolution} the spatiotemporal analysis of the evolution of the Run 1 NSR in terms of density, pressure, velocity and temperature. The NSR shell in Figure~\ref{Run 1 Spatiotemporal evolution} can be identified most clearly in the top left panel as the narrowing light green segment running from bottom left (at ${\sim}0.25$ parsec) to the top right. In addition, the boundary of the ejecta pile-up, separating the cavity and the ejecta pile-up region can be seen as the other apparent line left of the remnant shell, running from the bottom left to the top centre of the panel (this boundary can be seen most clearly in the bottom right panel showing temperature evolution). This radial evolution of the shell and ejecta pile-up boundary directly replicates those seen in the right-hand plot of Figure~\ref{1e-7 1ISM cool full}, however here we show how each parameter changes over the full simulation.

The average density of the early NSR shell is approximately $n\,{\simeq}\,6$ for the first $10^6$ years of growth (see top left panel of Figure~\ref{Run 1 Spatiotemporal evolution}). Beyond this epoch, we see the effect of radiative cooling as the NSR shell loses energy and is compressed by the surrounding ISM and incoming eruptions, thereby leading to an increase in the average density within the shell to $n\,{\simeq}\,36$ after ${\sim}3\times 10^7$ years. The average density within the ejecta pile-up region is much lower than the surrounding ISM and continuously decreases throughout the evolution, dropping as low as $n = 2.4 \times 10^{-4}$ by the final epoch. After $10^6$ years the mass of the shell is ${\sim}50 \ \text{M}_{\odot}$ but then substantially increases to $4 \times 10^3 \ \text{M}_{\odot}$ after $10^7$ years and ending with a mass of ${\sim}4 \times 10^4 \ \text{M}_{\odot}$ by the final epoch (${\sim}3\times 10^7$ years). This is consistent with the upper limiting shell masses derived from imaging and spectroscopy of the 12a NSR \citepalias[$7 \times 10^5 \ \text{M}_{\odot}$ and $10^6 \ \text{M}_{\odot}$ from assuming oblate and prolate geometries, respectively;][]{2019Natur.565..460D}.

As shown in the top right panel of Figure~\ref{Run 1 Spatiotemporal evolution}, the average pressure within the NSR shell is initially high as this thin high density region initially forms at high temperature. The pressure within the shell decreases until it matches the average pressure within the pile-up region after ${\sim}2 \times 10^7$ years. The outer edge of the shell remains at the same pressure for the remainder of the simulation. However, the pressure at the inner edge increases, creating a pressure gradient within the shell. With the average temperature of the ejecta pile-up region increasing monotonically throughout its evolution (see the bottom right panel of Figure~\ref{Run 1 Spatiotemporal evolution}), the pressure within follows the same trend once that region's size is established. The average pressure evolution illustrates how the NSR shell compression takes place during an intermediary period. The shell forms initially without compression, is then compressed as it is subjected to pressure gradients and after ${\sim}2 \times 10^7$ years, the thinner shell remains.

The average temperature of the Run 1 NSR shell falls as a direct result of cooling due to expansion and radiative losses, dropping from an initial $5 \times 10^3$ K to $40$ K after ${\sim}2.8 \times 10^7$ years before increasing modestly to $90$ K as later eruptions become more frequent and begin to impact the inner edge of the shell through the pile-up region, leading to compression and re-heating (see bottom right panel of Figure~\ref{Run 1 Spatiotemporal evolution}). On the other hand, the pile-up region begins with higher temperatures of ${\sim}1 \times 10^6$ K and continues to experience this temperature throughout before dramatically increasing to ${\sim}2.5 \times 10^8$ K after the full $3 \times 10^7$ years, maintaining these extremely high temperatures through shock-heating. 

The average velocity of the NSR shell, like the average temperature and average pressure, decreases throughout the evolution before a slight increase for the final $6 \times 10^6$ years (see bottom left panel of Figure~\ref{Run 1 Spatiotemporal evolution}). The velocity of the shell's outer edge at ${\sim}6 \times 10^3$ years is ${\sim}10 \ \text{km s}^{-1}$ and remains below this velocity throughout. However, the velocity of the inner edge does increase due to the more frequent collisions occurring within the pile-up region, leading to a small velocity gradient within the shell. The ejecta pile-up region follows a similar trend but with higher average velocities, a result of increasingly frequent and higher velocity ejecta impacting the ejecta pile-up boundary. As the cavity is essentially a vacuum, the increasing velocities within this region are directly reflecting the increasing velocities of the nova ejecta.

\subsection{Varying the ISM density}\label{The effect of ISM density}
\begin{figure}
\centering
\includegraphics[width=\columnwidth]{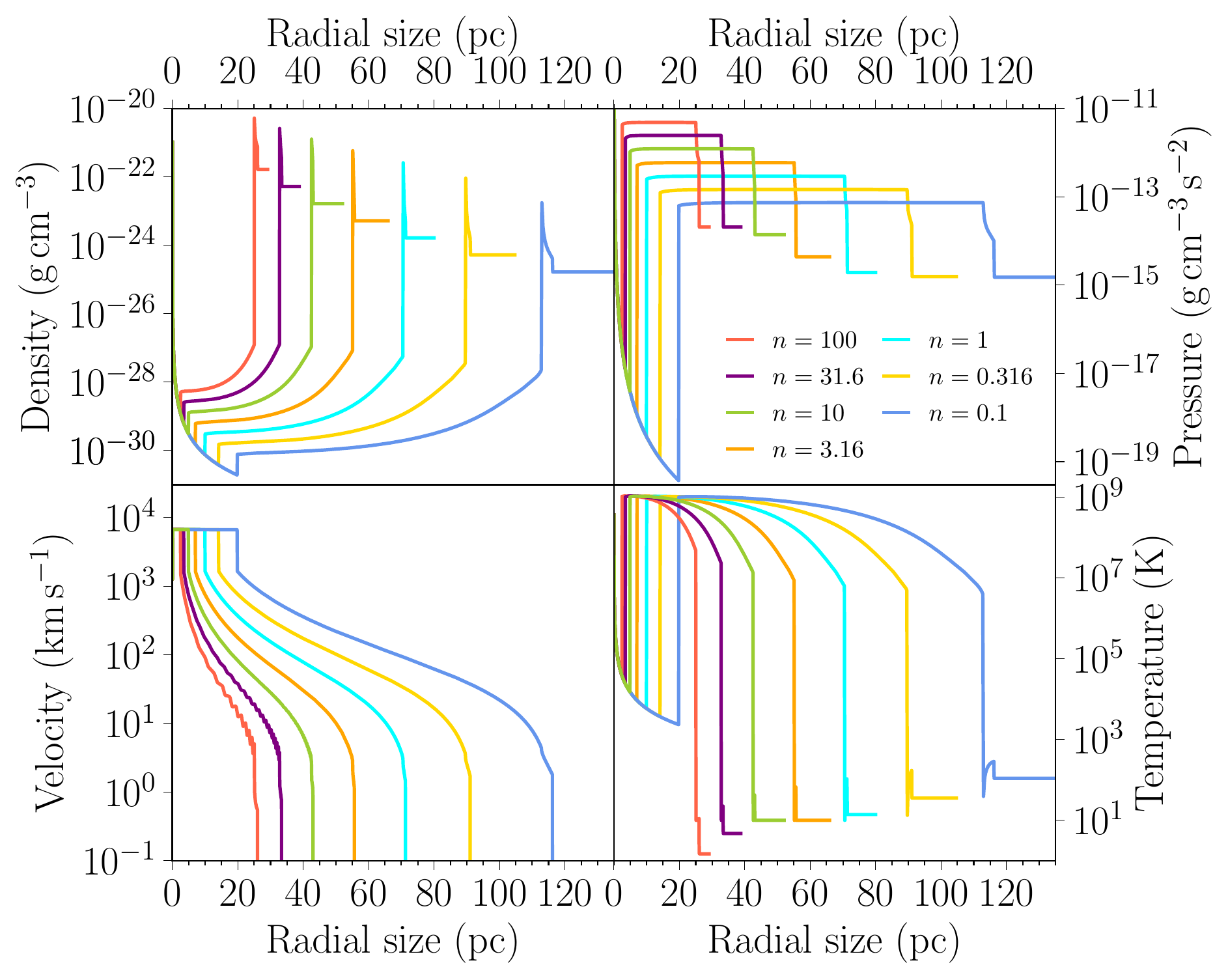}
\caption{Dynamics of Run 1 ($n = 1$) compared to Run 2 ($n = 0.1$), Run 3 ($n = 0.316$), Run 4 ($n = 3.16$), Run 5 ($n = 10$), Run 6 ($n = 31.6$) and Run 7 ($n = 100$).}
\label{reference sim all ISM}
\end{figure}
Here we consider the same nova system as Run 1 ($T_{\text{WD}} = 10^7$\,K; $\dot M = 10^{-7} \, \text{M}_{\odot} \, \text{yr}^{-1}$), but placed in lower and higher density surroundings. Run 2 is pre-populated by ISM with a lower density of $1.67 \times 10^{-25} \text{ g cm}^{-3}$ ($n = 0.1$) and the ISM density of Run 5 and Run 7 is $1.67 \times 10^{-23} \text{ g cm}^{-3}$ ($n = 10$) and $1.67 \times 10^{-22} \text{ g cm}^{-3}$ ($n = 100$), respectively. We also sampled between these ISM densities with Run 3 ($n = 10^{-0.5}\approx0.316$), Run 4 ($n = 10^{0.5}\approx3.16$) and Run 6 ($n = 10^{1.5}\approx31.6$). As illustrated in Figure~\ref{reference sim all ISM}, the full simulations extend progressively further as the ISM density is decreased (e.g., ${\sim}$116 pc, ${\sim}$43 pc and ${\sim}$26 pc for Run 2 ($n=0.1$), Run 5 ($n=10$) and Run 7 ($n=100$), respectively) and all maintain an exceptionally thin shell due to the suppression of the early shell formation, reminiscent of Run 1. Furthermore, the remnants grown in Run 1, 2, 5 and 7 with radiative cooling are 78.26\%, 63.29\%, 78.31\%, and 77.96\%, respectively, of the size of their counterpart without cooling (from Runs 1$^\star$, 2$^\star$, 5$^\star$ and 7$^\star$) as a direct result of radiative losses from cooling. The relative thickness of the NSR shell varies for each simulation but remains small ($\lesssim 4\%$) for all ISM densities, resulting from the same amount of work done by the same nova system on surroundings that present increasingly higher resistance.

\begin{figure*}
\includegraphics[width=\textwidth]{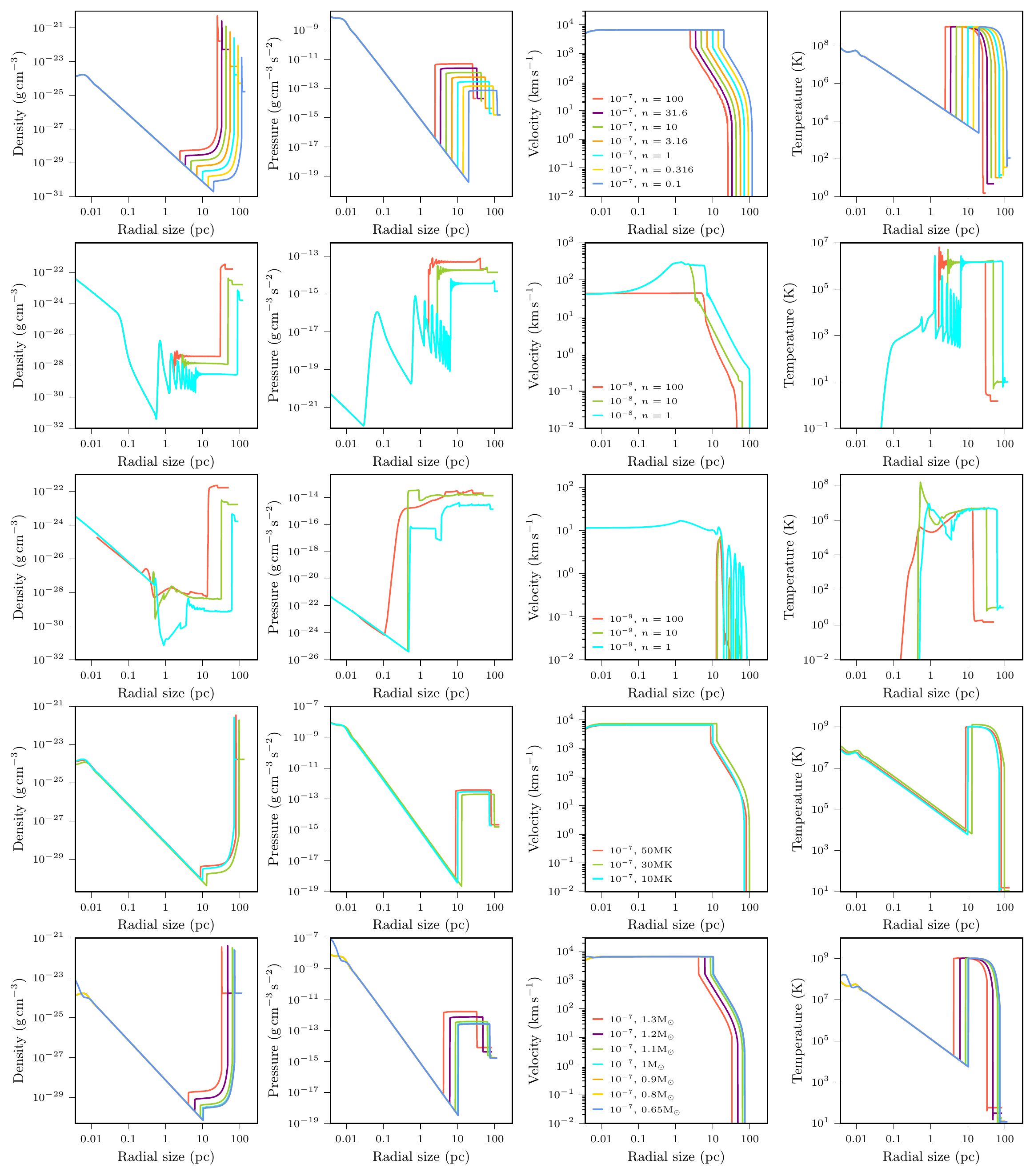}
\caption{End point dynamics of Runs 1--21. \textit{First row}: $\dot M = 10^{-7} \, \text{M}_{\odot} \, \text{yr}^{-1}$, $T_{\text{WD}}$ = 10\,MK, and $M_{\text{WD}} = 1 \, \text{M}_{\odot}$ for $n=0.1,0.316,1,3.16,10,31.6,100$ (Runs 2,3,1,4,5,6,7 respectively). \textit{Second row}: $\dot M = 10^{-8} \, \text{M}_{\odot} \, \text{yr}^{-1}$, $T_{\text{WD}}$ = 10\,MK, and $M_{\text{WD}} = 1 \, \text{M}_{\odot}$ for $n=1,10,100$ (Runs 8--10, respectively). \textit{Third row}: $\dot M = 10^{-9} \, \text{M}_{\odot} \, \text{yr}^{-1}$, $T_{\text{WD}}$ = 10\,MK and $M_{\text{WD}} = 1 \, \text{M}_{\odot}$ for $n=1,10,100$ (Runs 11--13, respectively). \textit{Fourth row}: $\dot M = 10^{-7} \, \text{M}_{\odot} \, \text{yr}^{-1}$, $n=1$ and $M_{\text{WD}} = 1 \, \text{M}_{\odot}$ for $T_{\text{WD}}$ = 10\,MK, 30\,MK, 50\,MK (Runs 1, 14, 15, respectively). \textit{Fifth row}: $\dot M = 10^{-7} \, \text{M}_{\odot} \, \text{yr}^{-1}$, $n=1$ and $T_{\text{WD}}$ = 10\,MK for $M_{\text{WD}} = 0.65 \, \text{M}_{\odot}$, $0.8 \, \text{M}_{\odot}$, $0.9 \, \text{M}_{\odot}$, $1 \, \text{M}_{\odot}$, $1.1 \ \text{M}_{\odot}$, $1.2 \ \text{M}_{\odot}$, $1.3 \ \text{M}_{\odot}$ (Runs 16--18, 1, 19--21, respectively).}
\label{All runs}
\end{figure*}

As expected, the density in the NSR cavity and pile-up region increases approximately in-line with ISM density. These regions are not only denser as a result of the ISM environment, but are also more compressed for higher $n$, leading to increased pressure. The velocity of material inside the NSR cavity from Runs 1--7 is identical as in all cases the ejecta are essentially undergoing free expansion. Also, temperatures in this region for each Runs 1--7 all reach the same extreme temperature of ${\sim}1 \times 10^{9}$ K, as nova ejecta expanding without resistance collide into earlier ejected matter in the pile-up region, before dropping away to $< 10$ K at the nova shell's inner edge (i.e., the properties in this region don't strongly depend upon $n$). The growth of the outer edge of the NSR shells within the $n = 10$ and $n = 100$ ISM follow a similar evolution as that of Run 1 (see the red line on the right plot of Figure~\ref{1e-7 1ISM cool full}).

We can summarise our findings for this section as follows: for a given total kinetic energy, an increase in local ISM density results in a smaller nova super-remnant.

\subsection{Varying the mass accretion rate}\label{The effect of mass accretion rate}

The next six simulations (Runs 8--13) explored NSR evolution while varying accretion rate. We considered a WD with a temperature $10^7$ K accreting hydrogen rich material at a rate of $\dot M = 10^{-8} \ \text{M}_{\odot} \ \text{yr}^{-1}$ as well as a nova with the same WD temperature but with a lower accretion rate of $\dot M = 10^{-9} \ \text{M}_{\odot} \ \text{yr}^{-1}$, placed within the three ISM densities used in Runs 1, 5 and 7, see Table~\ref{Runs}.

Runs 1--7 presumed that accretion was driven by the wind of a giant donor. We include mass loss from the donor between eruptions, although this has no impact upon the results (yet is computationally favourable, see Section~\ref{Hydrodynamical Simulations}). As such, we reduce the mass loss rate from the donor in line with any simulated changes to accretion rate for consistency and to ensure that the donor wind does not become important.

The WD growth models for $\dot M = 10^{-8} \ \text{M}_{\odot} \ \text{yr}^{-1}$ and $\dot M = 10^{-9} \ \text{M}_{\odot} \ \text{yr}^{-1}$ reveal that the WD loses mass with every eruption; it does not grow towards the Chandrasekhar limit, but is instead eroded. We therefore imposed a temporal upper limit of 100\,Myr for the $\dot M = 10^{-8} \ \text{M}_{\odot} \ \text{yr}^{-1}$ and $\dot M = 10^{-9} \ \text{M}_{\odot} \ \text{yr}^{-1}$ simulations. The WD growth models indicate that these systems require 40,343 eruptions and 2,094 eruptions, respectively, to reach the temporal upper limit. At which point, these systems would have a recurrence period of ${\sim}$3,000 years and ${\sim}$49,000 years, respectively.

\begin{figure}
\includegraphics[width=\columnwidth]{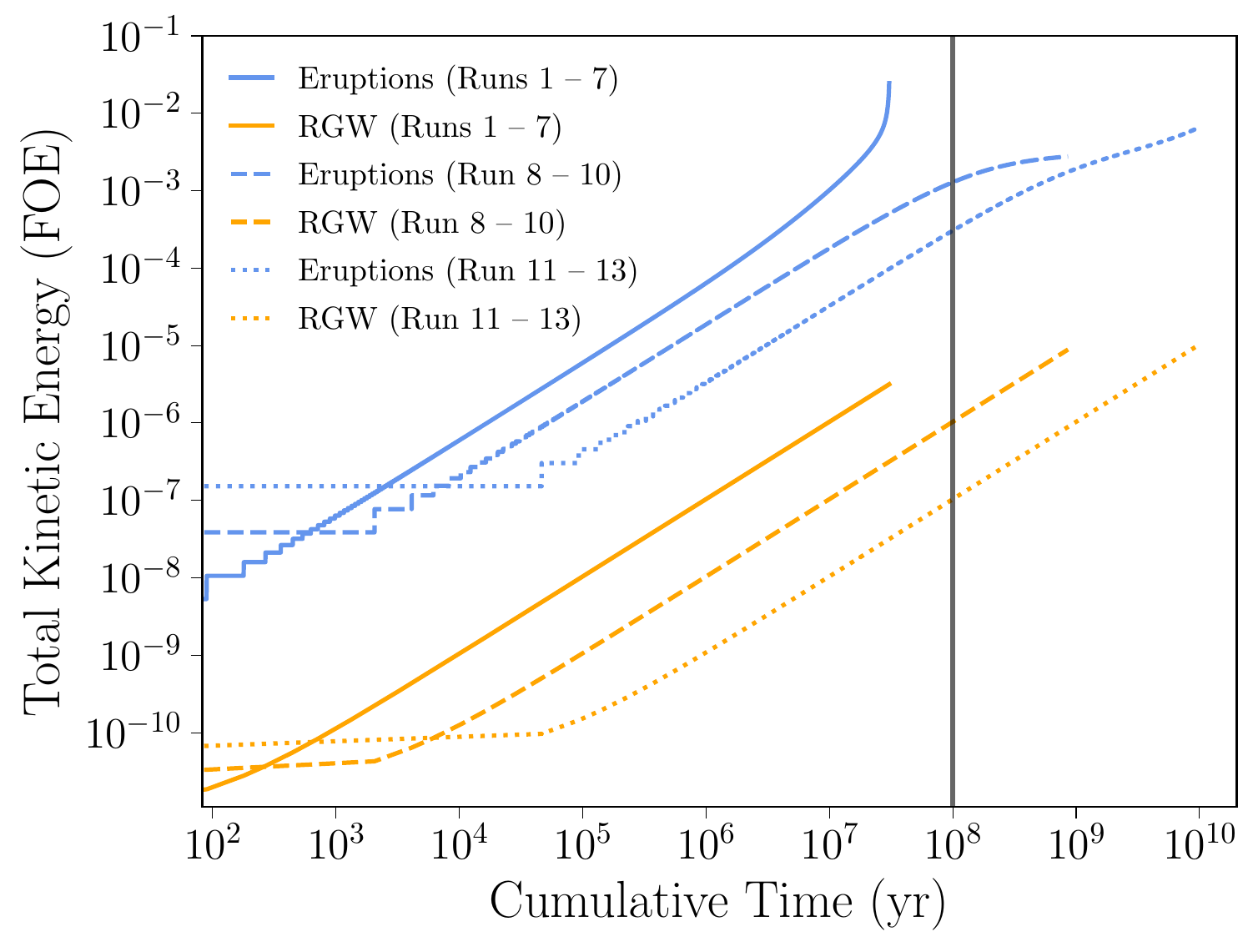}
\caption{Kinetic energy evolution from the simulated nova eruptions and red giant wind for Runs 1--13. The vertical black line represents the temporal cut-off point for Runs 8--13 i.e. the upper time limit for the simulations in which the WD is shrinking and therefore never reaches the Chandrasekhar limit.}
\label{KE evolution}
\end{figure}

Focusing on Runs 8--10 ($\dot M = 10^{-8} \ \text{M}_{\odot} \ \text{yr}^{-1}$) presented in the second row of Figure~\ref{All runs}, we find that the overall structure of the remnants are similar to those grown with higher accretion rate. The major difference is their much larger size and thicker shells. The shell grown in the lowest density ISM (Run 8; $n=1$) extends to ${\sim}$99 pc, with a shell thickness of ${\sim}$11\%, and Run 9 ($n = 10$) and Run 10 ($n = 100$) grow remnants with radial sizes of ${\sim}$62 pc and ${\sim}$40 pc, and shell thicknesses of ${\sim}$22\% and ${\sim}$25\%, respectively. These more extended shells are a consequence of the larger amount of kinetic energy ejected by the underlying system and the longer time over which it can act ($1 \times 10^8$ years compared to ${\sim}3.1 \times 10^7$ years in Run 1; see Figure~\ref{KE evolution}). The outer edge of the NSR shell follows the same evolutionary trend as seen in Runs 1--7 (in the same manner as the remnant in the right plot of Figure~\ref{1e-7 1ISM cool full}).

In Runs 11--13 ($\dot M = 10^{-9} \ \text{M}_{\odot} \ \text{yr}^{-1}$; $n = 1,10,100$, respectively), we see that the NSRs take the familiar shape seen in Runs 1--10 with a very low density cavity preceding a high density shell (see the third row of Figure~\ref{All runs}). The remnants grown in the Run 11 ($n = 1$), Run 12 ($n = 10$) and Run 13 ($n = 100$) extend to ${\sim}$75 pc, ${\sim}$48 pc and ${\sim}$26 pc, respectively, and have shell thicknesses of 17\%, 34\% and 39\%, respectively. Yet for each of these runs, the remnant shell is difficult to discern from the surroundings with the peak density within the NSR shell of Run 11, Run 12 and Run 13 reaching only 10.9\%, 1.9\% and 1.4\% beyond that of the prepopulated ISM density, respectively. As expected, the outer shells of the remnants grown in systems with the lower accretion rate ($\dot M = 10^{-9} \ \text{M}_{\odot} \ \text{yr}^{-1}$) follow the same growth curve over time as previous runs.

The nova eruptions from the systems in Run 11--13 occur infrequently for the vast majority of the evolution, starting with $P_\mathrm{rec}\sim$46,600 years when  $M_\mathrm{WD}=1\,\text{M}_{\odot}$ and increasing to ${\sim}49,000$ years after the full $1 \times 10^8$ years. Therefore, a combination of low energy eruptions and long recurrence period leads to a very broad, low-contrast shell as the ejecta individually dissipate into the surrounding ISM with minimal pile-up. Dynamically, such a NSR would be difficult to discern from the local environment. However, we would not expect this form of shell to exist around the known RNe as these systems would not (currently) be recognised as recurrent nova with their recurrence periods being $\gg 100$ years \citep[see, for example,][]{2020AdSpR..66.1147D}.

Equipped with the simulations of NSRs grown from systems with different accretion rates, we find that a lower accretion rate leads to more extended, but less well-defined, NSRs: a direct result of the longer evolutionary timescale.

\subsection{Varying the white dwarf temperature}\label{The effect of WD temperature}
The underlying WD temperature does not have a significant impact on the evolution of most of the various parameters given in Section~\ref{Growing a white dwarf}. For example, for $\dot M = 10^{-7} \ \text{M}_{\odot} \ \text{yr}^{-1}$, the evolution of each parameter is very similar throughout, regardless of the WD temperature. Yet, there is a moderate difference in the evolution of the mass accumulation efficiency for the different temperatures. This is also true for the total kinetic energy of the ejecta generated from the entirety of the nova eruptions whereby the 30\,MK and 50\,MK have approximately twice the kinetic energy output as the cooler 10\,MK WD. This is reflected in the set of simulations with the WD temperature varied from 10\,MK (Run 1) to 30\,MK (Run 14) to 50\,MK (Run 15) with $\dot M = 10^{-7} \ \text{M}_{\odot} \ \text{yr}^{-1}$ and $n = 1$. A comparison of the NSR shell, as shown in the fourth row of Figure~\ref{All runs} for the three different WD temperatures, reveals the overall structure of each to be similar, but with the 30\,MK WD remnant extending moderately further than the others. The outer edge of the remnant shell for the coolest WD is 71.3\,pc and the hottest WD leads to an outer edge of 79.7\,pc, whereas the outer edge of the 30\,MK WD remnant shell is 97.4\,pc. Yet, this informs us that, for the highest accretion rate we have considered, the WD temperature has a small impact on the large scale structure of the NSR in comparison to the effects of ISM density (Section~\ref{The effect of ISM density}) and mass accretion rate (Section~\ref{The effect of mass accretion rate}).

There are similarities with the evolution of the shell for each WD temperature and at each epoch the density and thickness of the shells are a close match. By analysing how the recurrence period and the total kinetic energy change as the NSR grows in each of these systems, it is apparent that the WD temperature only has a relatively small impact. This may be due to the system having a high accretion rate ($10^{-7} \, \text{M}_{\odot} \, \text{yr}^{-1}$), so being dominated by accretion heating\footnote{\citetalias{2005ApJ...623..398Y} accounted for accretion heating within their computations.}. Any influence of WD temperature may become more substantial as accretion rate decreases as accretion heating will become less severe and the WD would have more time to cool between eruptions.

A further consideration is that unlike accretion rate and ISM density, which were both varied by factors of 10 and 100, the WD temperatures considered here only vary by factors of 3 and 5. The range we use (10\,MK, 30\,MK and 50\,MK) was initially employed by \citet{1995ApJ...445..789P}\footnote{Before consequently being adopted by \citetalias{2005ApJ...623..398Y} with the incorporation of lower accretion rates for the cooler WDs.} and was chosen to represent two extremes and an intermediate WD core temperature; the lower limit was set as a colder WD delays hydrogen ignition leading to long accretion times (hence more substantial eruptions) and the upper limit accounts for hot WDs being able to quickly reach the conditions for TNR.

We can conclude, for the accretion rate and ISM density ($n$) sampled in Runs 1, 14, 15, that the expected variation in WD temperature has much less impact on NSR evolution than plausible variations in accretion rate or $n$.

\subsection{Varying the initial white dwarf mass}\label{The effect of initial WD mass}

So far we have considered nova eruptions generated by a WD growing from $1 \ \text{M}_{\odot}$ to $\text{M}_\mathrm{Ch}$. Here, we consider a number of different initial WD masses; $0.65 \ \text{M}_{\odot}$ in Run 16, $0.8 \ \text{M}_{\odot}$ in Run 17, $0.9 \ \text{M}_{\odot}$ in Run 18 and $1.1 \ \text{M}_{\odot}$ in Run 19 with $\dot M = 10^{-7} \ \text{M}_{\odot} \ \text{yr}^{-1}$ and $n = 1$. This upper initial mass is the upper formation limit for a CO WD \citep{1996ApJ...460..489R}. We also sample WDs with masses of $1.2 \ \text{M}_{\odot}$ in Run 20 and $1.3 \ \text{M}_{\odot}$ in Run 21. The number of eruptions appreciably increases as we lower the initial WD mass, as more eruptions are needed to reach $\text{M}_{\text{Ch}}$ (see Table~\ref{Runs}).

A comparison of the NSR shells from these runs, presented in the last row of Figure~\ref{All runs}, shows that each remnant becomes marginally larger as the initial WD mass is lowered, as more eruptions lead to more ejecta impacting the surrounding ISM over a longer period of time. The radial size of the NSRs in Runs 16, 17, 18 and 19 ($0.65 \ \text{M}_{\odot}$, $0.8 \ \text{M}_{\odot}$, $0.9 \ \text{M}_{\odot}$ and $1.1 \ \text{M}_{\odot}$) almost completely resemble that of the NSR from Run 1 ($1 \ \text{M}_{\odot}$) whereas starting with a WD mass $>1.1 \ \text{M}_{\odot}$ \citep[in the regime of ONe WDs;][]{1996ApJ...460..489R} such as simulated in Run 20 ($1.2 \ \text{M}_{\odot}$) and Run 21 ($1.3 \ \text{M}_{\odot}$) does make a difference to the radial size of the NSR. The structure of the shell for each NSR is remarkably similar, with the $0.65 \ \text{M}_{\odot}$ WD simulation finishing with a shell thickness of ${\sim}$1.1\% compared to ${\sim}$1.2\% for the $1.3 \ \text{M}_{\odot}$ WD. Each NSR shell also follows a very similar transition, with similar shell widths ratios at the same epochs. The radial growth curves of each simulation follow the same evolution with the $0.65 \ \text{M}_{\odot}$ WD taking ten times ($37$\,Myr) the time to reach $\text{M}_\mathrm{Ch}$ than the $1.3 \ \text{M}_{\odot}$ WD ($3.7$\,Myr).

We can therefore conclude, that the initial mass of the growing WD has little impact on the final structure of the NSR, much less than the prominent influence of the ISM density (Section~\ref{The effect of ISM density}) and accretion rate (Section~\ref{The effect of mass accretion rate}).

\section{Additional tests}\label{Additional tests}

In Section~\ref{Hydrodynamical Simulations}, we presented the full set of simulations. Here, we outline several tests of alternative models of the ejecta characteristics.

\subsection{Using broken fits to estimate system parameters}\label{Fitting a broken exponential to find system parameters}
\begin{figure*}
\includegraphics[width=\textwidth]{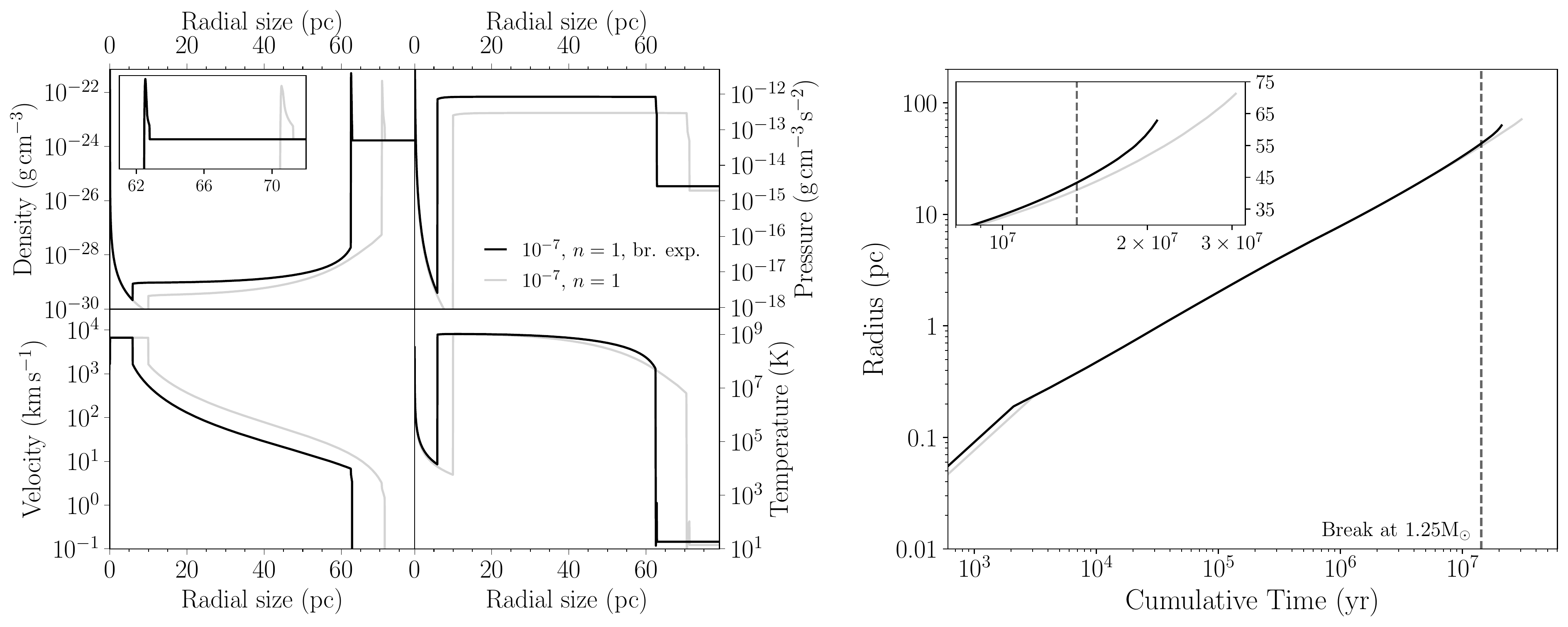}
\caption{As Figure~\ref{1e-7 1ISM cool full}, but comparing Run 1 (grey) to Run 1$^\dag$ (black), i.e., smooth versus broken exponential interpolation of the \citetalias{2005ApJ...623..398Y} relations. In the right panel, we indicate the point at which the break in the exponential fitting occurs (${\sim}1.4 \times 10^7$ years).}
\label{exponential fit run}
\end{figure*}

For Runs 1--21, we utilised ejecta characteristics determined from our WD growth model. This was based on interpolating between the results of multi-cycle nova evolutionary simulations by \citetalias{2005ApJ...623..398Y} (see Section~\ref{Parameter Space}). In our work, a smooth function asymptotically approaching $\text{M}_\mathrm{Ch}$ was fitted to the \citetalias{2005ApJ...623..398Y} grid.

An alternative way of interpolating between the \citetalias{2005ApJ...623..398Y} grid points is with a `knee' function \citep[e.g.,][their Figure 1]{2015A&A...583A.140S}, which we replicated by fitting two distinct exponentials (Figure~\ref{Five Relationships}). From here, we grew a $1 \ \text{M}_{\odot}$ WD with our model as outlined in Section~\ref{Growing a white dwarf}, but referring in this case to the broken exponential fits.

Eruption parameters evolve in the same way as those from the smooth function fitting, with the main difference being the abrupt `knee' at $1.25 \ \text{M}_{\odot}$. The total kinetic energy at the end of the WD growth is ${\sim}5.3 \times 10^{-2}$ foe ($10^{51}$ ergs). This is much greater than the total kinetic energy generated from our smooth fitting function in Section~\ref{Growing a white dwarf}; this ended with ${\sim}2.5 \times 10^{-2}$ foe. This reflects the more extreme eruptions later on in this system's evolution as a direct result of the higher ejecta velocities after the WD has surpassed 1.25\,$\text{M}_{\odot}$.

We ran a simulation (Run 1$^\dag$) of nova eruptions generated from the two distinct exponential fits, with the same parameters as our reference simulation (Run 1) including $\dot M =10^{-7} \ \text{M}_{\odot} \ \text{yr}^{-1}$ and a WD temperature of $10^7$ K and $n = 1$ ISM (see Table~\ref{Runs}). As can be seen in Figure~\ref{exponential fit run}, the shell grown from the broken exponential fitting does not grow as large as the shell grown from the smooth fitting. This is a consequence of the much higher mass accumulation efficiency between $1 \ \text{M}_{\odot}$ and $1.25 \ \text{M}_{\odot}$ (see Figure~\ref{Five Relationships}) for the broken exponential fit, resulting in lower levels of ejecta and substantially less kinetic energy during the early stages of NSR growth; this period has a major impact on the proceeding evolution. Beyond $1.25 \ \text{M}_{\odot}$, the radial growth curve of the Run 1$^\dag$ shell deviates from that of Run 1 (at approximately $1.4 \times 10^7$ years; see the inset of the right panel in Figure~\ref{exponential fit run}) as a result of the later eruptions becoming more extreme.

As shown in the inset of the left panel in Figure~\ref{exponential fit run}, both the shells in Run 1 and Run 1$^\dag$ have a similar structure however the shell in Run 1$^\dag$ is thinner, and consequently, comprises a higher density inner edge. This also has a greater impact on the temperature gradient of the shell in Run 1$^\dag$, with the outer edge being much hotter than the inner edge, unlike Run 1.

It is clear that using an alternative interpolation to the values given in \citetalias{2005ApJ...623..398Y} does have an effect on the final simulated NSR. In the case of the Run 1$^\dag$, the shell width is approximately half the shell width of the remnant in Run 1 plus the size of the remnant decreases by a factor of ${\sim}$12\%. Whilst a non-negligible difference, we consider the more realistic smooth evolution of system parameters adopted for our study to be a truer representation for NSR simulations. Nevertheless, this does indicate the need for more finely sampled nova model grids.

\subsection{Eruption characteristics}\label{Eruption characteristics}

Although we are predominantly concerned with the long term evolution (and therefore large scale structure) of a NSR, we explored several eruption characteristics to observe how NSR evolution is affected. Firstly, we know that the timescale of the nova eruption can vary as we see a wide range of SSS periods \citep[see, e.g.][]{2014A&A...563A...2H}. Secondly, shocks play a key role within the nova ejecta, and instead of material being ejected in one event, the eruption contains a number of components with varying masses and velocities \citep{2014MNRAS.442..713M,2020NatAs...4..776A,2020ApJ...905...62A,2022MNRAS.514.6183M}.

\subsubsection{Eruption duration}\label{Eruptions timescales}

As an extension to the Run 0 tests (\citetalias{2019Natur.565..460D}), to determine if the duration of a nova eruption affects NSR large scale structure we ran high resolution (${\sim}$4 AU/cell) simulations, each with 1000 eruptions utilising the Run 0 setup with a range of eruption durations: 0.07\,d, 0.7\,d, 7\,d, 70\,d and 350\,d. For each test, the eruption duration plus the quiescent period match the recurrence period (350 days; e.g., 349.03d\,+\,0.07d or 343d\,+\,7d), with a fixed ejecta velocity of 3000 km\,s$^{-1}$. We required each test to inject the same total kinetic energy, so the eruption mass-loss rate was decreased to account for the longer timescales. After around 100 eruptions, the inner and outer edges of the NSR shell followed the same evolutionary trend regardless of eruption duration, and even though the NSR pile-up fluctuates more than the shell, they again settle into similar growth rates. This removes eruption duration dependency and indicates that our NSR results are not sensitive to any assumptions made about eruption time-scales.

\subsubsection{Intra-eruption shocks}\label{Intra-eruption shocks}

We also wanted to test whether having a non-uniform ejection of material from the nova would affect the large scale structure of the shell. For this, we considered the composition of a classical nova whereby the eruption takes place over a certain timescale and over that time period, the speed of ejection increases \citep{1989Sci...246..136B,1994MNRAS.271..155O,2014MNRAS.442..713M,2020NatAs...4..776A,2020ApJ...905...62A}. This implies that the outburst is comprised of a slow wind followed by a faster wind, creating a shock within the ejecta \citep{1994MNRAS.271..155O,2014MNRAS.442..713M,2020NatAs...4..776A,2020ApJ...905...62A}.

We ran a Run 0-based simulation following 1000 eruptions with a 7 day duration. To incorporate intra-ejecta shocks, we split the ejecta into two separate components. For moderate-speed novae, the ejecta velocities range from 500--2000 km s$^{-1}$ but for fast novae, this range is 1000--4000 km s$^{-1}$ \citep{1994MNRAS.271..155O}. As we are considering recurrent nova eruptions and therefore dealing with fast novae, we used the latter range of velocities for this test. We ejected half the mass at 1000 km s$^{-1}$ over 3.5 days immediately followed by half of the mass at 4123 km s$^{-1}$ over the next 3.5 days, such that the total kinetic energy matched that of a 7 day eruption with an ejecta velocity of 3000 km\,s$^{-1}$. As the second half of the mass is ejected at a higher velocity than the first, we see intra-ejecta shocks as the later ejecta overtakes and interacts with the earlier ejecta. Again, after around 100 years, the inner and outer edge of the NSR shells created from ejecta with and without intra-eruption shocks follow the same evolutionary trend. 

In Sections~\ref{Eruptions timescales} and \ref{Intra-eruption shocks}, we have demonstrated that the long term evolution of nova ejecta is not affected by the nova eruption duration nor by the presence of intra-ejecta shocks, and consequently, neither is any NSR. NSR evolution only depends upon the total kinetic energy of the ejecta and the surrounding medium\footnote{Here, we are considering a pure adiabatic scenario.}.

\section{Observational predictions}\label{Observational predictions}

Here, we investigate the evolution of NSR observables, derived from Run 1 (Section~\ref{Reference simulation}), in part to inform any NSR follow-up observations or searches. The simplest and computationally cheapest way to predict the emission over a full simulation of a NSR is by assuming a pure hydrogen environment. We can thus compute the ionisation fraction ($f$), emission measure (EM), recombination time-scale, X-ray luminosity, and H$\alpha$ emission. In general, an assumption of pure hydrogen provides a good estimate of $f$ throughout the NSR.

\subsection{Evolution of emission measure}\label{Evolution of the ionisation fraction and emission measure}

Assuming pure hydrogen, we employed the \citet{1921RSPSA..99..135S} equation to compute $f$ for each NSR cell across all epochs.  As the number of free protons in a medium of fully ionised hydrogen is equal to the number of electrons, we define the EM in each NSR cell as the square of electron density ($n_\mathrm{e}^2$) integrated over the volume of the spherical shell represented by each cell.

The EM from the different Run 1 NSR regions (cavity, ejecta pile-up, shell, and the entire NSR) at each epoch were calculated by integrating over all shells within each region. The mean ionisation fraction ($\bar{f}$) in each region, per epoch, was computed in a similar fashion while also weighting each shell by density.

\begin{figure}
\centering
\includegraphics[width=\columnwidth]{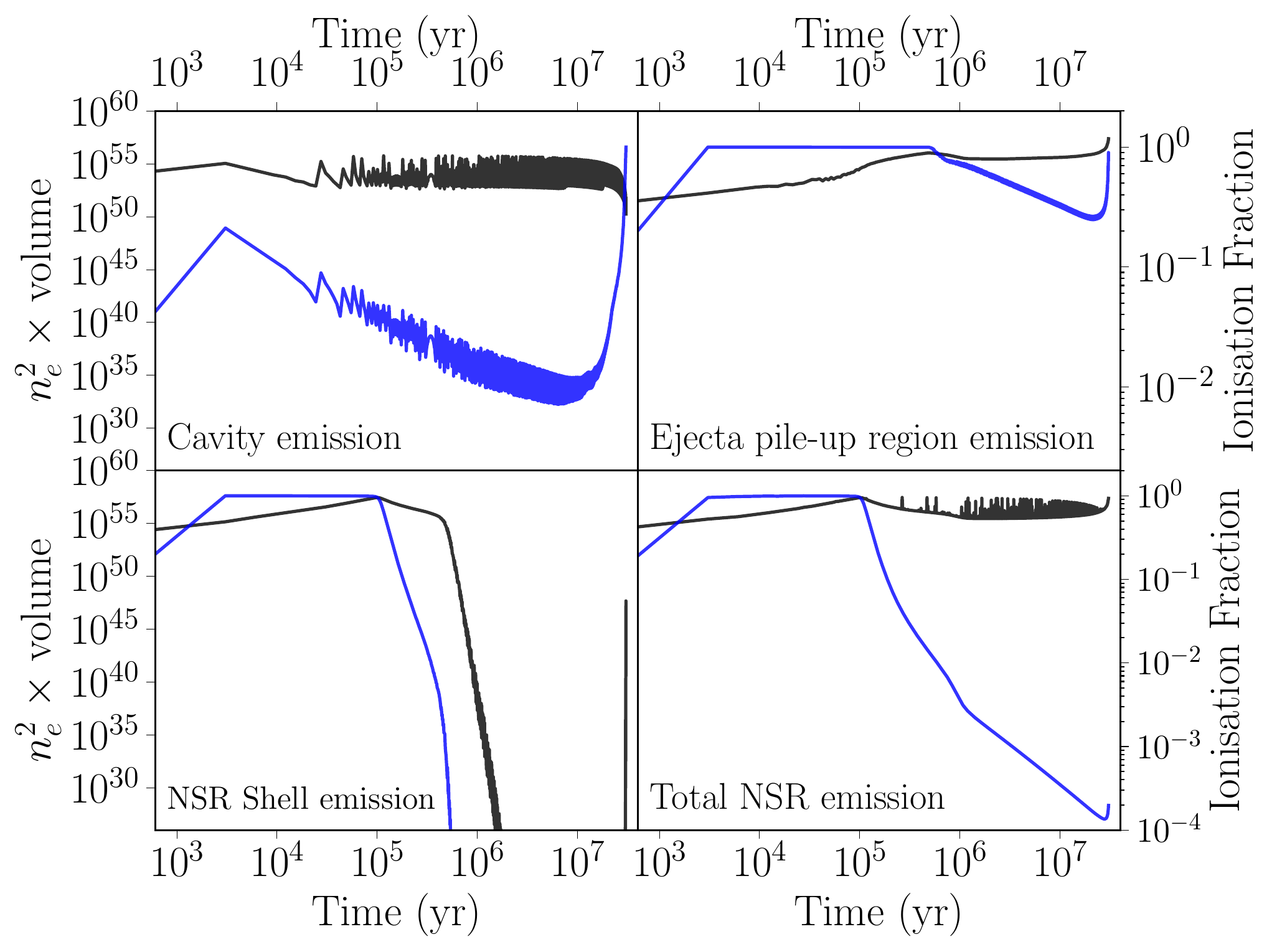}
\caption{Run 1 ionisation fraction (blue) and emission measure (black) evolution within the cavity, the pile-up region, shell and the entire NSR. The `bump' in the cavity emission measure at ${\sim} 3 \times 10^5$ years is an artefact of the temporal sampling. Note that the cavity and ejecta pile-up region panels have different ionisation fraction limits to the NSR shell and total NSR panels.}
\label{Evolution of ionisation frac and emission measure}
\end{figure}

\begin{figure}
\centering
\includegraphics[width=\columnwidth]{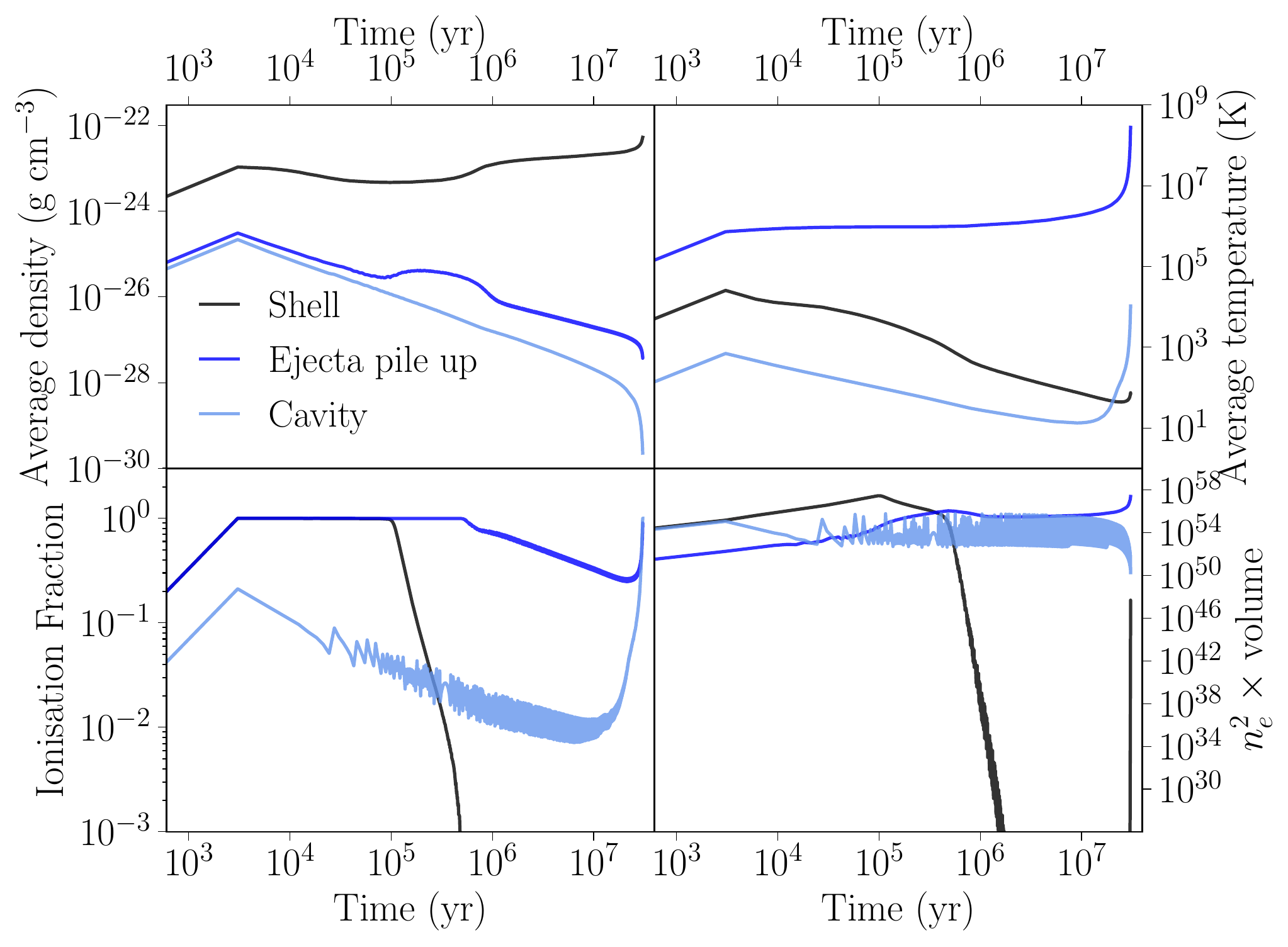}
\caption{Run 1 average density, average temperature, ionisation fraction, and emission measure evolution within the cavity (light blue), the pile-up region (dark blue) and the NSR shell (black).}
\label{d_T_ionfrac_emission}
\end{figure}

The evolution of $\bar{f}$ and the total EM for each region is shown in Figure~\ref{Evolution of ionisation frac and emission measure}. In Figure~\ref{d_T_ionfrac_emission}, we show the evolution of $\bar{f}$ and EM for the cavity, ejecta pile-up region and shell alongside the evolution of the mean density and temperature.

As illustrated in Figure~\ref{d_T_ionfrac_emission}, the mean temperature of the pile-up region is approximately a few $\times10^6$ K for ${\sim}2.7 \times 10^7$ years (except during the initial eruption) and begins to increase toward ${\sim}2.8 \times 10^8$ K during the next ${\sim}3.5 \times 10^6$ years of the NSR evolution. The density in this region decreases by over a factor of 2 as it grows but the extremely high temperatures maintains $\bar{f}\gtrsim25\%$. As a result, the EM from this region remains high. Within the cavity, $\bar{f}\gtrsim1\%$, and so even with density decreasing over time, the emission from this region remains a contributing, albeit fluctuating, factor, until latter stages of NSR evolution.

If we focus on $\bar{f}$ within the NSR shell in Figure~\ref{d_T_ionfrac_emission}, we see the effect of recombination as a result of the high densities and cooling. For the first $10^5$ years, the NSR shell is fully ionised, here the shell EM is high and the dominant source. After this, $\bar{f}$ in the shell decreases to negligible levels as the material recombines and remains neutral for the majority of the NSR lifetime (from ${\sim}10^5$ years to ${\sim}3 \times 10^7$ years) which, combined with an almost constant mean density during this period, leads to a drop in EM to effectively zero. However, as with the other regions, the late-time frequent highly energetic eruptions begin to re-heat the NSR shell, increasing $\bar{f}$ marginally. The high NSR shell density at this time leads to the NSR shell again contributing to the EM at the end of the simulation.

The evolution of the total NSR EM is shown in the bottom-right panel of Figure~\ref{Evolution of ionisation frac and emission measure}. The NSR shell initially dominates the EM as this high density region begins to sweep up ISM. After ${\sim}5 \times 10^5$ years, the average temperature within the shell has decreased enough for the material to recombine, resulting in a dramatic reduction in EM from this region. As a result, the total EM from the NSR becomes dominated by the pile-up region between ${\sim}5 \times 10^5$ years and ${\sim}3 \times 10^7$ years, with additional contribution from the fluctuating cavity emission throughout (originating from the eruptions themselves). Once the later stages have been reached (the last ${\sim}5 \times 10^5$ years), with frequent highly energetic ejecta, the rate of ionisation within the very high density shell (particularly at the inner edge)  leads to a substantial increase in EM from this region. However, unlike at early times when EM was dominated by the entire NSR shell, the emission at these later times emanates exclusively from the pile-up region and the inner edge of the shell.

\subsection{Evolution of recombination time}\label{Evolution of recombination time}

The \texttt{Morpheus} code only informs as to the ionisation state of the material based upon the dynamics of the simulation; it does not include radiative transfer. As such, when considering the emission from simulated NSRs, and indeed their observability, we must also take account of recombination timescales ($t_\mathrm{recomb}$).

As recombination time depends upon the relative abundances of the gas, from this point on we assume that all material is of Solar composition. While this will be a good approximation for the ISM it will be less so for the ejecta. However, the NSR is predominantly swept up ISM. Abundances from \citet{2000ApJ...542..914W} were utilised to determine $f$ for H, He, C, N, O, Ne, Na, Mg, Al, Si, P, S, Cl, Ar, Ca, Ti, Cr, Mn, Fe, Co, and Ni within the NSR. We compute the minimum recombination time for all cells of Run 1 by assuming the NSR is fully ionised, thus providing a lower limit on recombination time for each cell.

\begin{figure}
\centering
\includegraphics[width=\columnwidth]{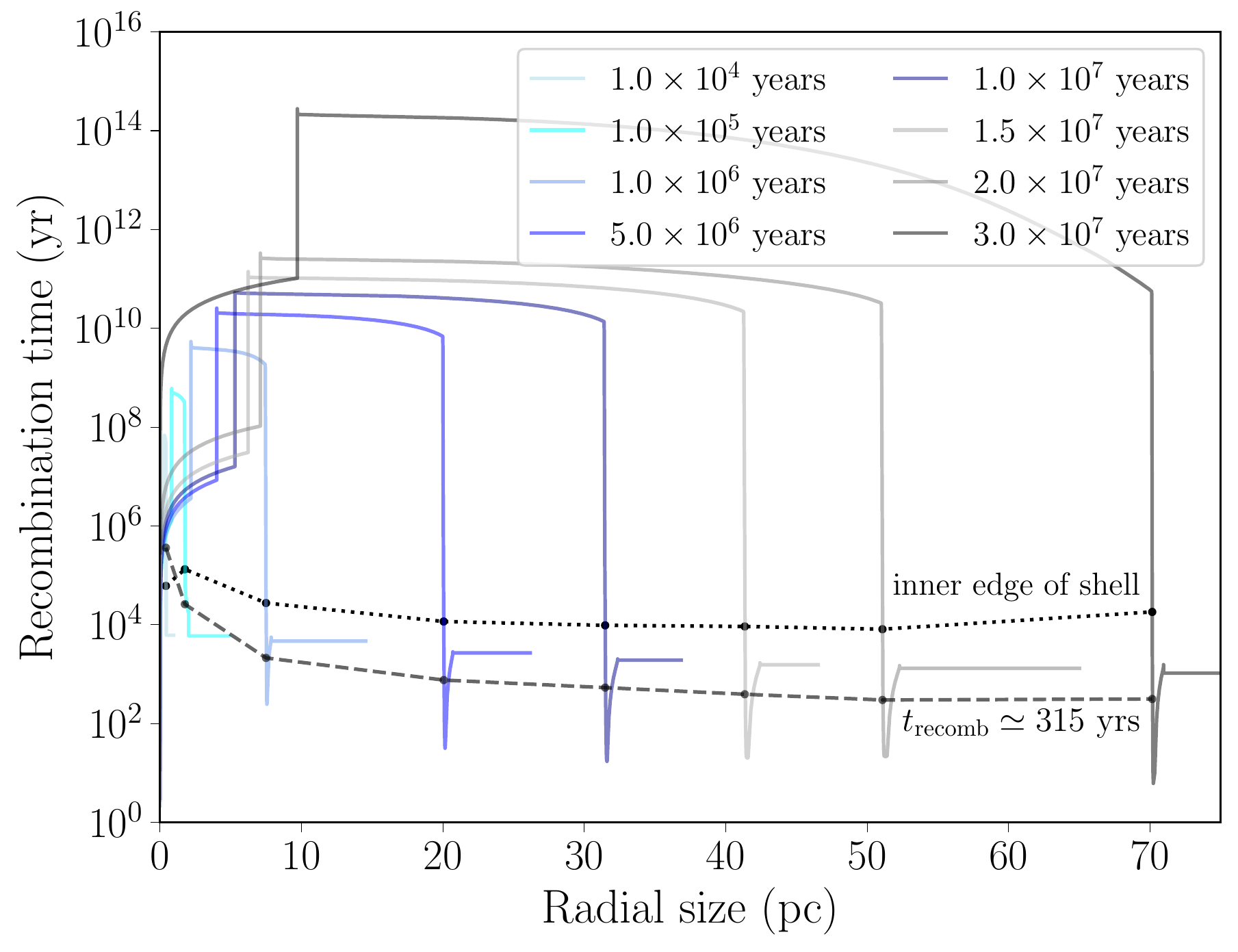}
\caption{Run 1 recombination time evolution at various epochs when assuming that all material is completely ionised. The median mass weighted recombination time at each epoch is represented with the dashed line. The thick black dotted line traces the inner edge of the NSR shell.}
\label{recomb time evolution}
\end{figure}

The recombination time evolution across the entire Run 1 NSR remnant is shown in Figure~\ref{recomb time evolution}. Here, we see that maximum recombination time within the NSR shell (except for the first epoch considered) is always $\lesssim 3 \times 10^4$ years and with the peak always being at the inner edge of the shell, the minimum recombination time of the shell approximately corresponds with the peak density. As the evolving WD approaches $\text{M}_\mathrm{Ch}$, the amount of ionised mass within the NSR ejecta pile-up region (effectively the entire NSR), reaches ${\sim}10\,\mathrm{M}_\odot$, as gas within the pile-up region is heated by the late-time frequent and energetic eruptions. This is once again reflected in the moderate rise of the recombination time at the inner edge of the shell in Figure~\ref{recomb time evolution} (the thick black dotted line tracing the NSR shell inner edge). Notably, the mass weighted median recombination time (indicated by the dashed line) remains essentially constant throughout after ${\sim}5 \times 10^6$ years, hence we adopt $t_\mathrm{recomb}=315$\,yr throughout the Run 1 NSR shell (the mass weighted median recombination time during the epoch when $P_\mathrm{rec}=1$\,yr).

\subsection{Evolution of X-ray luminosity}\label{Evolution of X-ray luminosity}

Following \citet{Vaytet_Thesis}, \citet{2011ApJ...740....5V} and \citetalias{2019Natur.565..460D}, we compute the EM contribution from each Run 1 NSR spherical shell (as defined by the simulation cells) and then bin the EM contribution into 95 logarithmically divided temperature bins ranging from 149 K to ${\sim}3.9 \times 10^9$ K (based on the shell/cell temperature). The temperature-binned EMs are used as inputs to \texttt{XSPEC}. Within \texttt{XSPEC}, we utilise the APEC \citep{2001ApJ...556L..91S} model which computes an emission spectrum containing lines for H, He, C, N, O, Ne, Mg, Al, Si, S, Ar, Ca, Fe, and Ni with Solar abundances (He fixed at cosmic) from a collisionally-ionised diffuse gas.

The EM histograms can also be used to broadly explore the evolution of NSR emission as a function of photon energy and hence wavelength. Tracking the emission evolution for the Run 0 NSR \citepalias[see Extended Data Figure 7 in][]{2019Natur.565..460D} reveals that it starts off at high temperatures, emitting mostly in X-rays at ${\sim}$1\,keV as in Run 0, as the eruptions are immediately frequent and highly energetic. But, as the NSR shell grows and cools, the EM peak moves toward lower energies, ending in the optical/NIR region (${\sim}2 \times 10^{-3}$\,keV) after the full $10^5$ eruptions. A logarithmic extrapolation of the EM indicates that the present day peak might be in the infrared, around 12--13\,$\mu$m, and could be a potential target for {\it JWST} \citepalias{2019Natur.565..460D}.

On the other hand, the Run 1 NSR begins with the peak EM at low energies (optical/NIR) due to the long period between the initial low energy eruptions allowing the NSR to cool. The temperature of the NSR {\it as a whole} remains low throughout the evolution and the EM peak remains at low energies through all 1,900,750 eruptions.

Separating the EM evolution into the component NSR parts, namely the cavity, pile-up, and shell, provides the contributions from each of these regions. The cavity emission remains relatively low compared to other regions throughout the full evolution. For the first ${\sim}10^4$ eruptions, the cavity emits in the optical/NIR regime. However, when the recurrence period approaches one year, the contribution from the cavity, albeit small, branches across to higher energies. This may be attributed to the ejected material colliding with the inner edge of the pile-up region.

Emission levels from the pile-up region are considerably higher than from the cavity and contribute more to the X-ray emission at later times as incoming ejecta continuously shock-heat this region. In fact, after the full 1,900,750 eruptions, a portion of the pile-up region emits in excess of 100\,keV. In contrast, the NSR shell emits mostly in the optical at early times before peaking after only $10^3$ eruptions, when the majority of the emission lies in the NIR. Beyond this epoch, for the entire evolution of the NSR, the shell contributes a negligible amount of emission and it remains the coolest part of the NSR, largely shielded from the highly energetic material.

\begin{figure*}
\centering
\includegraphics[width=0.45\textwidth]{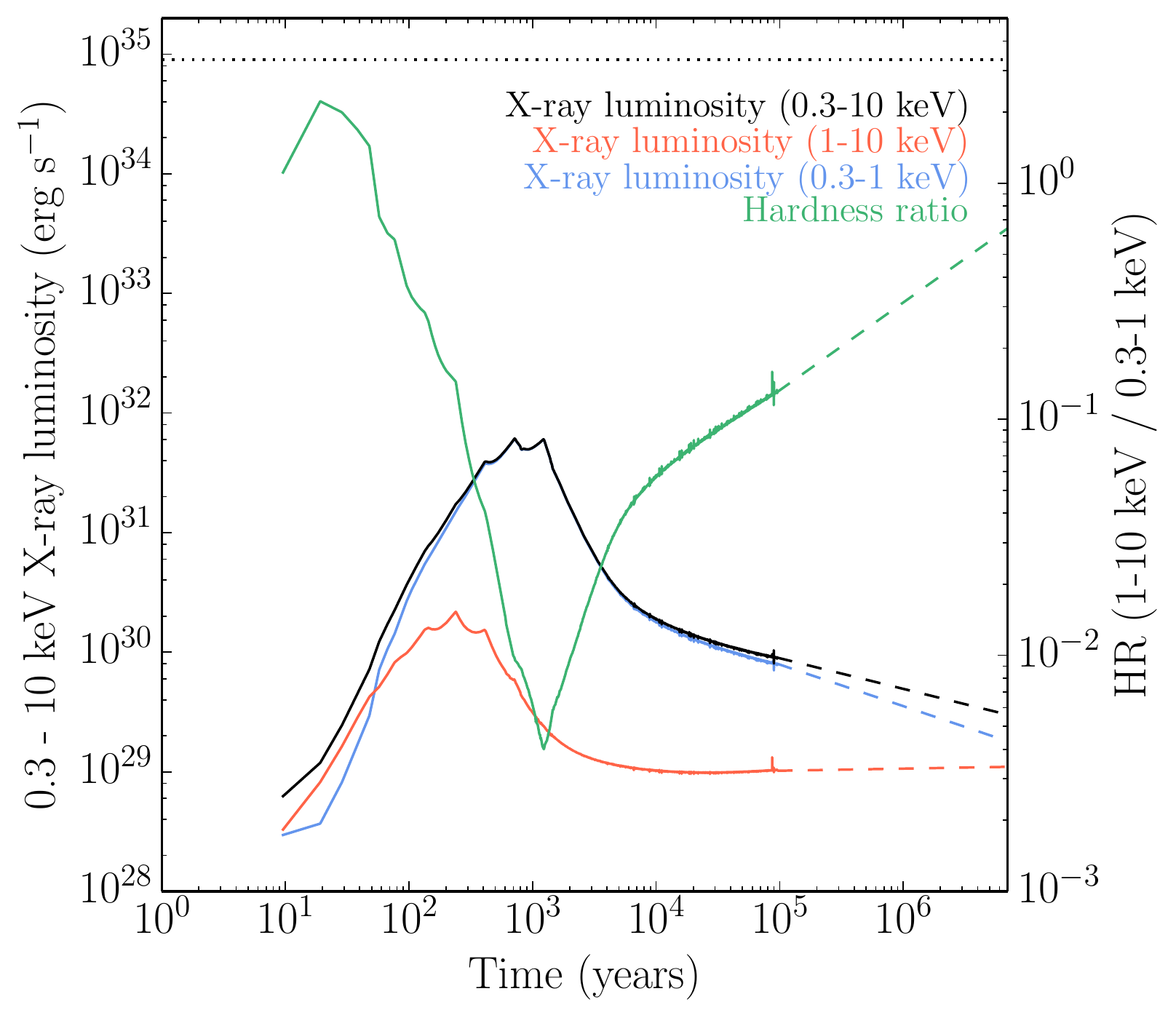}
\centering
\includegraphics[width=0.51\textwidth]{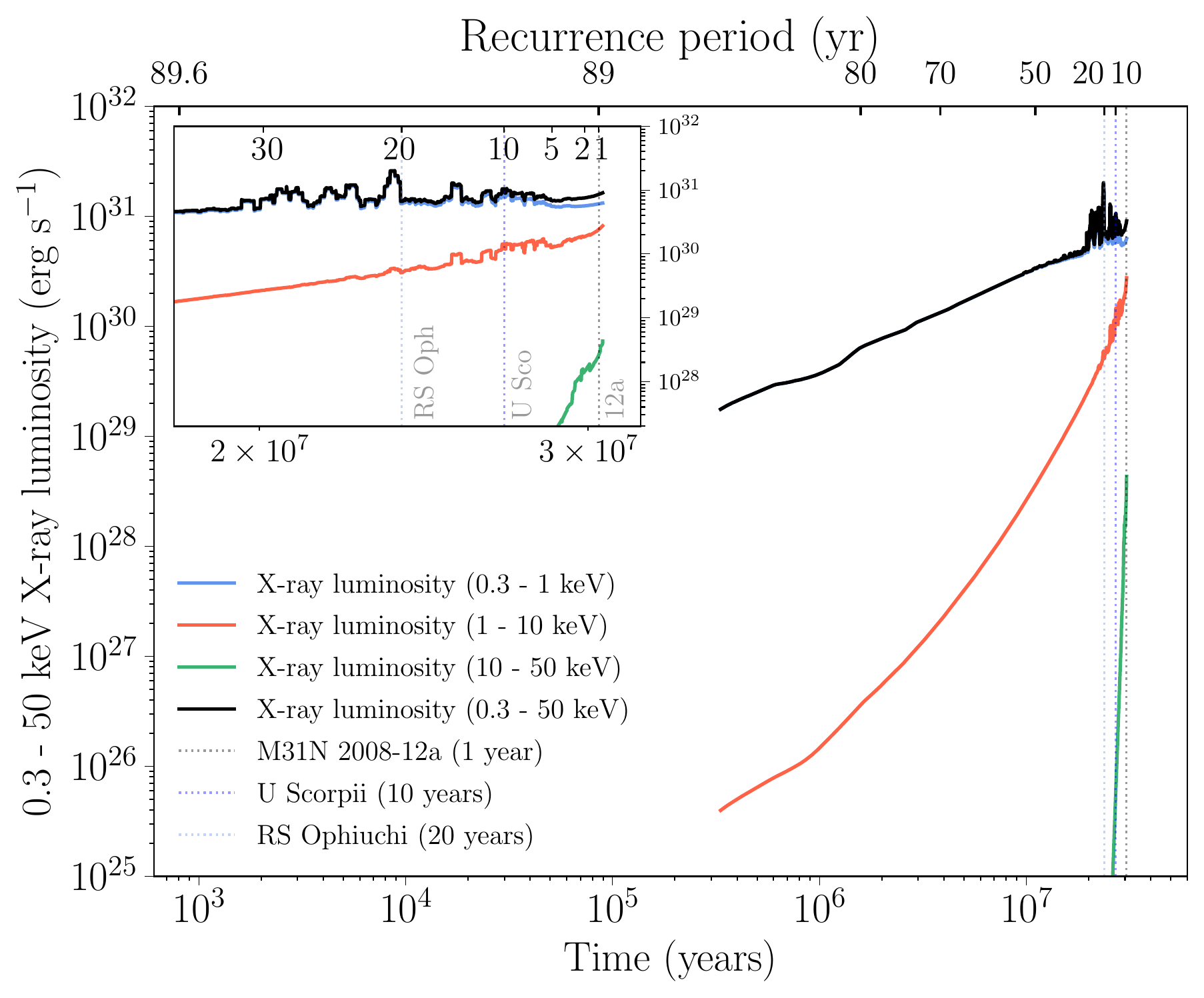}
\caption{Left: Reproduced from \citetalias{2019Natur.565..460D} (see their Figure 5) showing the evolution of the Run 0 synthetic X-ray luminosity. The soft (0.3--1\,keV; blue), hard (1--10\,keV; red) and total X-ray luminosity (0.3--10\,keV; black) are shown alongside the hardness ratio (hard/soft; green). The horizontal dashed line indicates the 3$\sigma$ upper limit derived from extensive and deep XMM-Newton observations \citepalias[see][for more details]{2019Natur.565..460D}. Right: Evolution of the Run 1 synthetic X-ray (0.3--50\,keV) luminosity with respect to elapsed time (bottom abscissa) and recurrence period (top abscissa). The soft (0.3--1\,keV; blue), hard (1--10\,keV; red), harder X-rays (10--50\,keV; green) are shown alongside the total X-ray luminosity (0.3--50\,keV; black). Recurrence period for the RNe 12a, U\,Scorpii and RS\,Ophiuchi are shown by vertical grey lines and the inset zooms in on the X-ray luminosity from $1.8 \times 10^7$ years to the end of the evolution.}
\label{X-ray evolution comparison}
\end{figure*}

We use the EM to predict the evolution of the Run 1 X-ray luminosity. We assumed that our simulated NSR is at a distance of 778 kpc \citep[][i.e., within M\,31]{1998ApJ...503L.131S}. To remove the impact of single eruptions, we re-bin to a lower temporal resolution. This is illustrated in Figure~\ref{X-ray evolution comparison} with comparison to the X-ray luminosity evolution from the NSR created in Run 0.

As shown in the left plot of Figure~\ref{X-ray evolution comparison}, for the Run 0 NSR, the X-ray luminosity peaks at ${\sim}6 \times 10^{31}\,\text{erg}\,\text{s}^{-1}$ after approximately $10^3$ years (equivalent to $10^3$ eruptions for Run 0). This luminosity then fades to ${\sim}9 \times 10^{29}\,\text{erg}\,\text{s}^{-1}$ after $10^5$ years/eruptions and with a power-law extrapolation to the latest time, representing present day in \citetalias{2019Natur.565..460D}, the total X-ray luminosity drops to ${\sim}3 \times 10^{29}\,\text{erg}\,\text{s}^{-1}$. As detailed in \citetalias{2019Natur.565..460D}, the X-ray luminosities predicted for the entire NSR evolution lie well below the $3\sigma$ upper limiting luminosity of ${\sim}9 \times 10^{34}\,\text{erg}\,\text{s}^{-1}$ constrained by archival X-ray observations (see horizontal dotted line in the left plot of Figure~\ref{X-ray evolution comparison}).

The Run 1 NSR X-ray luminosity follows an entirely different evolution from Run 0 (see right plot of Figure~\ref{X-ray evolution comparison}). While X-ray emission is predicted from the onset of Run 0, we predict negligible X-ray emission from the Run 1 NSR until $3 \times 10^5$ years; $P_\mathrm{rec}\lesssim85$\,yr. From that point, the X-ray luminosity rises as the recurrence period falls and the ejecta become more energetic, increasing significantly during the final ${\sim}10^7$ years. Starting at ${\sim}2 \times 10^{29}\,\text{erg}\,\text{s}^{-1}$ after ${\sim}3 \times 10^5$ years, the initial X-ray luminosity is dominated by soft emission between 0.3--1\,keV. 

The influence of the more frequent and energetic eruptions becomes evident over the next 26 Myr as harder emission from shock-heating, with energies between 1--10\,keV, reaches ${\sim}1.5 \times 10^{30}\,\text{erg}\,\text{s}^{-1}$ after ${\sim}2.7 \times 10^7$ years (see inset of the right plot in Figure~\ref{X-ray evolution comparison}), contributing greatly to the total X-ray luminosity of ${\sim}1 \times 10^{31}\,\text{erg}\,\text{s}^{-1}$ at this epoch. However, this is still much fainter than typical nova X-ray luminosities such as, for example, M31N\,2004-01b, 2005-02a, and 2006-06b with $L_{\text{X}} = (11.1 \pm 1.6) \times 10^{36}\,\text{erg} \,\text{s}^{-1}$, $2.6 \times 10^{37}\,\text{erg}\,\text{s}^{-1}$ and $(3.6 \pm 0.3) \times 10^{36}\,\text{erg}\,\text{s}^{-1}$, respectively\footnote{Unabsorbed luminosity between 0.2--10\,keV.} \citep[see][for a large sample of M\,31 CNe X-ray luminosities]{2010A&A...523A..89H,2011A&A...533A..52H}. Instead, this X-ray luminosity is more akin to that seen in quiescent novae such as ${\sim}6 \times 10^{31} \ \text{erg} \ \text{s}^{-1}$ for RS Ophiuchi \citep{2022arXiv220503232P}.

The NSR X-ray luminosity then continues to increase for the remainder of the evolution, ending with a luminosity of ${\sim}1 \times 10^{31}\,\text{erg}\,\text{s}^{-1}$. This is due to hard emission (1--10\,keV) becoming increasingly significant, with harder emission between 10--50\,keV appearing in the final $4 \times 10^6$ years. If we consider the $P_\mathrm{rec}=1$\,yr epoch, the Run 1 NSR X-ray luminosity is ${\sim}9 \times 10^{30}\,\text{erg}\,\text{s}^{-1}$ (see inset of the right plot in Figure~\ref{X-ray evolution comparison}). This is $30 \times$ greater than the present day extrapolated luminosity from Run 0.

\subsection{Evolution of H\texorpdfstring{\boldmath{$\alpha$}}{alpha} flux}\label{Evolution of Halpha flux}

\begin{figure}
\centering
\includegraphics[width=\columnwidth]{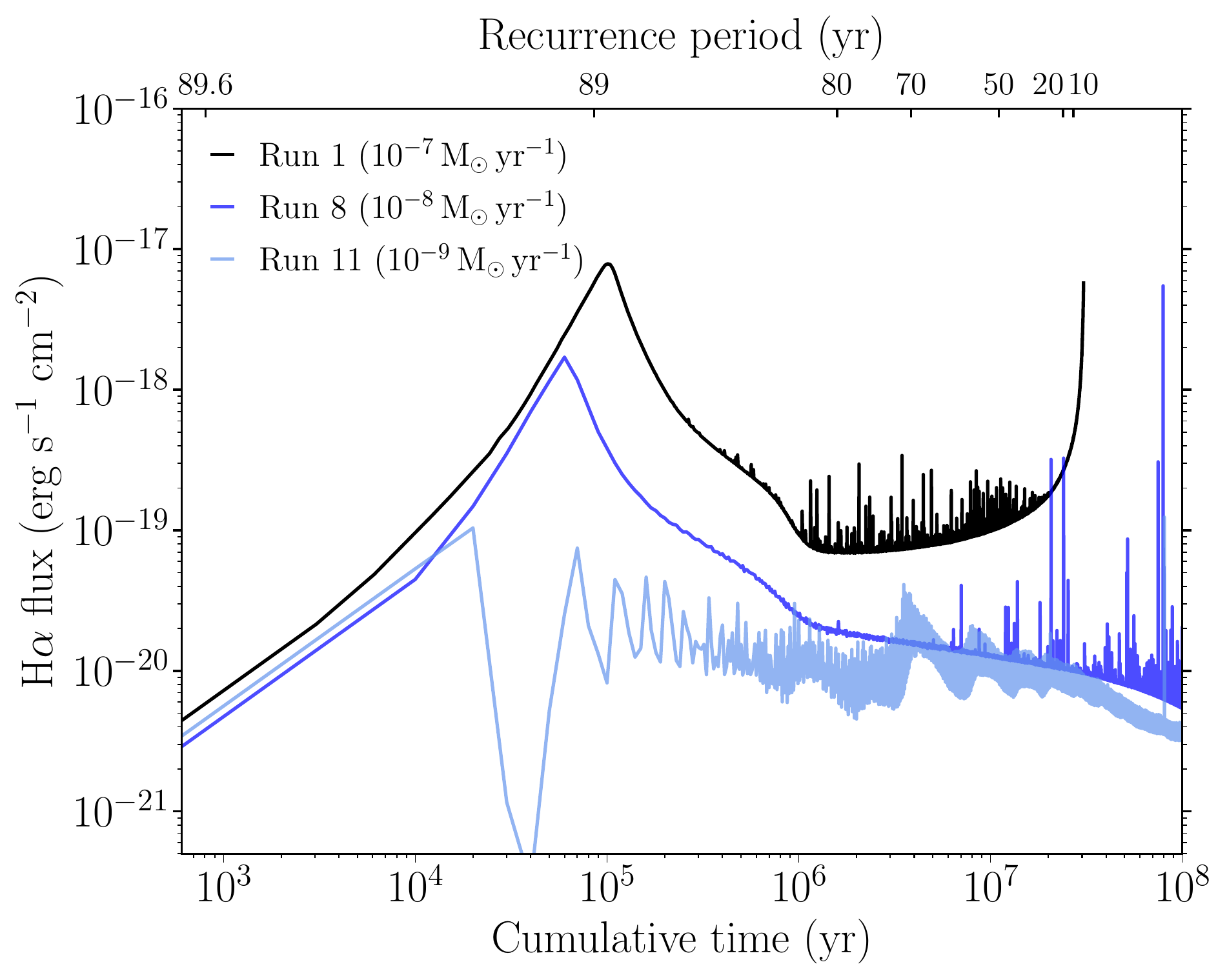}
\caption{Run 1, 8 and 11 H$\alpha$ flux evolution from a NSR at the distance of M\,31 with respect to elapsed time. The top abscissa indicates the associated recurrence period {\it only} for Run 1. The nova systems within Run 8 and Run 11 have very different recurrence periods after the same cumulative time. Individual eruptions are responsible for the `spikes' in each run.}
\label{Halpha luminosity evolution}
\end{figure}

From observations of the 12a NSR, we know such structures should be visible through their H$\alpha$ emission \citepalias{2019Natur.565..460D}. As such, we utilised Run 1 to predict the evolution of H$\alpha$ emission from a NSR in a similar manner to that described in \citet{AnderssonThesis}. The H$\alpha$ luminosity was calculated by convolving the EM histograms with the appropriate temperature-dependent recombination coefficient for the given temperature \citep[from][]{1991A&A...251..680P}. The NSR was placed at the distance of M\,31 and we applied extinction of $A_{\text{H}_{\alpha}} = 0.253$ to find the H$\alpha$ flux across the simulated NSR. The evolution of the Run 1 NSR H$\alpha$ flux is presented in Figure~\ref{Halpha luminosity evolution}.

The Run 1 NSR H$\alpha$ evolution broadly follows the EM evolution (cf.\ Figures~\ref{Evolution of ionisation frac and emission measure} and \ref{d_T_ionfrac_emission}). Initially, as the early NSR shell sweeps into the ISM, the H$\alpha$ emission (predominantly emanating from the shell) follows a roughly power-law increase, reaching a peak of ${\sim}8 \times 10^{-18}\,\text{erg s}^{-1}\,\text{cm}^{-2}$ after $10^5$\,yr. Beyond this time however, the shell temperature decreases, allowing for recombination and a consequent (power-law-like) drop in H$\alpha$ emission. As described in Section~\ref{Evolution of the ionisation fraction and emission measure}, between ${\sim}10^5$ yrs and ${\sim}3 \times 10^7$ yrs, the main sources of H$\alpha$ emission are the pile-up region and cavity. The cavity contribution can be seen as the numerous spikes in H$\alpha$ flux, with the later energetic eruptions from the nova colliding with the sparse material within that region. As shown in Figure~\ref{Halpha luminosity evolution}, the last ${\sim}8 \times 10^6$ years then see a dramatic increase in H$\alpha$ emission, almost exclusively coming from the highly energetic eruptions at this stage impacting the high density inner edge of the formed NSR shell and pile-up region, reaching a maximum of ${\sim}6 \times 10^{-18}\,\text{erg s}^{-1}\,\text{cm}^{-2}$ after the full $3.1 \times 10^7$ years.

As shown in Figure~\ref{Halpha luminosity evolution}, we also modelled the NSR H$\alpha$ flux evolution for Run 8 ($\dot M = 10^{-8} \, \text{M}_{\odot} \, \text{yr}^{-1}$) and Run 11 ($\dot M = 10^{-9} \, \text{M}_{\odot} \, \text{yr}^{-1}$) to explore the impact of mass accretion rate on H$\alpha$ observability. Early in their evolution, H$\alpha$ emission from these NSRs follow a similar, but much fainter, evolution to the NSR emission in Run 1 (with $\dot M = 10^{-7} \, \text{M}_{\odot} \, \text{yr}^{-1}$). However, unlike in Run 1 where the H$\alpha$ flux begins to increase beyond ${\sim}10^6$ years, the emission in Run 8 and Run 11 drops away, and continues to do so for the rest of the NSR's growth. In both Run 8 and Run 11 the H$\alpha$ flux drops to ${\sim}2 \times 10^{-20}\,\text{erg s}^{-1}\,\text{cm}^{-2}$ after $1 \times 10^7$ years (compared to ${\sim}1 \times 10^{-19}\,\text{erg s}^{-1}\,\text{cm}^{-2}$ in Run 1) and ends with ${\sim}6 \times 10^{-21}\,\text{erg s}^{-1}\,\text{cm}^{-2}$ and ${\sim}4 \times 10^{-21}\,\text{erg s}^{-1}\,\text{cm}^{-2}$, respectively, after $1 \times 10^8$ years.

We can tentatively conclude from these models, that NSR H$\alpha$ emission for systems with high accretion rates is significant early on in NSR growth (younger RNe systems) and again late on in the NSR evolution, from older RNe systems such as the RRNe. Furthermore, the brightest NSRs are the systems containing near-Chandrasekhar mass WDs. However, for systems with lower accretion rates, in which the WD is eroding, the H$\alpha$ emission at latter stages of evolution is orders of magnitude fainter than observed in high accretion systems.

\section{Comparing simulations and observations}\label{Comparing simulations and observations}

\subsection{Run 1 versus the 12a nova super-remnant:\ dynamics}\label{Reference simulations vs 12a NSR}

To determine how well these simulations recreate properties of the only known NSR, we compare them to observations of the 12a NSR. For this, we will consider the simulated NSR grown from a nova with parameters that most resemble 12a. The 12a mass accretion rate derived from observations is $(6 - 14) \times 10^{-7}\,\mathrm{M}_{\odot} \, \mathrm{yr}^{-1}$, the closest accretion rate we were able to consider is $10^{-7} \, \mathrm{M}_{\odot} \, \mathrm{yr}^{-1}$, within Runs 1--7. The 12a $P_\mathrm{rec}=1$\,yr, therefore we compare with simulations at this recurrence period (${\sim}$99.54\% through the simulations). At this point, the simulated WD mass is ${\sim}1.396 \ \mathrm{M}_{\odot}$.

The most immediate difference we see between observations and the simulations is the NSR radial size and the shell thickness. Within the reference simulation (Run 1; $n=1$), the NSR extends to ${\sim}71.3$\,pc compared to the observed 67\,pc \citepalias{2019Natur.565..460D}. Furthermore, \citetalias{2019Natur.565..460D} assumed that 12a is located within a high density environment, which leads to a smaller NSR, more closely resembling the Run 7 NSR ($n = 100$). The shell thickness of the Run 1 NSR is ${\sim}1\%$, dramatically smaller than the 22\% derived from observations of the inner and outer edges of the 12a NSR \citepalias{2019Natur.565..460D}.

As with the first simulation (Run 0) of a NSR, the general shell structure of the NSR in Runs 1 -- 7 is reminiscent of the observed shell. They all have a very low density central cavity (not apparent in observations) with freely expanding high velocity ejecta leading up to a very hot pile-up region. Spectroscopy of an inner `knot' in the 12a NSR reveals strong [\ion{O}{iii}] emission, indicative of higher temperatures closer to the 12a system \citepalias{2019Natur.565..460D}. In the 12a observations, we see evidence for a high density shell sweeping up the surrounding ISM, which is replicated in Runs 1 -- 7. The lack of [\ion{O}{iii}] emission in the 12a shell demonstrates that the shell has cooled below the ionisation temperature of O$^{+}$ \citepalias{2019Natur.565..460D}.

We can conclude that the simulations that most resemble the 12a NSR, in terms of accretion rate and ISM density (Runs 1 -- 7), can replicate the radial size of the NSR that is observed, but not its shell thickness. As a result, we can only conclude that there must be other contributing factors in the evolution and shaping (or geometry) of these structures that we have not yet considered. In particular, we wish to explore the impact of early helium flashes as well as a non-fixed accretion rate on NSR evolution in future work. Additionally, the simulations presented in this work are one-dimensional and so are not susceptible to Rayleigh-Taylor or Richtmyer–Meshkov instabilities. This additional physics is likely to influence the dynamics of the growing shell, through for example, shell fragmentation as seen in \citet{2013ApJ...768...48T}.

But importantly, our models only simulate the dynamically grown structure, and associated emission of a NSR, they do not (yet) consider additional effects that photoionisation may have on any observed NSR (see Section~\ref{Photoionisation region?}).

\subsection{Run 1 versus the 12a nova super-remnant:\ emission}\label{Simulated NSR Emission at one year recurrence period}

We again explore the epoch of Run 1 that coincides with $P_\mathrm{rec}=1$\,yr (after $3.04 \times 10^7$\,yr) to predict the X-ray luminosity and H$\alpha$ flux, as in Section~\ref{Observational predictions}, to directly compare to the emission from 12a's NSR. 

\subsubsection{Emission measure at one year recurrence period}\label{Emission measure at one year recurrence period}

\begin{figure}
\centering
\includegraphics[width=\columnwidth]{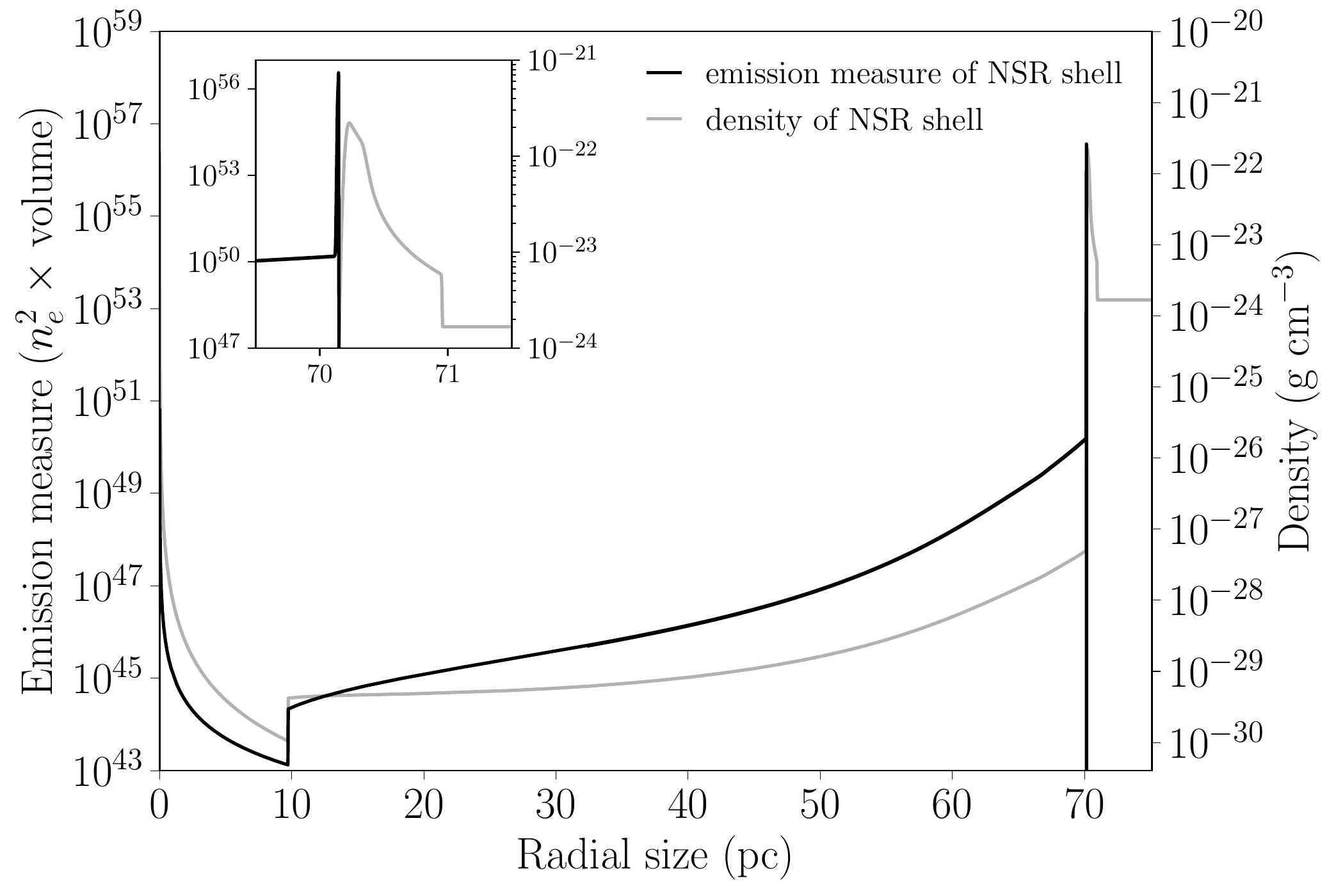}
\caption{Run 1 NSR emission measure (black) and density (grey) distribution for $P_{\mathrm{rec}} = 1$\,year. Inset focuses on the NSR shell emission peak.}
\label{Reference sim 1 year emission}
\end{figure}

We follow the procedures in Section~\ref{Evolution of the ionisation fraction and emission measure} to compute the ionisation fraction ($f$) and emission measure (EM) for the NSR at $3.04 \times 10^7$\,yr (see Figure~\ref{Reference sim 1 year emission}). Here, the entire NSR, up to the inner edge of the shell is fully ionised ($f = 1$). The ionisation decreases dramatically, to negligible values, within the shell. This fully ionised state within the cavity (up to ${\sim}10$\,pc) can be attributed to the ejecta interaction with the RGW and subsequent free expansion. Within the pile up region (between ${\sim}10 - 70.2$\,pc), gas is continuously impacted by incoming eruptions and shocks resulting in collisional excitation and, consequently, $f = 1$. Shocks are also present at the inner edge of the NSR shell (${\sim}70.2$\,pc) as gas flows through the pile-up region into the swept up shell. However, further into the shell, toward the outer edge (${\sim}71$\,pc), the gas is dynamically shielded from incoming shocks and does not experience a high level of ionisation. 

\subsubsection{Recombination time at one year recurrence period}\label{Recombination time of NSR at one year recurrence period}

In Section~\ref{Evolution of recombination time}, we computed minimum recombination times throughout the NSR evolution by considering the recombination time for a hypothetical fully ionised NSR. For the epoch of this simulation where $P_{\text{rec}}=1$\,yr, we also compute recombination time for the NSR given the $f$ predicted by the dynamic growth.

\begin{figure}
\centering
\includegraphics[width=\columnwidth]{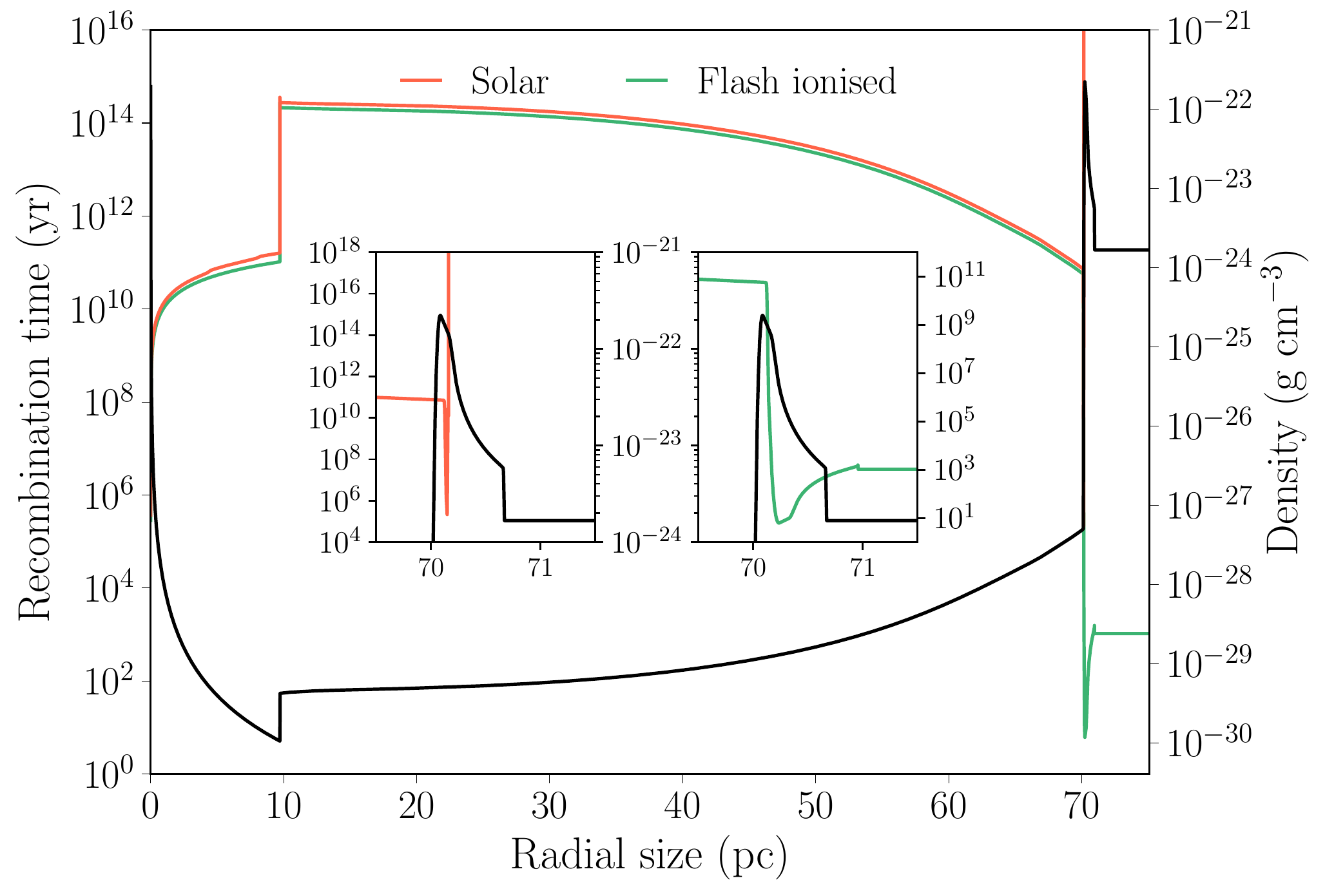}
\caption{Run 1 recombination timescale distribution for simulated Solar material (red) and completely ionised material (green) at $P_{\mathrm{rec}} = 1$ year. The black line is the corresponding density distribution. The left and right inset zoom in on the NSR shell for the simulated Solar material and the completely ionised case, respectively.}
\label{Reference sim 1 year recombination time}
\end{figure}

The recombination times for a NSR dominated by Solar material are illustrated in Figure~\ref{Reference sim 1 year recombination time} with the red line. Recombination times throughout the NSR are extremely long, owing to the extremely low density and continuous ejecta--RGW shocks within the cavity (up to ${\sim}10$\,pc). Within the pile-up region (${\sim}10-70.2$\,pc) the continual shock-heating from colliding ejecta drives the recombination time high. At the inner edge of the NSR shell, where the gas density dramatically increases, we see the recombination time drop to a $2\times10^5$\,yrs. Beyond the inner edge (at the front end of the shell), cooler neutral gas forces the recombination time to increase substantially. When considering an already fully-ionised NSR, we still see extremely long recombination times within the cavity and pile-up regions. However, we do see a significant difference within the NSR shell. As before, the recombination time drops dramatically at the inner edge yet now we see $t_\mathrm{recomb}{\sim}10$\,yr at the inner edge, rising to ${\sim}10^4$\,yrs at the outer edge.

As a result of the high recombination times within cavity and pile-up regions of the NSR and recombination times in the shell on a par with the travel time for nova ejecta to cross the NSR (${\sim}3.4 \times 10^4$\,yrs for ejecta travelling at ${\sim}2000 \, \text{km} \, \text{s}^{-1}$), the NSR shell may exhibit emission induced by photoionisation from the nova eruptions. Furthermore, if the ISM density is low enough (see Section~\ref{Photoionisation region?}), then ionising radiation from the central source might traverse the (fully collisionally ionised) inner regions of the NSR with the ability to potentially create an ionised region beyond (or within the shell of) the dynamically grown NSR.

\subsubsection{X-ray luminosity at one year recurrence period}\label{X-ray luminosity at one year recurrence period}

\begin{figure}
\centering
\includegraphics[width=\columnwidth]{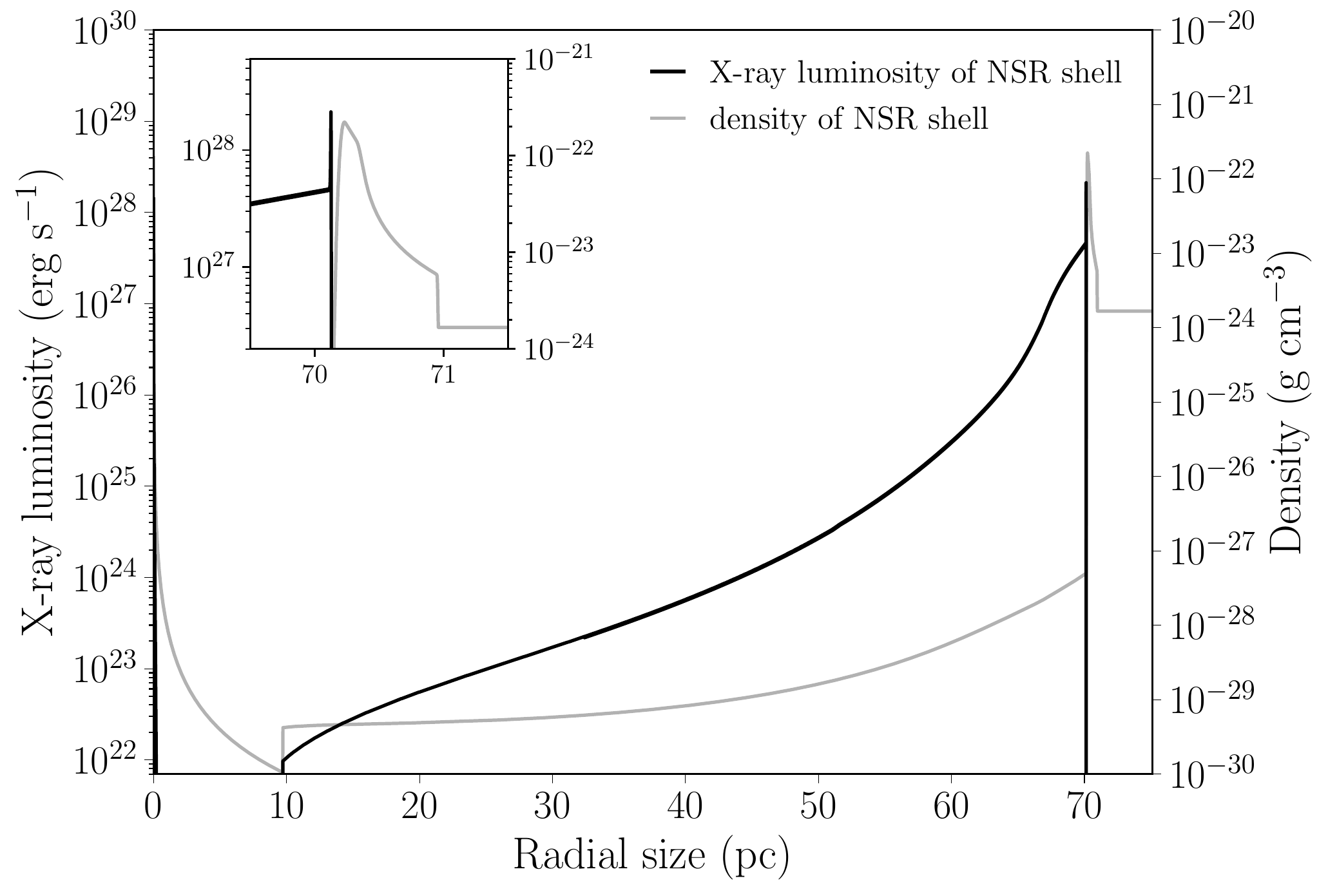}
\caption{Run 1 synthetic X-ray luminosity (without absorption; black) for $P_{\mathrm{rec}} = 1$ year, along with the NSR density distribution (grey). The inset zooms in on the NSR shell to illustrate the peak X-ray luminosity.}
\label{x-ray luminosity of 12a NSR}
\end{figure}

The output from the Run 1 NSR at the epoch coinciding with $P_\mathrm{rec}=1$\,yr was processed and passed to \texttt{XSPEC}. The X-ray luminosity as a function of radius was calculated using the APEC model (without the incorporation of absorption) and is shown in Figure~\ref{x-ray luminosity of 12a NSR}. 

At the centre of the remnant, there is a high X-ray luminosity from the underlying system due to the nova eruptions, however this is then followed by negligible emission from the cavity as the ejecta are in free expansion. Beyond this cavity, the ejecta begins to impact the higher density pile-up region (up to ${\sim}10$\,pc), leading to a significant jump in the X-ray luminosity (${\sim}1 \times 10^{22} \, \text{erg} \, \text{s}^{-1}$). As more and more ejecta contribute toward shock-heating the pile-up region further from the centre, we see a continuous increase in X-ray emission up to the inner edge of the NSR shell at ${\sim}70.2$\,pc, where $\text{L}_{\text{X-ray}} \simeq 4 \ \times 10^{27}\, \text{erg} \, \text{s}^{-1}$. The total predicted X-ray luminosity from the NSR at this epoch is ${\sim}1 \times 10^{31}\, \text{erg} \, \text{s}^{-1}$  (see Figure~\ref{X-ray evolution comparison}). This is consistent with the unabsorbed luminosity upper limit of the NSR associated with 12a derived from archival {\it XMM-Newton} observations \citepalias[${<}9 \ \times 10^{34}\ \text{erg} \ \text{s}^{-1}$;][]{2019Natur.565..460D}.

\subsubsection{H$\alpha$ flux at one year recurrence period}\label{Halpha luminosity at one year recurrence period}

\begin{figure}
\centering
\includegraphics[width=\columnwidth]{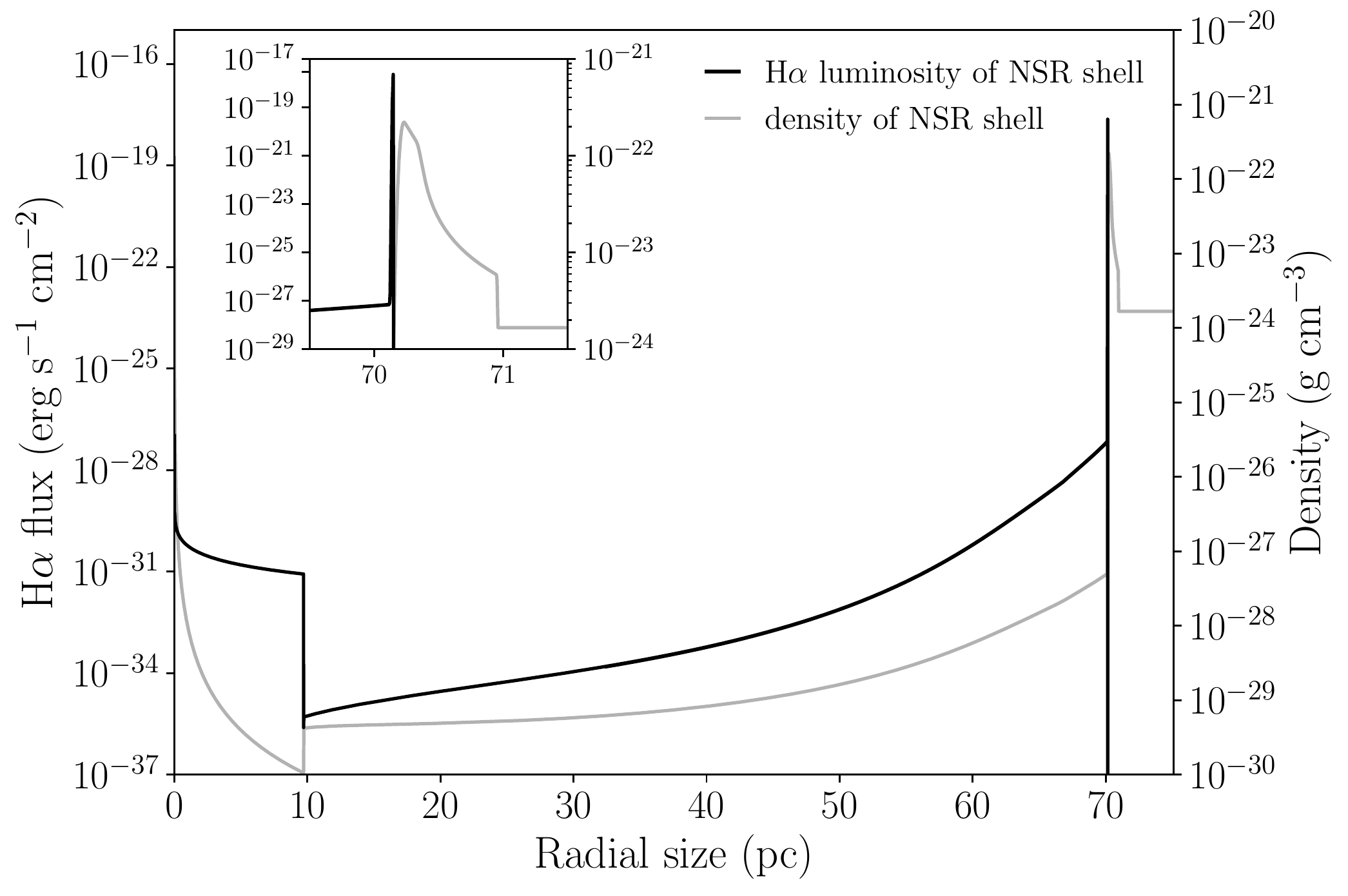}
\caption{As Figure~\ref{x-ray luminosity of 12a NSR} but for the simulated H$\alpha$ flux.}
\label{Halpha luminosity of 12a NSR}
\end{figure}

We applied the technique set out in Section~\ref{Evolution of Halpha flux} to Run 1 at the epoch corresponding to $P_{\mathrm{rec}} = 1$\,yr to compare the predicted H$\alpha$ emission to that from the 12a NSR (see Figure~\ref{Halpha luminosity of 12a NSR}). Here, we see that there is H$\alpha$ emission from the cavity and increasingly from the ejecta pile-up region, yet this always remains below ${\sim}10^{-27} \, \text{erg s}^{-1} \, \text{cm}^{-2}$. However, as is the case for X-ray emission, the majority of H$\alpha$ flux originates at the inner edge of the NSR shell. Here, the density of hydrogen is extremely high compared to the rest of the NSR and so the large amount of collisional excitation from the impacting ejecta results in high levels of recombination and H$\alpha$ emission of ${\sim}3 \times 10^{-18} \, \text{erg s}^{-1} \, \text{cm}^{-2}$, many orders of magnitudes higher than anywhere else across the NSR. The total predicted H$\alpha$ luminosity from the NSR at this epoch is $\text{L}_{\text{H}\alpha} \simeq 3.6 \ \times 10^{32}\, \text{erg} \, \text{s}^{-1}$.

\subsection{A photoionisation remnant?}\label{Photoionisation region?}

As stated in the previous section, the dynamic simulations in this work, with parameters most similar to 12a, replicate the broad observed structure, but not the shell thickness, or potentially observability, of the 12a NSR. But so far, we have only considered the growth and emission of the dynamically formed NSR. However, a proportion of the NSR will be exposed to photoionisation directly from the central system, the accretion disk, the eruptions, as well as any shock emission. As such, we consider here the formation and radial size of the photoionisation remnant, and any dependence upon ISM density. We will assume that material inwards from the NSR shell is fully ionised throughout the evolution as discussed in Section~\ref{Evolution of the ionisation fraction and emission measure} and shown in Figure~\ref{d_T_ionfrac_emission}.

\begin{figure}
\centering
\includegraphics[width=\columnwidth]{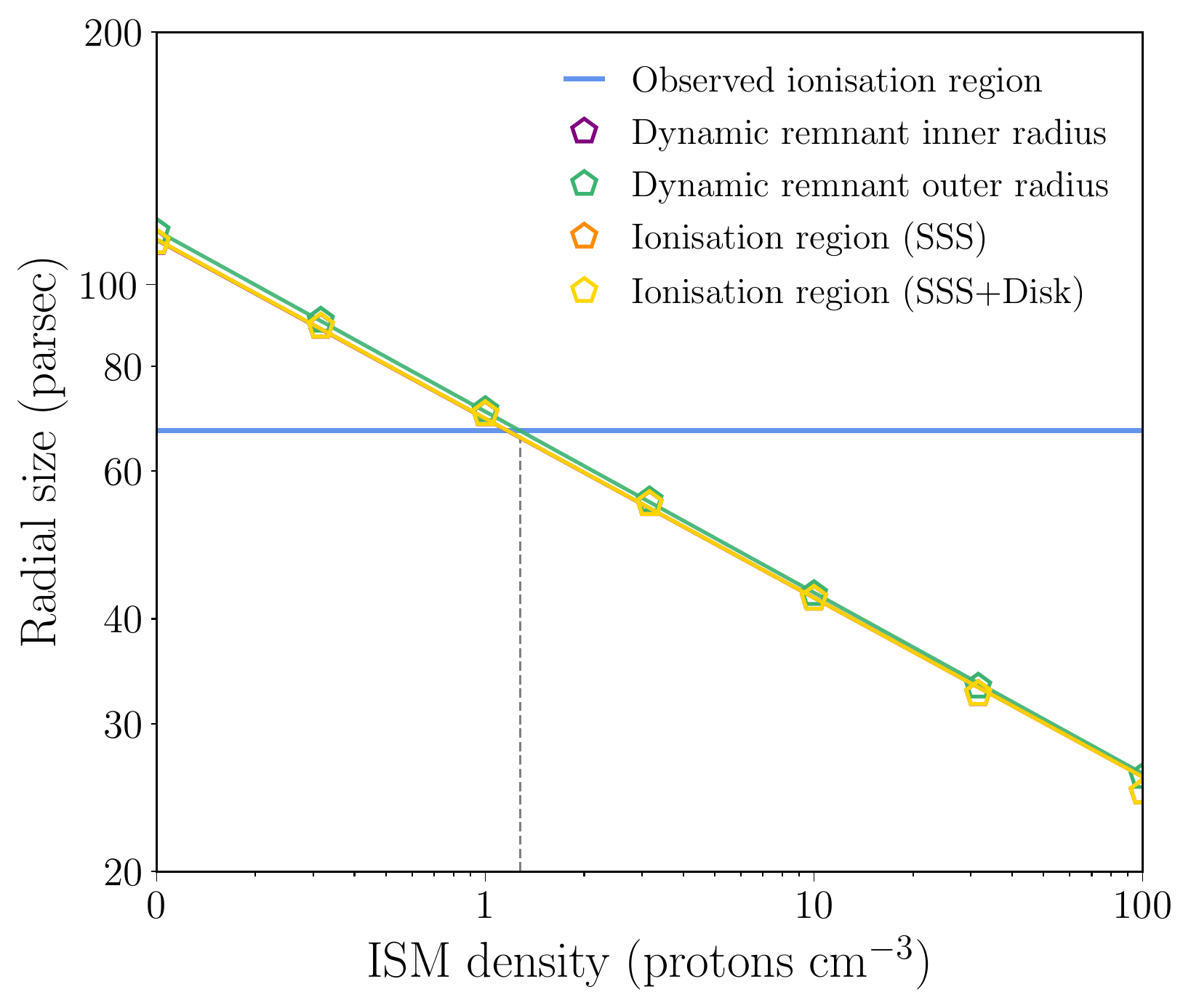}
\caption{Top: Low ISM density interpolation of Run 1--7 photoionisation regions at $P_\mathrm{rec}=1$\,yr. The purple and green lines indicate the interpolated fits to the inner and outer radii (purple and green points) of the NSR dynamical shells and the horizontal blue line is the observed outer radius of the 12a NSR emission. The outer edge of any photoionised region created by the nova emission or the combined nova and accretion emission are indicated by the orange and yellow points. The interpolation fitted to these points are shown with orange and yellow lines, respectively. The dashed vertical line indicates the ISM density at which the extrapolated outer radius fitting would equal the outer radius of the observed NSR around M\,31N 2008-12a.}
\label{12a photoionisation region ISM}
\end{figure}

We show in Figure~\ref{12a photoionisation region ISM}, the dynamical remnant inner (purple) and outer (green) radii for Runs 1--7 with respect to ISM density at the epoch when each of the runs have recurrence periods of one year and assuming that the mass accretion rate is $10^{-7} \ \text{M}_{\odot} \ \text{yr}^{-1}$. We then interpolated these points with a power-law fit.

To estimate the size of any photoionisation region generated by the nova eruptions, we can perform a Str\"{o}mgren-like analysis as $t_\mathrm{recomb}\gg P_\mathrm{rec}$ within the NSR shell and the ISM. However, because all material inwards of the NSR shell is always fully shock ionised, instead of a Str\"{o}mgren sphere, we will have a Str\"{o}mgren shell. Consequently, the photoionisation region can be estimated thus:

\begin{equation}
\label{eq:stromgren torus}
r_\mathrm{out}^3 = \frac{3S_{\star}}{4\pi n(r)^{2} \beta(r,T)}+r_{\textrm{in}}^3,
\end{equation}

\noindent where $r_\mathrm{out}$ is the outer radius of the photoionised region, $S_{\star}$ is the ionising luminosity from the source, $n(r)$ is the number density of the medium, $\beta(r,T)$ is the total recombination rate for Case B recombination \citep[see, e.g.,][]{1980pim..book.....D}, and $r_{\text{in}}$ is assumed to be the outer edge of the fully ionised region of the NSR. This $r_{\text{in}}$ was determined to be the first point from the center of the NSR in which the ionisation fraction ($\bar{f}$) falls below 100\%.

We will take the ionising luminosity from the nova eruptions (or the SSS emission) as the Eddington luminosity of a $1.396\,\text{M}_{\odot}$ WD (the mass of the WD in our models at the time when $P_{\text{rec}} = 1$\,yr) minus the observed luminosity of the 12a SSS, such that $\text{L}_{\text{Edd}} - \text{L}_{\text{obs}} \approx 41,400 \ \text{L}_{\odot}$ for two weeks \citep[the SSS timescale of each eruption;][]{2018ApJ...857...68H} and assume a spectrum of 15\,eV photons, giving a time averaged $S_{\star,\text{SSS}} = 6.6 \times 10^{48} \, \mathrm{photons} \, \mathrm{s}^{-1}$. Substituting this into the equation \ref{eq:stromgren torus} along with varying values for $n$ (ISM density) provides us with the width of the ionisation region for Runs 1--7 (see the orange points in Figure~\ref{12a photoionisation region ISM}). We also calculated a similar ionisation region but with the inclusion of the disk luminosity ($5910 \ \mathrm{L}_{\odot}$) such that $S_{\star,\text{disk}} = 9.4 \times 10^{47} \ \mathrm{photons} \ \mathrm{s}^{-1}$ (with this emission present at all times), alongside the SSS emission (see the yellow points in Figure~\ref{12a photoionisation region ISM}). Again, we assumed a $1.396\,\text{M}_{\odot}$ WD, $\dot{M}=10^{-7} \, \text{M}_{\odot} \, \text{yr}^{-1}$, and a spectrum of 15\,eV photons to estimate the disk luminosity using $\mathrm{L}_{\mathrm{disk}} = (G\dot{m}\mathrm{M}_{\mathrm{WD}})/  \mathrm{R}_{\mathrm{WD}}$. We also considered the contribution of ionising photons from shocks, by computing the shock emission within \texttt{XSPEC} when $P_{\text{rec}} = 1$\,yr, yielding $S_{\star,\text{shock}} = 2.9 \times 10^{41} \ \mathrm{photons} \ \mathrm{s}^{-1}$. But this is many orders of magnitude less than $S_{\star,\text{SSS}}$ and $S_{\star,\text{disk}}$ and so was not considered further. 

With the two luminosities we do consider, the widths of the ionisation regions (SSS or SSS+disk) produced can be found and are shown in Figure~\ref{12a photoionisation region ISM} with orange (SSS) and yellow (SSS+disk) points. For all of the NSRs grown in Runs 1--7, the emission from the nova system (eruptions and disk) cannot ionise the NSR shell and so the ionisation regions are fully contained within the remnant shell. This suggests that observations of NSRs should exhibit emission at the inner edge of the NSR shell. 

To test this, we created a synthetic sky image (with the inclusion of seeing) to directly compare with observations of the NSR surrounding M\,31N 2008-12a. Utilising the outer edge fitting presented in Figure~\ref{12a photoionisation region ISM}, we determined the ISM density required to grow a NSR with the same radial extent as the observed NSR around M\,31N 2008-12a (67\,pc) to be $n=1.278$. With this, we ran another simulation (Run 22) with the same system parameters as Run 1 but with an ISM density of $n=1.278$. The outer edge of the NSR grown within Run 22 extends to ${\sim}67.4$ pc (as expected) and exhibits a shell thickness of $1.1\%$. The boundary between the cavity and ejecta pile-up region is located at ${\sim}9$ pc and the inner edge of the NSR shell is at ${\sim}66.6$ pc, with a density of $2.6 \times 10^{-22}$ ($n \simeq 122$).

\begin{figure}
\centering
\includegraphics[width=\columnwidth]{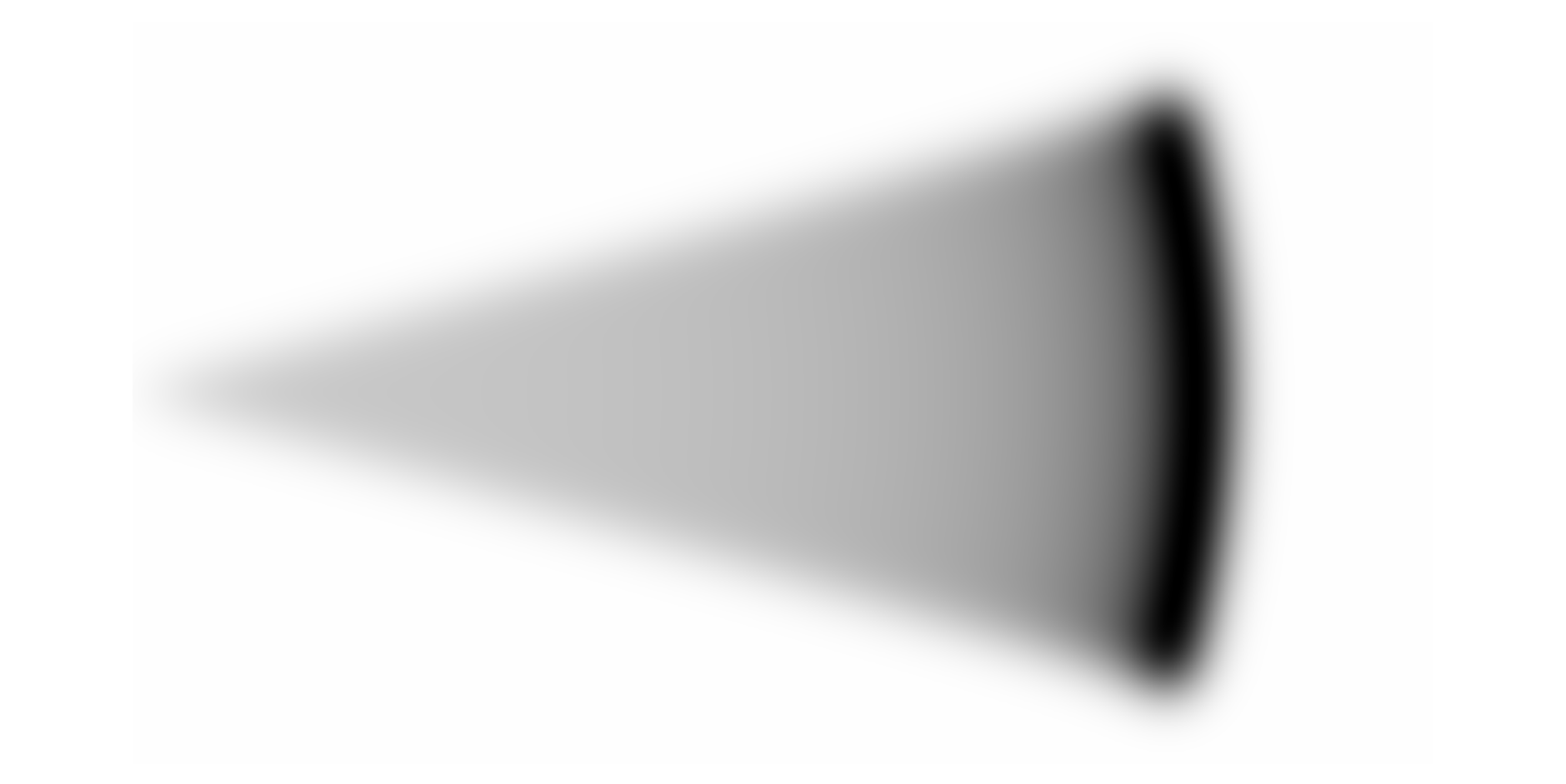}
\caption{Synthetic image (at 1 arcsecond seeing) showing a portion of the predicted H$\alpha$ emission from Run 22 at the epoch when $P_\mathrm{rec}=1$\,yr. Chosen grey scale shows linear changes in H$\alpha$ flux.}
\label{Halpha sky image}
\end{figure}

Using the same technique as described in Sections~\ref{Evolution of Halpha flux} and \ref{Halpha luminosity at one year recurrence period}, we took the Run 22 NSR at the epoch when $P_{\text{rec}} = 1$ yr and predicted its H$\alpha$ emission profile. We then generated a synthetic sky image of H$\alpha$ emission for Run 22 by integrating this H$\alpha$ emission radial profile over the volume of a sphere, collapsing this sphere along one axis into a two-dimensional image before convolving this with a Gaussian with a width of 1 arcsecond to represent the typical seeing at the Liverpool Telescope \citep[LT;][]{2004SPIE.5489..679S}. A wedge of this spherical NSR is shown in Figure~\ref{Halpha sky image}.

As can be seen in Figure~\ref{Halpha sky image}, the structure does resemble the structure of the observed remnant around M\,31N 2008-12a, as seen from the ground with the LT \citep[see Figure 8 in][]{2015A&A...580A..45D}. Specifically, we can see a negligible measure near the origin of the NSR (light grey) and a very low measure at the transitionary ejecta pile-up region (same light grey section), mimicking the LT observations. Then, at the inner edge of the shell, we see a vastly significant increase in the emission measure (dark grey band) as the ejecta that traversed the pile-up region collides with the extremely high density remnant shell. There is, however, a geometrical difference between the full synthetic sky image which uses a spherically symmetric model and the observed remnant around 12a which is elliptical, likely from an inclined torus or barrel-like structure. 

As well as replicating the 12a NSR on the sky, this is the type of structure we would also expect to observe around other novae hosting NSRs, using ground-based facilities. Based on our full suite of simulations of NSRs, we find that detectable remnants can form around novae with very different system parameters and so should actively be searched for around all types of novae, not just those with very short recurrence periods.

\section{Conclusions}\label{Conclusions}
We have presented a suite of hydrodynamical simulations of recurrent nova eruptions to determine how system parameters such as accretion rate, ISM density, WD temperature and initial WD mass affect the growth of a nova super-remnant. We follow the evolution of the WD from its formation mass up to either the Chandrasekhar mass (for high accretion rate systems) or the mass at a temporal upper limit (for lower accretion rate systems), and evolve the eruption properties as the mass changes. We utilised these simulations to predict the observational signatures associated with NSRs such as X-ray and H$\alpha$ emission, before comparing our simulations with the NSR observed around 12a, including the generation of a synthetic sky image. Here, we summarise the key results:

\begin{enumerate}
\item Dynamic nova super-remnants (NSR) should be found around all RNe, including those with long recurrence periods and lengthy evolutionary times, as the nova eruptions naturally drive their creation.
\item Unlike the \citetalias{2019Natur.565..460D} study, we find that radiative cooling plays a key part in the formation of dynamic NSRs, and significantly alters the density and thickness of the outer dynamic shell.
\item The creation of a dynamic NSR occurs whether the WD mass is increasing or decreasing, indicating that NSRs also exist around old novae with low mass WDs.
\item The evolving eruptions create NSRs many parsecs in radius comprising a very low density cavity, bordered by a very hot pile-up region, and surrounded by a cool, thin, high density shell.
\item A high density ISM restricts the NSR size, as does a high accretion rate; these parameters have the largest effect on NSR size.
\item The temperature of the WD and initial WD mass may have much less impact on NSR size, however NSRs grown from ONe WDs (${>}1.1 \ \text{M}_{\odot}$) are significantly reduced.
\item The simulated NSRs can replicate the size of the 12a NSR and can reproduce the associated structure of H$\alpha$ emission.
\item Only NSRs grown from systems with high accretion rates will currently be observable.
\end{enumerate}

NSR structures may have been overlooked within the Milky Way as they will extend across large regions of sky, far beyond their central RN. Ultimately though, the discovery of a second NSR surrounding another RN would provide strong evidence for an association between RNe and NSRs. NSRs also offer an opportunity to find unknown/unconfirmed RNe, and have the potential to point to `extinct' novae where the donor has been exhausted \citep{2021gacv.workE..44D}. Additionally, with the WD in a proportion of these systems being close to $\mathrm{M}_\mathrm{Ch}$ with the real possibility to explode as a SN\,Ia, these phenomena can also provide ``a clear and {\it persistent signpost} to the progenitor-type of that SN Ia''  \citep{2021gacv.workE..44D}, and provide a mechanism for the removal of hydrogen from the immediate vicinity of a single-degenerate SN\,Ia \citep[removing ${\sim}10^6 \, \mathrm{ M}_{\odot}$ of gas tens of parsecs from the central system;][]{2016sros.confE.137H,2021gacv.workE..44D}.

\section*{Acknowledgements}
The authors would like to gratefully thank our reviewer, Michael Shara, for his insightful suggestions that helped to improve our study and strengthened our conclusions. MWH-K acknowledges a PDRA position funded by the UK Science and Technology Facilities Council (STFC). MWH-K, MJD, EJH and PAJ receive funding from STFC grant number ST/S505559/1. This work made use of the high performance computing facilities at Liverpool John Moores University, partly funded by LJMU’s Faculty of Engineering and Technology and by the Royal Society.

\section*{Data Availability}
The data in this study can be shared on reasonable request to the corresponding author. This work was conducted with the \texttt{Morpheus} \citep{2007ApJ...665..654V} program and analysed using the Python libraries: Numpy \citep{harris2020array} and Matplotlib \citep{Hunter:2007}.

\balance
\bibliographystyle{mnras}
\bibliography{On_the_Observability_of_Recurrent_Nova_Super-Remnants}

\begin{thebibliography}{}
\makeatletter
\relax
\def\mn@urlcharsother{\let\do\@makeother \do\$\do\&\do\#\do\^\do\_\do\%\do\~}
\def\mn@doi{\begingroup\mn@urlcharsother \@ifnextchar [ {\mn@doi@}
  {\mn@doi@[]}}
\def\mn@doi@[#1]#2{\def\@tempa{#1}\ifx\@tempa\@empty \href
  {http://dx.doi.org/#2} {doi:#2}\else \href {http://dx.doi.org/#2} {#1}\fi
  \endgroup}
\def\mn@eprint#1#2{\mn@eprint@#1:#2::\@nil}
\def\mn@eprint@arXiv#1{\href {http://arxiv.org/abs/#1} {{\tt arXiv:#1}}}
\def\mn@eprint@dblp#1{\href {http://dblp.uni-trier.de/rec/bibtex/#1.xml}
  {dblp:#1}}
\def\mn@eprint@#1:#2:#3:#4\@nil{\def\@tempa {#1}\def\@tempb {#2}\def\@tempc
  {#3}\ifx \@tempc \@empty \let \@tempc \@tempb \let \@tempb \@tempa \fi \ifx
  \@tempb \@empty \def\@tempb {arXiv}\fi \@ifundefined
  {mn@eprint@\@tempb}{\@tempb:\@tempc}{\expandafter \expandafter \csname
  mn@eprint@\@tempb\endcsname \expandafter{\@tempc}}}

\bibitem[\protect\citeauthoryear{Andersson}{Andersson}{2021}]{AnderssonThesis}
Andersson T.,  2021, Master's Thesis, Liverpool John Moores University

\bibitem[\protect\citeauthoryear{{Aydi} et~al.,}{{Aydi}
  et~al.}{2020a}]{2020NatAs...4..776A}
{Aydi} E.,  et~al., 2020a, \mn@doi [Nature Astronomy]
  {10.1038/s41550-020-1070-y}, \href
  {https://ui.adsabs.harvard.edu/abs/2020NatAs...4..776A} {4, 776}

\bibitem[\protect\citeauthoryear{{Aydi} et~al.,}{{Aydi}
  et~al.}{2020b}]{2020ApJ...905...62A}
{Aydi} E.,  et~al., 2020b, \mn@doi [\apj] {10.3847/1538-4357/abc3bb}, \href
  {https://ui.adsabs.harvard.edu/abs/2020ApJ...905...62A} {905, 62}

\bibitem[\protect\citeauthoryear{{Bode} \& {Evans}}{{Bode} \&
  {Evans}}{1989}]{1989Sci...246..136B}
{Bode} M.~F.,  {Evans} A.,  1989, Science, \href
  {https://ui.adsabs.harvard.edu/abs/1989Sci...246..136B} {246, 136}

\bibitem[\protect\citeauthoryear{{Bode} \& {Kahn}}{{Bode} \&
  {Kahn}}{1985}]{1985MNRAS.217..205B}
{Bode} M.~F.,  {Kahn} F.~D.,  1985, \mn@doi [\mnras] {10.1093/mnras/217.1.205},
  \href {https://ui.adsabs.harvard.edu/abs/1985MNRAS.217..205B} {217, 205}

\bibitem[\protect\citeauthoryear{{Bode}, {O'Brien}  \& {Simpson}}{{Bode}
  et~al.}{2004}]{2004ApJ...600L..63B}
{Bode} M.~F.,  {O'Brien} T.~J.,   {Simpson} M.,  2004, \mn@doi [\apjl]
  {10.1086/381529}, \href
  {https://ui.adsabs.harvard.edu/abs/2004ApJ...600L..63B} {600, L63}

\bibitem[\protect\citeauthoryear{{Chandrasekhar}}{{Chandrasekhar}}{1931}]{1931ApJ....74...81C}
{Chandrasekhar} S.,  1931, \mn@doi [\apj] {10.1086/143324}, \href
  {https://ui.adsabs.harvard.edu/abs/1931ApJ....74...81C} {74, 81}

\bibitem[\protect\citeauthoryear{{Darnley}}{{Darnley}}{2021}]{2021gacv.workE..44D}
{Darnley} M.~J.,  2021, in The Golden Age of Cataclysmic Variables and Related
  Objects V. p.~44 (\mn@eprint {arXiv} {1912.13209})

\bibitem[\protect\citeauthoryear{{Darnley} \& {Henze}}{{Darnley} \&
  {Henze}}{2020}]{2020AdSpR..66.1147D}
{Darnley} M.~J.,  {Henze} M.,  2020, \mn@doi [Advances in Space Research]
  {10.1016/j.asr.2019.09.044}, \href
  {https://ui.adsabs.harvard.edu/abs/2020AdSpR..66.1147D} {66, 1147}

\bibitem[\protect\citeauthoryear{{Darnley}, {Ribeiro}, {Bode}, {Hounsell}  \&
  {Williams}}{{Darnley} et~al.}{2012}]{2012ApJ...746...61D}
{Darnley} M.~J.,  {Ribeiro} V.~A.~R.~M.,  {Bode} M.~F.,  {Hounsell} R.~A.,
  {Williams} R.~P.,  2012, \mn@doi [\apj] {10.1088/0004-637X/746/1/61}, \href
  {https://ui.adsabs.harvard.edu/abs/2012ApJ...746...61D} {746, 61}

\bibitem[\protect\citeauthoryear{{Darnley}, {Williams}, {Bode}, {Henze},
  {Ness}, {Shafter}, {Hornoch}  \& {Votruba}}{{Darnley}
  et~al.}{2014}]{2014A&A...563L...9D}
{Darnley} M.~J.,  {Williams} S.~C.,  {Bode} M.~F.,  {Henze} M.,  {Ness} J.-U.,
  {Shafter} A.~W.,  {Hornoch} K.,   {Votruba} V.,  2014, \aap, 563, L9

\bibitem[\protect\citeauthoryear{{Darnley} et~al.,}{{Darnley}
  et~al.}{2015}]{2015A&A...580A..45D}
{Darnley} M.~J.,  et~al., 2015, \mn@doi [\aap] {10.1051/0004-6361/201526027},
  \href {https://ui.adsabs.harvard.edu/abs/2015A&A...580A..45D} {580, A45}

\bibitem[\protect\citeauthoryear{{Darnley} et~al.,}{{Darnley}
  et~al.}{2016}]{2016ApJ...833..149D}
{Darnley} M.~J.,  et~al., 2016, \apj, 833, 149

\bibitem[\protect\citeauthoryear{{Darnley} et~al.,}{{Darnley}
  et~al.}{2017a}]{2017ApJ...847...35D}
{Darnley} M.~J.,  et~al., 2017a, \mn@doi [\apj] {10.3847/1538-4357/aa8867},
  \href {https://ui.adsabs.harvard.edu/abs/2017ApJ...847...35D} {847, 35}

\bibitem[\protect\citeauthoryear{{Darnley} et~al.,}{{Darnley}
  et~al.}{2017b}]{2017ApJ...849...96D}
{Darnley} M.~J.,  et~al., 2017b, \mn@doi [\apj] {10.3847/1538-4357/aa9062},
  \href {https://ui.adsabs.harvard.edu/abs/2017ApJ...849...96D} {849, 96}

\bibitem[\protect\citeauthoryear{{Darnley} et~al.,}{{Darnley}
  et~al.}{2019}]{2019Natur.565..460D}
{Darnley} M.~J.,  et~al., 2019, \mn@doi [\nat] {10.1038/s41586-018-0825-4},
  \href {https://ui.adsabs.harvard.edu/abs/2019Natur.565..460D} {565, 460}

\bibitem[\protect\citeauthoryear{{Drake} \& {Orlando}}{{Drake} \&
  {Orlando}}{2010}]{2010ApJ...720L.195D}
{Drake} J.~J.,  {Orlando} S.,  2010, \mn@doi [\apjl]
  {10.1088/2041-8205/720/2/L195}, \href
  {https://ui.adsabs.harvard.edu/abs/2010ApJ...720L.195D} {720, L195}

\bibitem[\protect\citeauthoryear{{Dyson} \& {Williams}}{{Dyson} \&
  {Williams}}{1980}]{1980pim..book.....D}
{Dyson} J.~E.,  {Williams} D.~A.,  1980, {Physics of the interstellar medium}.
{Manchester: University Press, Manchester, UK}

\bibitem[\protect\citeauthoryear{{Figueira}, {Jos{\'e}}, {Garc{\'{\i}}a-Berro},
  {Campbell}, {Garc{\'{\i}}a-Senz}  \& {Mohamed}}{{Figueira}
  et~al.}{2018}]{2018A&A...613A...8F}
{Figueira} J.,  {Jos{\'e}} J.,  {Garc{\'{\i}}a-Berro} E.,  {Campbell} S.~W.,
  {Garc{\'{\i}}a-Senz} D.,   {Mohamed} S.,  2018, \aap, 613, A8

\bibitem[\protect\citeauthoryear{{Gill} \& {O'Brien}}{{Gill} \&
  {O'Brien}}{1998}]{1998MNRAS.300..221G}
{Gill} C.~D.,  {O'Brien} T.~J.,  1998, \mn@doi [\mnras]
  {10.1046/j.1365-8711.1998.01899.x}, \href
  {https://ui.adsabs.harvard.edu/abs/1998MNRAS.300..221G} {300, 221}

\bibitem[\protect\citeauthoryear{{Hachisu}, {Kato}, {Nomoto}  \&
  {Umeda}}{{Hachisu} et~al.}{1999a}]{1999ApJ...519..314H}
{Hachisu} I.,  {Kato} M.,  {Nomoto} K.,   {Umeda} H.,  1999a, \mn@doi [\apj]
  {10.1086/307370}, \href
  {https://ui.adsabs.harvard.edu/abs/1999ApJ...519..314H} {519, 314}

\bibitem[\protect\citeauthoryear{{Hachisu}, {Kato}  \& {Nomoto}}{{Hachisu}
  et~al.}{1999b}]{1999ApJ...522..487H}
{Hachisu} I.,  {Kato} M.,   {Nomoto} K.,  1999b, \mn@doi [\apj]
  {10.1086/307608}, \href
  {https://ui.adsabs.harvard.edu/abs/1999ApJ...522..487H} {522, 487}

\bibitem[\protect\citeauthoryear{{Hachisu}, {Kato}  \& {Luna}}{{Hachisu}
  et~al.}{2007}]{2007ApJ...659L.153H}
{Hachisu} I.,  {Kato} M.,   {Luna} G. J.~M.,  2007, \mn@doi [\apjl]
  {10.1086/516838}, \href
  {https://ui.adsabs.harvard.edu/abs/2007ApJ...659L.153H} {659, L153}

\bibitem[\protect\citeauthoryear{{Harman} \& {O'Brien}}{{Harman} \&
  {O'Brien}}{2003}]{2003MNRAS.344.1219H}
{Harman} D.~J.,  {O'Brien} T.~J.,  2003, \mn@doi [\mnras]
  {10.1046/j.1365-8711.2003.06906.x}, \href
  {https://ui.adsabs.harvard.edu/abs/2003MNRAS.344.1219H} {344, 1219}

\bibitem[\protect\citeauthoryear{Harris et~al.,}{Harris
  et~al.}{2020}]{harris2020array}
Harris C.~R.,  et~al., 2020, \mn@doi [Nature] {10.1038/s41586-020-2649-2}, 585,
  357

\bibitem[\protect\citeauthoryear{{Harvey}, {Redman}, {Boumis}, {Kopsacheili},
  {Akras}, {Sabin}  \& {Jurkic}}{{Harvey} et~al.}{2016a}]{2016sros.confE.137H}
{Harvey} E.,  {Redman} M.~P.,  {Boumis} P.,  {Kopsacheili} M.,  {Akras} S.,
  {Sabin} L.,   {Jurkic} T.,  2016a, in Supernova Remnants: An Odyssey in Space
  after Stellar Death. p.~137 (\mn@eprint {arXiv} {1608.05016})

\bibitem[\protect\citeauthoryear{{Harvey}, {Redman}, {Boumis}  \&
  {Akras}}{{Harvey} et~al.}{2016b}]{2016A&A...595A..64H}
{Harvey} E.,  {Redman} M.~P.,  {Boumis} P.,   {Akras} S.,  2016b, \mn@doi
  [\aap] {10.1051/0004-6361/201628132}, \href
  {https://ui.adsabs.harvard.edu/abs/2016A&A...595A..64H} {595, A64}

\bibitem[\protect\citeauthoryear{{Harvey} et~al.,}{{Harvey}
  et~al.}{2020}]{2020MNRAS.499.2959H}
{Harvey} E.~J.,  et~al., 2020, \mn@doi [\mnras] {10.1093/mnras/staa2896}, \href
  {https://ui.adsabs.harvard.edu/abs/2020MNRAS.499.2959H} {499, 2959}

\bibitem[\protect\citeauthoryear{{Henze} et~al.,}{{Henze}
  et~al.}{2010}]{2010A&A...523A..89H}
{Henze} M.,  et~al., 2010, \mn@doi [\aap] {10.1051/0004-6361/201014710}, \href
  {https://ui.adsabs.harvard.edu/abs/2010A&A...523A..89H} {523, A89}

\bibitem[\protect\citeauthoryear{{Henze} et~al.,}{{Henze}
  et~al.}{2011}]{2011A&A...533A..52H}
{Henze} M.,  et~al., 2011, \aap, 533, A52

\bibitem[\protect\citeauthoryear{{Henze} et~al.,}{{Henze}
  et~al.}{2014}]{2014A&A...563A...2H}
{Henze} M.,  et~al., 2014, \mn@doi [\aap] {10.1051/0004-6361/201322426}, \href
  {https://ui.adsabs.harvard.edu/abs/2014A&A...563A...2H} {563, A2}

\bibitem[\protect\citeauthoryear{{Henze}, {Darnley}, {Kabashima}, {Nishiyama},
  {Itagaki}  \& {Gao}}{{Henze} et~al.}{2015}]{2015A&A...582L...8H}
{Henze} M.,  {Darnley} M.~J.,  {Kabashima} F.,  {Nishiyama} K.,  {Itagaki} K.,
   {Gao} X.,  2015, \mn@doi [\aap] {10.1051/0004-6361/201527168}, \href
  {https://ui.adsabs.harvard.edu/abs/2015A&A...582L...8H} {582, L8}

\bibitem[\protect\citeauthoryear{{Henze} et~al.,}{{Henze}
  et~al.}{2018}]{2018ApJ...857...68H}
{Henze} M.,  et~al., 2018, \mn@doi [\apj] {10.3847/1538-4357/aab6a6}, \href
  {https://ui.adsabs.harvard.edu/abs/2018ApJ...857...68H} {857, 68}

\bibitem[\protect\citeauthoryear{{Hillebrandt} \& {Niemeyer}}{{Hillebrandt} \&
  {Niemeyer}}{2000}]{2000ARA&A..38..191H}
{Hillebrandt} W.,  {Niemeyer} J.~C.,  2000, \mn@doi [\araa]
  {10.1146/annurev.astro.38.1.191}, \href
  {https://ui.adsabs.harvard.edu/abs/2000ARA&A..38..191H} {38, 191}

\bibitem[\protect\citeauthoryear{{Hillman}, {Prialnik}, {Kovetz}  \&
  {Shara}}{{Hillman} et~al.}{2015}]{2015MNRAS.446.1924H}
{Hillman} Y.,  {Prialnik} D.,  {Kovetz} A.,   {Shara} M.~M.,  2015, \mn@doi
  [\mnras] {10.1093/mnras/stu2235}, \href
  {https://ui.adsabs.harvard.edu/abs/2015MNRAS.446.1924H} {446, 1924}

\bibitem[\protect\citeauthoryear{{Hillman}, {Prialnik}, {Kovetz}  \&
  {Shara}}{{Hillman} et~al.}{2016}]{2016ApJ...819..168H}
{Hillman} Y.,  {Prialnik} D.,  {Kovetz} A.,   {Shara} M.~M.,  2016, \mn@doi
  [\apj] {10.3847/0004-637X/819/2/168}, \href
  {https://ui.adsabs.harvard.edu/abs/2016ApJ...819..168H} {819, 168}

\bibitem[\protect\citeauthoryear{Hunter}{Hunter}{2007}]{Hunter:2007}
Hunter J.~D.,  2007, \mn@doi [Computing in Science \& Engineering]
  {10.1109/MCSE.2007.55}, 9, 90

\bibitem[\protect\citeauthoryear{{Hutchings}}{{Hutchings}}{1972}]{1972MNRAS.158..177H}
{Hutchings} J.~B.,  1972, \mn@doi [\mnras] {10.1093/mnras/158.2.177}, \href
  {https://ui.adsabs.harvard.edu/abs/1972MNRAS.158..177H} {158, 177}

\bibitem[\protect\citeauthoryear{{Iben} \& {Tutukov}}{{Iben} \&
  {Tutukov}}{1984}]{1984ApJS...54..335I}
{Iben} I. J.,  {Tutukov} A.~V.,  1984, \mn@doi [\apjs] {10.1086/190932}, \href
  {https://ui.adsabs.harvard.edu/abs/1984ApJS...54..335I} {54, 335}

\bibitem[\protect\citeauthoryear{{Kato}, {Saio}  \& {Hachisu}}{{Kato}
  et~al.}{2015}]{2015ApJ...808...52K}
{Kato} M.,  {Saio} H.,   {Hachisu} I.,  2015, \mn@doi [\apj]
  {10.1088/0004-637X/808/1/52}, \href
  {https://ui.adsabs.harvard.edu/abs/2015ApJ...808...52K} {808, 52}

\bibitem[\protect\citeauthoryear{{Kato}, {Saio}  \& {Hachisu}}{{Kato}
  et~al.}{2017}]{2017ApJ...838..153K}
{Kato} M.,  {Saio} H.,   {Hachisu} I.,  2017, \mn@doi [\apj]
  {10.3847/1538-4357/838/2/153}, \href
  {https://ui.adsabs.harvard.edu/abs/2017ApJ...838..153K} {838, 153}

\bibitem[\protect\citeauthoryear{{Lloyd}, {O'Brien}  \& {Bode}}{{Lloyd}
  et~al.}{1997}]{1997MNRAS.284..137L}
{Lloyd} H.~M.,  {O'Brien} T.~J.,   {Bode} M.~F.,  1997, \mn@doi [\mnras]
  {10.1093/mnras/284.1.137}, \href
  {https://ui.adsabs.harvard.edu/abs/1997MNRAS.284..137L} {284, 137}

\bibitem[\protect\citeauthoryear{{Metzger}, {Hasco{\"e}t}, {Vurm},
  {Beloborodov}, {Chomiuk}, {Sokoloski}  \& {Nelson}}{{Metzger}
  et~al.}{2014}]{2014MNRAS.442..713M}
{Metzger} B.~D.,  {Hasco{\"e}t} R.,  {Vurm} I.,  {Beloborodov} A.~M.,
  {Chomiuk} L.,  {Sokoloski} J.~L.,   {Nelson} T.,  2014, \mn@doi [\mnras]
  {10.1093/mnras/stu844}, \href
  {https://ui.adsabs.harvard.edu/abs/2014MNRAS.442..713M} {442, 713}

\bibitem[\protect\citeauthoryear{{Mohamed}, {Booth}  \&
  {Podsiadlowski}}{{Mohamed} et~al.}{2013}]{2013ASPC..469..323M}
{Mohamed} S.,  {Booth} R.,   {Podsiadlowski} P.,  2013, in {Krzesi{\'n}ski} J.,
   {Stachowski} G.,  {Moskalik} P.,   {Bajan} K.,  eds,  Astronomical Society
  of the Pacific Conference Series Vol. 469, 18th European White Dwarf
  Workshop.. p.~323

\bibitem[\protect\citeauthoryear{{Murphy-Glaysher} et~al.,}{{Murphy-Glaysher}
  et~al.}{2022}]{2022MNRAS.514.6183M}
{Murphy-Glaysher} F.~J.,  et~al., 2022, \mn@doi [\mnras]
  {10.1093/mnras/stac1577}, \href
  {https://ui.adsabs.harvard.edu/abs/2022MNRAS.514.6183M} {514, 6183}

\bibitem[\protect\citeauthoryear{{O'Brien}, {Bode}  \& {Kahn}}{{O'Brien}
  et~al.}{1992}]{1992MNRAS.255..683O}
{O'Brien} T.~J.,  {Bode} M.~F.,   {Kahn} F.~D.,  1992, \mn@doi [\mnras]
  {10.1093/mnras/255.4.683}, \href
  {https://ui.adsabs.harvard.edu/abs/1992MNRAS.255..683O} {255, 683}

\bibitem[\protect\citeauthoryear{{O'Brien}, {Lloyd}  \& {Bode}}{{O'Brien}
  et~al.}{1994}]{1994MNRAS.271..155O}
{O'Brien} T.~J.,  {Lloyd} H.~M.,   {Bode} M.~F.,  1994, \mn@doi [\mnras]
  {10.1093/mnras/271.1.155}, \href
  {https://ui.adsabs.harvard.edu/abs/1994MNRAS.271..155O} {271, 155}

\bibitem[\protect\citeauthoryear{{O'Brien}, {Davis}, {Bode}, {Eyres}  \&
  {Porter}}{{O'Brien} et~al.}{2001}]{2001IAUS..205..260O}
{O'Brien} T.~J.,  {Davis} R.~J.,  {Bode} M.~F.,  {Eyres} S.~P.~S.,   {Porter}
  J.~M.,  2001, in {Schilizzi} R.~T.,  ed.,  International Astronomical Union
  Vol. 205, Galaxies and their Constituents at the Highest Angular Resolutions.
  p.~260

\bibitem[\protect\citeauthoryear{{Page} et~al.,}{{Page}
  et~al.}{2022}]{2022arXiv220503232P}
{Page} K.~L.,  et~al., 2022, arXiv e-prints, \href
  {https://ui.adsabs.harvard.edu/abs/2022arXiv220503232P} {p. arXiv:2205.03232}

\bibitem[\protect\citeauthoryear{{Pagnotta}, {Schaefer}, {Xiao}, {Collazzi}  \&
  {Kroll}}{{Pagnotta} et~al.}{2009}]{2009AJ....138.1230P}
{Pagnotta} A.,  {Schaefer} B.~E.,  {Xiao} L.,  {Collazzi} A.~C.,   {Kroll} P.,
  2009, \mn@doi [\aj] {10.1088/0004-6256/138/5/1230}, \href
  {https://ui.adsabs.harvard.edu/abs/2009AJ....138.1230P} {138, 1230}

\bibitem[\protect\citeauthoryear{{Pequignot}, {Petitjean}  \&
  {Boisson}}{{Pequignot} et~al.}{1991}]{1991A&A...251..680P}
{Pequignot} D.,  {Petitjean} P.,   {Boisson} C.,  1991, \aap, \href
  {https://ui.adsabs.harvard.edu/abs/1991A&A...251..680P} {251, 680}

\bibitem[\protect\citeauthoryear{{Prialnik} \& {Kovetz}}{{Prialnik} \&
  {Kovetz}}{1995}]{1995ApJ...445..789P}
{Prialnik} D.,  {Kovetz} A.,  1995, \mn@doi [\apj] {10.1086/175741}, \href
  {https://ui.adsabs.harvard.edu/abs/1995ApJ...445..789P} {445, 789}

\bibitem[\protect\citeauthoryear{{Raymond}, {Cox}  \& {Smith}}{{Raymond}
  et~al.}{1976}]{1976ApJ...204..290R}
{Raymond} J.~C.,  {Cox} D.~P.,   {Smith} B.~W.,  1976, \mn@doi [\apj]
  {10.1086/154170}, \href
  {https://ui.adsabs.harvard.edu/abs/1976ApJ...204..290R} {204, 290}

\bibitem[\protect\citeauthoryear{{Ritossa}, {Garcia-Berro}  \&
  {Iben}}{{Ritossa} et~al.}{1996}]{1996ApJ...460..489R}
{Ritossa} C.,  {Garcia-Berro} E.,   {Iben} Icko J.,  1996, \mn@doi [\apj]
  {10.1086/176987}, \href
  {https://ui.adsabs.harvard.edu/abs/1996ApJ...460..489R} {460, 489}

\bibitem[\protect\citeauthoryear{{Saha}}{{Saha}}{1921}]{1921RSPSA..99..135S}
{Saha} M.~N.,  1921, \mn@doi [Proceedings of the Royal Society of London Series
  A] {10.1098/rspa.1921.0029}, \href
  {https://ui.adsabs.harvard.edu/abs/1921RSPSA..99..135S} {99, 135}

\bibitem[\protect\citeauthoryear{{Santamar{\'\i}a}, {Guerrero}, {Ramos-Larios},
  {Sabin}, {V{\'a}zquez}, {G{\'o}mez-Mu{\~n}oz}  \&
  {Toal{\'a}}}{{Santamar{\'\i}a} et~al.}{2019}]{2019MNRAS.483.3773S}
{Santamar{\'\i}a} E.,  {Guerrero} M.~A.,  {Ramos-Larios} G.,  {Sabin} L.,
  {V{\'a}zquez} R.,  {G{\'o}mez-Mu{\~n}oz} M.~A.,   {Toal{\'a}} J.~A.,  2019,
  \mn@doi [\mnras] {10.1093/mnras/sty3364}, \href
  {https://ui.adsabs.harvard.edu/abs/2019MNRAS.483.3773S} {483, 3773}

\bibitem[\protect\citeauthoryear{{Santamar{\'\i}a}, {Guerrero}, {Ramos-Larios},
  {Toal{\'a}}, {Sabin}, {Rubio}  \& {Quino-Mendoza}}{{Santamar{\'\i}a}
  et~al.}{2020}]{2020ApJ...892...60S}
{Santamar{\'\i}a} E.,  {Guerrero} M.~A.,  {Ramos-Larios} G.,  {Toal{\'a}}
  J.~A.,  {Sabin} L.,  {Rubio} G.,   {Quino-Mendoza} J.~A.,  2020, \mn@doi
  [\apj] {10.3847/1538-4357/ab76c5}, \href
  {https://ui.adsabs.harvard.edu/abs/2020ApJ...892...60S} {892, 60}

\bibitem[\protect\citeauthoryear{{Santamar{\'\i}a}, {Guerrero}, {Zavala},
  {Ramos-Larios}, {Toal{\'a}}  \& {Sabin}}{{Santamar{\'\i}a}
  et~al.}{2022}]{2022arXiv220213946S}
{Santamar{\'\i}a} E.,  {Guerrero} M.~A.,  {Zavala} S.,  {Ramos-Larios} G.,
  {Toal{\'a}} J.~A.,   {Sabin} L.,  2022, arXiv e-prints, \href
  {https://ui.adsabs.harvard.edu/abs/2022arXiv220213946S} {p. arXiv:2202.13946}

\bibitem[\protect\citeauthoryear{{Schaefer}}{{Schaefer}}{2010}]{2010ApJS..187..275S}
{Schaefer} B.~E.,  2010, \mn@doi [\apjs] {10.1088/0067-0049/187/2/275}, \href
  {https://ui.adsabs.harvard.edu/abs/2010ApJS..187..275S} {187, 275}

\bibitem[\protect\citeauthoryear{{Shara}, {Zurek}, {Williams}, {Prialnik},
  {Gilmozzi}  \& {Moffat}}{{Shara} et~al.}{1997}]{1997AJ....114..258S}
{Shara} M.~M.,  {Zurek} D.~R.,  {Williams} R.~E.,  {Prialnik} D.,  {Gilmozzi}
  R.,   {Moffat} A. F.~J.,  1997, \mn@doi [\aj] {10.1086/118470}, \href
  {https://ui.adsabs.harvard.edu/abs/1997AJ....114..258S} {114, 258}

\bibitem[\protect\citeauthoryear{{Shara} et~al.,}{{Shara}
  et~al.}{2007}]{2007Natur.446..159S}
{Shara} M.~M.,  et~al., 2007, \mn@doi [\nat] {10.1038/nature05576}, \href
  {https://ui.adsabs.harvard.edu/abs/2007Natur.446..159S} {446, 159}

\bibitem[\protect\citeauthoryear{{Shara}, {Mizusawa}, {Wehinger}, {Zurek},
  {Martin}, {Neill}, {Forster}  \& {Seibert}}{{Shara}
  et~al.}{2012}]{2012ApJ...758..121S}
{Shara} M.~M.,  {Mizusawa} T.,  {Wehinger} P.,  {Zurek} D.,  {Martin} C.~D.,
  {Neill} J.~D.,  {Forster} K.,   {Seibert} M.,  2012, \mn@doi [\apj]
  {10.1088/0004-637X/758/2/121}, \href
  {https://ui.adsabs.harvard.edu/abs/2012ApJ...758..121S} {758, 121}

\bibitem[\protect\citeauthoryear{{Slavin}, {O'Brien}  \& {Dunlop}}{{Slavin}
  et~al.}{1995}]{1995MNRAS.276..353S}
{Slavin} A.~J.,  {O'Brien} T.~J.,   {Dunlop} J.~S.,  1995, \mn@doi [\mnras]
  {10.1093/mnras/276.2.353}, \href
  {https://ui.adsabs.harvard.edu/abs/1995MNRAS.276..353S} {276, 353}

\bibitem[\protect\citeauthoryear{{Smith}, {Brickhouse}, {Liedahl}  \&
  {Raymond}}{{Smith} et~al.}{2001}]{2001ApJ...556L..91S}
{Smith} R.~K.,  {Brickhouse} N.~S.,  {Liedahl} D.~A.,   {Raymond} J.~C.,  2001,
  \mn@doi [\apjl] {10.1086/322992}, \href
  {https://ui.adsabs.harvard.edu/abs/2001ApJ...556L..91S} {556, L91}

\bibitem[\protect\citeauthoryear{{Soraisam} \& {Gilfanov}}{{Soraisam} \&
  {Gilfanov}}{2015}]{2015A&A...583A.140S}
{Soraisam} M.~D.,  {Gilfanov} M.,  2015, \mn@doi [\aap]
  {10.1051/0004-6361/201424118}, \href
  {https://ui.adsabs.harvard.edu/abs/2015A&A...583A.140S} {583, A140}

\bibitem[\protect\citeauthoryear{{Stanek} \& {Garnavich}}{{Stanek} \&
  {Garnavich}}{1998}]{1998ApJ...503L.131S}
{Stanek} K.~Z.,  {Garnavich} P.~M.,  1998, \mn@doi [\apjl] {10.1086/311539},
  \href {https://ui.adsabs.harvard.edu/abs/1998ApJ...503L.131S} {503, L131}

\bibitem[\protect\citeauthoryear{{Starrfield}, {Truran}, {Sparks}  \&
  {Kutter}}{{Starrfield} et~al.}{1972}]{1972ApJ...176..169S}
{Starrfield} S.,  {Truran} J.~W.,  {Sparks} W.~M.,   {Kutter} G.~S.,  1972,
  \mn@doi [\apj] {10.1086/151619}, \href
  {https://ui.adsabs.harvard.edu/abs/1972ApJ...176..169S} {176, 169}

\bibitem[\protect\citeauthoryear{{Starrfield}, {Sparks}  \&
  {Truran}}{{Starrfield} et~al.}{1976}]{1976IAUS...73..155S}
{Starrfield} S.,  {Sparks} W.~M.,   {Truran} J.~W.,  1976, in {Eggleton} P.,
  {Mitton} S.,   {Whelan} J.,  eds,  IAU Symposium Vol. 73, Structure and
  Evolution of Close Binary Systems. p.~155

\bibitem[\protect\citeauthoryear{{Starrfield}, {Sparks}  \&
  {Shaviv}}{{Starrfield} et~al.}{1988}]{1988ApJ...325L..35S}
{Starrfield} S.,  {Sparks} W.~M.,   {Shaviv} G.,  1988, \mn@doi [\apjl]
  {10.1086/185105}, \href
  {https://ui.adsabs.harvard.edu/abs/1988ApJ...325L..35S} {325, L35}

\bibitem[\protect\citeauthoryear{{Starrfield}, {Bose}, {Iliadis}, {Hix},
  {Woodward}  \& {Wagner}}{{Starrfield} et~al.}{2020}]{2020ApJ...895...70S}
{Starrfield} S.,  {Bose} M.,  {Iliadis} C.,  {Hix} W.~R.,  {Woodward} C.~E.,
  {Wagner} R.~M.,  2020, \mn@doi [\apj] {10.3847/1538-4357/ab8d23}, \href
  {https://ui.adsabs.harvard.edu/abs/2020ApJ...895...70S} {895, 70}

\bibitem[\protect\citeauthoryear{{Starrfield}, {Bose}, {Iliadis}, {Hix},
  {Woodward}  \& {Wagner}}{{Starrfield} et~al.}{2021}]{2021gacv.workE..30S}
{Starrfield} S.,  {Bose} M.,  {Iliadis} C.,  {Hix} W.~R.,  {Woodward} C.~E.,
  {Wagner} R.~M.,  2021, in The Golden Age of Cataclysmic Variables and Related
  Objects V. p.~30 (\mn@eprint {arXiv} {2006.01827})

\bibitem[\protect\citeauthoryear{{Steele} et~al.,}{{Steele}
  et~al.}{2004}]{2004SPIE.5489..679S}
{Steele} I.~A.,  et~al., 2004, in {Oschmann} Jr. J.~M.,  ed.,  \procspie Vol.
  5489, Ground-based Telescopes. pp 679--692

\bibitem[\protect\citeauthoryear{{Takeda}, {Diaz}, {Campbell}, {Lyke},
  {Lawrence}, {Linford}  \& {Sokolovsky}}{{Takeda}
  et~al.}{2022}]{2022MNRAS.511.1591T}
{Takeda} L.,  {Diaz} M.,  {Campbell} R.~D.,  {Lyke} J.~E.,  {Lawrence} S.~S.,
  {Linford} J.~D.,   {Sokolovsky} K.~V.,  2022, \mn@doi [\mnras]
  {10.1093/mnras/stac097}, \href
  {https://ui.adsabs.harvard.edu/abs/2022MNRAS.511.1591T} {511, 1591}

\bibitem[\protect\citeauthoryear{{Toraskar}, {Mac Low}, {Shara}  \&
  {Zurek}}{{Toraskar} et~al.}{2013}]{2013ApJ...768...48T}
{Toraskar} J.,  {Mac Low} M.-M.,  {Shara} M.~M.,   {Zurek} D.~R.,  2013,
  \mn@doi [\apj] {10.1088/0004-637X/768/1/48}, \href
  {https://ui.adsabs.harvard.edu/abs/2013ApJ...768...48T} {768, 48}

\bibitem[\protect\citeauthoryear{{Vaytet}}{{Vaytet}}{2009}]{Vaytet_Thesis}
{Vaytet} N.~M.~H.,  2009, PhD thesis,
  \url{https://www.nbi.dk/~nvaytet/documents/thesis.pdf}

\bibitem[\protect\citeauthoryear{{Vaytet}, {O'Brien}  \& {Bode}}{{Vaytet}
  et~al.}{2007}]{2007ApJ...665..654V}
{Vaytet} N.~M.~H.,  {O'Brien} T.~J.,   {Bode} M.~F.,  2007, \mn@doi [\apj]
  {10.1086/519000}, \href
  {https://ui.adsabs.harvard.edu/abs/2007ApJ...665..654V} {665, 654}

\bibitem[\protect\citeauthoryear{{Vaytet}, {O'Brien}, {Page}, {Bode}, {Lloyd}
  \& {Beardmore}}{{Vaytet} et~al.}{2011}]{2011ApJ...740....5V}
{Vaytet} N.~M.~H.,  {O'Brien} T.~J.,  {Page} K.~L.,  {Bode} M.~F.,  {Lloyd} M.,
    {Beardmore} A.~P.,  2011, \mn@doi [\apj] {10.1088/0004-637X/740/1/5}, \href
  {https://ui.adsabs.harvard.edu/abs/2011ApJ...740....5V} {740, 5}

\bibitem[\protect\citeauthoryear{{Wade}}{{Wade}}{1990}]{1990LNP...369..179W}
{Wade} R.~A.,  1990, in {Cassatella} A.,  {Viotti} R.,  eds, , Vol.~369, IAU
  Colloq. 122: Physics of Classical Novae.
{Springer-Verlag, Berlin, Germany; New York, NY}, p.~179,
  \mn@doi{10.1007/3-540-53500-4_118}

\bibitem[\protect\citeauthoryear{{Walker}}{{Walker}}{1954}]{1954PASP...66..230W}
{Walker} M.~F.,  1954, \mn@doi [\pasp] {10.1086/126703}, \href
  {https://ui.adsabs.harvard.edu/abs/1954PASP...66..230W} {66, 230}

\bibitem[\protect\citeauthoryear{{Wareing}, {O'Brien}, {Zijlstra}, {Kwitter},
  {Irwin}, {Wright}, {Greimel}  \& {Drew}}{{Wareing}
  et~al.}{2006}]{2006MNRAS.366..387W}
{Wareing} C.~J.,  {O'Brien} T.~J.,  {Zijlstra} A.~A.,  {Kwitter} K.~B.,
  {Irwin} J.,  {Wright} N.,  {Greimel} R.,   {Drew} J.~E.,  2006, \mn@doi
  [\mnras] {10.1111/j.1365-2966.2005.09875.x}, \href
  {https://ui.adsabs.harvard.edu/abs/2006MNRAS.366..387W} {366, 387}

\bibitem[\protect\citeauthoryear{{Warner}}{{Warner}}{1995}]{1995CAS....28.....W}
{Warner} B.,  1995, Cambridge Astrophysics Series, \href
  {http://adsabs.harvard.edu/abs/1995CAS....28.....W} {28}

\bibitem[\protect\citeauthoryear{{Webbink}}{{Webbink}}{1984}]{1984ApJ...277..355W}
{Webbink} R.~F.,  1984, \mn@doi [\apj] {10.1086/161701}, \href
  {https://ui.adsabs.harvard.edu/abs/1984ApJ...277..355W} {277, 355}

\bibitem[\protect\citeauthoryear{{Whelan} \& {Iben}}{{Whelan} \&
  {Iben}}{1973}]{1973ApJ...186.1007W}
{Whelan} J.,  {Iben} Icko J.,  1973, \mn@doi [\apj] {10.1086/152565}, \href
  {https://ui.adsabs.harvard.edu/abs/1973ApJ...186.1007W} {186, 1007}

\bibitem[\protect\citeauthoryear{{Williams}, {Woolf}, {Hege}, {Moore}  \&
  {Kopriva}}{{Williams} et~al.}{1978}]{1978ApJ...224..171W}
{Williams} R.~E.,  {Woolf} N.~J.,  {Hege} E.~K.,  {Moore} R.~L.,   {Kopriva}
  D.~A.,  1978, \mn@doi [\apj] {10.1086/156362}, \href
  {https://ui.adsabs.harvard.edu/abs/1978ApJ...224..171W} {224, 171}

\bibitem[\protect\citeauthoryear{{Wilms}, {Allen}  \& {McCray}}{{Wilms}
  et~al.}{2000}]{2000ApJ...542..914W}
{Wilms} J.,  {Allen} A.,   {McCray} R.,  2000, \mn@doi [\apj] {10.1086/317016},
  \href {https://ui.adsabs.harvard.edu/abs/2000ApJ...542..914W} {542, 914}

\bibitem[\protect\citeauthoryear{{Worters}, {Eyres}, {Bromage}  \&
  {Osborne}}{{Worters} et~al.}{2007}]{2007MNRAS.379.1557W}
{Worters} H.~L.,  {Eyres} S.~P.~S.,  {Bromage} G.~E.,   {Osborne} J.~P.,  2007,
  \mn@doi [\mnras] {10.1111/j.1365-2966.2007.12066.x}, \href
  {https://ui.adsabs.harvard.edu/abs/2007MNRAS.379.1557W} {379, 1557}

\bibitem[\protect\citeauthoryear{{Woudt} \& {Ribeiro}}{{Woudt} \&
  {Ribeiro}}{2014}]{2014ASPC..490.....W}
{Woudt} P.~A.,  {Ribeiro} V. A. R.~M.,  eds, 2014, {Stella Novae: Past and
  Future Decades}  Astronomical Society of the Pacific Conference Series Vol.
  490

\bibitem[\protect\citeauthoryear{{Yaron}, {Prialnik}, {Shara}  \&
  {Kovetz}}{{Yaron} et~al.}{2005}]{2005ApJ...623..398Y}
{Yaron} O.,  {Prialnik} D.,  {Shara} M.~M.,   {Kovetz} A.,  2005, \mn@doi
  [\apj] {10.1086/428435}, \href
  {https://ui.adsabs.harvard.edu/abs/2005ApJ...623..398Y} {623, 398}

\makeatother
\end{thebibliography}
\bsp

\label{lastpage}
\end{document}